\let\accentvec\vec
\documentclass[english]{aa} 
\usepackage[utf8]{inputenc}
\usepackage[T1]{fontenc}
\usepackage[]{natbib}
\usepackage{lmodern} 
\usepackage{graphicx}
\usepackage{float}
\usepackage{tabularx}
\usepackage{color}
\usepackage{ulem}
\usepackage{booktabs}
\usepackage{placeins}
\usepackage[squaren, Gray, cdot]{SIunits}

\makeatletter

\let\vec\accentvec

\usepackage{natbib,twoopt}
\usepackage[breaklinks=true]{hyperref} 
\bibpunct{(}{)}{;}{a}{}{,}             
\makeatletter
  \newcommandtwoopt{\citeads}[3][][]{\href{http://adsabs.harvard.edu/abs/#3}%
    {\def\hyper@linkstart##1##2{}%
     \let\hyper@linkend\@empty\citealp[#1][#2]{#3}}}
  \newcommandtwoopt{\citepads}[3][][]{\href{http://adsabs.harvard.edu/abs/#3}%
    {\def\hyper@linkstart##1##2{}%
     \let\hyper@linkend\@empty\citep[#1][#2]{#3}}}
  \newcommandtwoopt{\citetads}[3][][]{\href{http://adsabs.harvard.edu/abs/#3}%
    {\def\hyper@linkstart##1##2{}%
     \let\hyper@linkend\@empty\citet[#1][#2]{#3}}}
  \newcommandtwoopt{\citeyearads}[3][][]%
    {\href{http://adsabs.harvard.edu/abs/#3}
    {\def\hyper@linkstart##1##2{}%
     \let\hyper@linkend\@empty\citeyear[#1][#2]{#3}}}
\makeatother

\makeatletter

\providecommand{\LyX}{L\kern-.1667em\lower.25em\hbox{Y}\kern-.125emX\@}

\DeclareRobustCommand*{\lyxarrow}{%
\@ifstar
{\leavevmode\,$\triangleleft$\,\allowbreak}
{\leavevmode\,$\triangleright$\,\allowbreak}}

\makeatother

\begin{document}

\title{The theoretical pulsation spectra of hot B subdwarfs}
\subtitle{Static and evolutionary STELUM models}

\author{
N. Guyot\inst{1}
\and V. Van Grootel\inst{1}
\and S. Charpinet\inst{2}
\and M. Farnir\inst{1}
\and M.-A. Dupret\inst{1}
\and P. Brassard\inst{3}
}

\institute{Space sciences, Technologies and Astrophysics Research (STAR) Institute, Universit\' e
  de Li\` ege, 19C All\'ee du 6 Ao\^ ut, B-4000 Li\` ege, Belgium \\
\email{valerie.vangrootel@uliege.be}
\and Institut de Recherche en Astrophysique et Plan\'etologie, CNRS, Universit\'e de Toulouse, CNES, 14 Avenue Edouard Belin, 31400 Toulouse, France
\and D\'epartement de Physique, Universit\'e de Montr\'eal, Qu\'ebec H3C 3J7, Canada
}

\offprints{V. Van Grootel}

\date{Received ...; Accepted...}

\abstract
{The {\sl Kepler} and {\sl TESS} space missions have revealed the rich gravity (g-)mode pulsation spectra of many hot subdwarf B (sdB) stars in detail. These spectra exhibit complex behaviors, with some stars exhibiting trapped modes interposing in the asymptotic period sequences of regular period spacing, while others do not.}
{We aim to thoroughly compute theoretical g-mode pulsation spectra, using our current sdB models, useful for future reference when comparing to observations. This also enables us to explore relationships with features of the internal structure of these stars. Such studies provide guidance in conducting future asteroseismic analyses of these pulsators and insights on how to interpret their outcomes.}
{We used our STELlar modeling from the Université de Montréal (STELUM) code to compute static (parametric) and evolutionary models of sdB stars, with different prescriptions for their chemical and  thermal structures. We used our adiabatic  PULSE code to compute the theoretical spectra of g-mode pulsations for degrees of $\ell=1$ to 4 and for periods between 1000 s and 15,000 s, amply covering the range of observed g-modes in these stars.}
{We show that g-mode pulsation spectra and, in particular, the appearance of trapped modes are highly dependent on the chemical and thermal structures in the models as the star evolves,  particularly in the region just above the He-burning core. Depending on the prescriptions and specific evolutionary stage, we observe mainly three types of spectra for mid to high radial-order g-modes (the ones observed in sdB stars): ``flat'' spectra of nearly constant period spacing; spectra with deep minima of the period spacing interposing between modes with more regular spacing (which correspond to trapped modes); and spectra showing a ``wavy'' pattern in period spacing. For the two latter cases, we have identified the region where the modes are trapped in the star.} 
{Detailed comparisons with observed g-mode spectra ought to be carried out next to progress on this issue and constrain the internal structure of core-He burning stars via asteroseismology, in particular, for the region above the He-burning core.}


\titlerunning{The theoretical g-mode pulsation spectra of sdB stars}
\authorrunning{Guyot et al.}

\maketitle

\section{Introduction}
\label{intro}
The subdwarf B (sdB) spectral type refers to stars showing strong and large Balmer lines, no or weak He I lines, and no He II lines. It corresponds to effective temperatures ($T_{\rm eff}$) between 20 000 and 40 000 K and surface gravities (in logarithmic scale; log $g$) in the 5.0 -- 6.2 range \citep[see, e.g.,][]{Saffer1994, 2008ASPC..392...75G,2016PASP..128h2001H}. These stars are identified, from an evolutionary standpoint, to the so-called extreme horizontal branch (EHB), an evolutionary stage of core He-burning (CHeB) stars (\citealt{1986A&A...155...33H}, and references therein). The hot effective temperatures of sdB stars are directly connected to the thinness of their H-rich envelopes ($M_{\rm env} \lesssim 0.02 M_{\odot}$), which were presumably almost entirely lost close to the tip of the red giant branch (RGB) primarily due to binary interactions \citep{2002MNRAS.336..449H,2003MNRAS.341..669H}. The mass distribution of sdB stars is strongly peaked around $\sim 0.47$ M$_{\odot}$ \citep{2012A&A...539A..12F}, which corresponds to the canonical mass for core He-burning ignition through the He-flash at the RGB tip. A dispersion exists however around this value \citep{2022A&A...666A.182S}, which may indicate origins from more massive progenitors \citep[see, e.g.,][]{2019A&A...632A..90C} or more exotic formation channels. The residual envelope of sdB stars is not massive enough to durably sustain H-shell burning, unlike other CHeB stars. Consequently, these stars do not ascend the asymptotic giant branch (AGB) after the CHeB phase. They remain hot and compact, directly evolving towards the white dwarf stage \citep{1993ApJ...419..596D}. This channel is expected to produce a small population of relatively low-mass ($\sim0.4 - 0.5 M_\odot$) C/O core white dwarfs \citep{Saffer1994}.

The sdB stars may exhibit, mainly depending on their $T_{\rm eff}$ and log $g$, two flavors of nonradial pulsations. The first type of pulsations have short periods in the range 60 s to 600 s and amplitudes of a few milli-magnitudes that are identified to low-degree, low-radial-order acoustic (p-)modes. The second type of pulsations have much longer periods, ranging from 45 min to a few hours, usually with lower amplitudes, which corresponds to low-degree, mid- to high-radial-order gravity (g-)modes. The hotter and most compact sdB pulsators preferentially exhibit short-period pulsations (this group is named the V361 Hya or sdBV$_{\rm r}$ stars, originally the EC 14026 stars; \citealt{1997MNRAS.285..640K}), while the cooler and less compact sdB stars show mostly long-period pulsations. This second group, discovered by \citet{2003ApJ...583L..31G}, is referred to as V1093 Her, or sdBV$_{\rm s}$ stars. The sdB stars lying at the ``boundary'' between the two groups in a $\log g - T_{\rm eff}$ diagram (see, e.g., Fig.~1 of  \citealt{2014ASPC..481...19F}) usually show both short and long-period pulsations \citep{2005MNRAS.360..737B,2006A&A...445L..31S} and are known as ``hybrid'' or sdBV$_{\rm rs}$ pulsators. Whilst firm statistics still do not exist on that matter, the vast majority of sdB stars with appropriate $T_{\rm eff}$ and log $g$ do pulsate with g-modes, while less than 10\% of hotter and more compact sdB stars do pulsate with p-modes \citep{2010A&A...513A...6O}. We note, however, that ultra high-precision photometric space-borne observations have revealed that very low-amplitude short periods in cool sdB stars or long periods in hot sdBs are possible \citep[e.g.,][]{2011A&A...530A...3C,2023A&A...669A..48B,2024A&A...686A..65B}, although the dominant flavor of oscillations depends well on the $T_{\rm eff}$/log $g$ of the star. In sdB stars, short-period p-mode pulsations have significant amplitudes mainly in the outermost layers, while long-period g-modes propagate in deep regions of the star, down to the convective He-burning core \citep{2000ApJS..131..223C,2002ApJS..139..487C}. Both p- and g-mode pulsations are thought to be driven by the $\kappa$-mechanism powered by local envelope accumulations of heavy elements, such as iron, due to radiative levitation (\citealt{1996ApJ...471L.103C,1997ApJ...483L.123C,2001PASP..113..775C}; \citealt{2003ApJ...597..518F}; see also \citealt{Jeffery2006,Jeffery2007,2011MNRAS.418..195H,2014A&A...569A.123B}). 

During the nominal {\sl Kepler} mission \citep{2010Sci...327..977B}, 18 pulsating sdB stars were monitored continuously (after the asteroseismology survey phase of Q1-Q5) at a one-minute cadence. Out of these 18 stars, 16 turned out to be g-mode pulsators (\citealt{2010MNRAS.409.1470O,2011MNRAS.414.2860O,2010MNRAS.409.1496R,2011MNRAS.414.2871B}). Asymptotic sequences of nearly equally spaced periods have been reported in 13 of them \citep{2011MNRAS.414.2885R}. As is well known from stellar pulsation theory, the period spacing between g-modes of consecutive radial orders $k$ (and of same degree $\ell$) becomes approximately constant in the limit of high radial orders, hence, the name ``asymptotic theory'' \citep{1980ApJS...43..469T}. This asymptotic theory is valid, stricto sensu, for radiative and chemically homogeneous stars (more accurately, these conditions should apply in the cavity where g-modes propagate). For four g-mode sdB pulsators observed with K2, such asymptotic sequences were also detected, allowing for an identification of nearly all observed g-modes \citep{2021MNRAS.507.4178R}. Nearly constant period spacings were also detected in five g-mode sdB pulsators observed with the Transiting Exoplanet Survey Satellite (TESS) \citep{2021A&A...651A.121U}. In a series of cases, however, isolated modes do not belong to the $\ell=1$ or $\ell=2$ asymptotic sequences and could be interpreted as higher degree modes or as so-called trapped modes. In a sequence of g-modes of same degree $\ell$, such trapped modes typically show a much smaller period difference with their neighbors of adjacent radial orders, leaving a minimum in the period spacing sequence. Formally identifying trapped modes is difficult, as it requires the detection of many modes with contiguous radial orders, which is rarely achieved. It has however been possible for a handful of sdB g-mode pulsators monitored by {\sl Kepler}, in which several trapped modes have been formally identified:  KIC 10553698A \citep{2014A&A...569A..15O}, KIC 10001893 \citep{2017MNRAS.472..700U}, KIC 11558725 \citep{2018MNRAS.474.4709K}, and EPIC 211779126 \citep{2017A&A...597A..95B}. Trapped modes correspond to modes that propagate in a smaller cavity (in which they are ``trapped'') than normal modes. They are affected, through partial wave reflection, by rapid structural or chemical variations (sometimes called structural glitches), for example chemical discontinuities \citep{1992ApJS...80..369B}. This occurs when the structural changes take place on scales comparable to (or smaller than) the local wavelength of the mode (otherwise the variation is not seen as a discontinuity by the wave; \citealt{Charpinet2014a,2015ApJ...805..127C}). Trapped modes therefore carry specific information on the stellar internal structure, in particular on the region where they are trapped and on structural changes that generate this trapping. The regions where g-modes could be trapped in sdB stars have yet to be formally identified. In an early development, \citet{2000ApJS..131..223C,2002ApJS..139..487C} identified  the main culprit for g-mode trapping as the He/H chemical transition zone between the H-rich envelope and the He radiative mantle. However, only g-modes with periods up to 3 000 s (corresponding to rather low radial orders) were investigated at that time in a purely academic exercise. This was before the discovery of actual g-modes pulsators \citep{2003ApJ...583L..31G} showing mid- to high-radial-order modes of longer periods. Extensions of this work to higher radial order g-modes were initiated later-on by \citet{Charpinet2013,Charpinet2014a,Charpinet2014b}, showing the influence of other sources of g-mode trapping due, in particular, to the transition between the He-burning C/O-enriched core and the surrounding He-mantle, or to a potential composition glitch left in the He-mantle by a former He-flash episode \citep{Hu2009}. Concerning other CHeB stars, and in particular red clump giants, an abundant collection of literature, starting with \citet{2015ApJ...805..127C}, has developed around the investigation of the behavior of the period spacings of mixed modes observed in these stars as a function of the structural glitches near the He-burning convective core \citep[and references therein]{Vrard2022}.

\begin{figure}[t]
\centering
\includegraphics[width=0.49\textwidth]{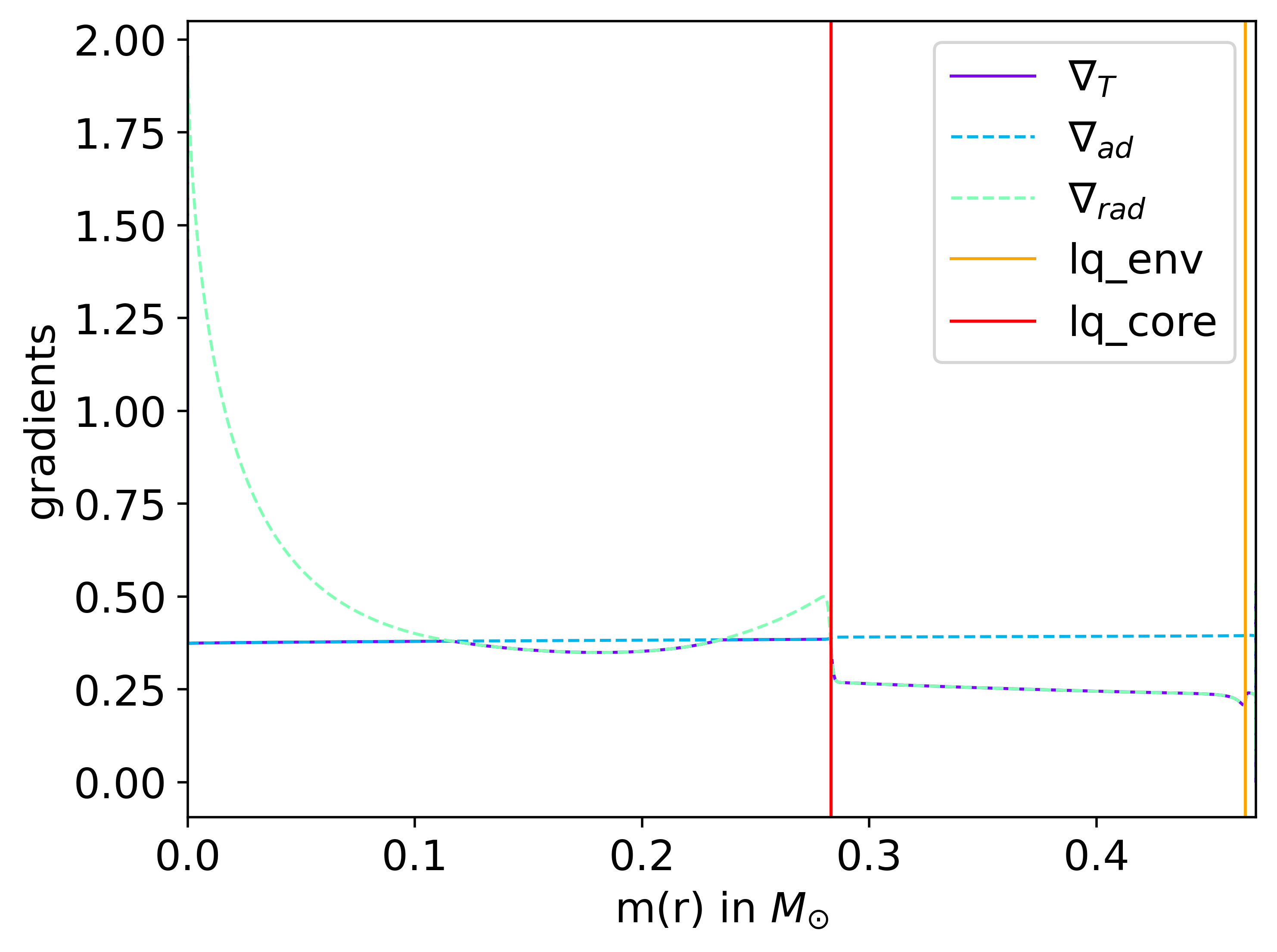}
\caption{Temperature gradients: actual ($\nabla_T$), adiabatic ($\nabla_{\rm ad}$), and radiative ($\nabla_{\rm rad}$) as a function of $m(r)$, the mass contained in a sphere of radius $r$, in a sdB 4G static model with 0.47 $M_{\odot}$, lq\_core~=$-$0.4, lq\_env~=$-$2.0, and X(He)$_{\rm core}=0.05$. The red vertical line depicts the transition lq\_core from the mixed He-C-O core to the radiative He mantle and the orange vertical line indicates the transition lq\_env from the latter to the envelope.} 
\label{fig:sc_pb}
\end{figure}

The thermal and chemical structure of He-burning cores, particularly at their upper boundary, is an longstanding and complex problem of stellar physics \citep[e.g.,][]{1971Ap&SS..10..340C,Dorman1993b}. \citet{2015MNRAS.452..123C} and \citet{2024MNRAS.527.4847B} have provided an excellent description of the problem in their introduction, which we summarize here. At the center, the core is convective due to the very high temperature dependence of the nuclear reactions that fuse He into C and O (the radiative temperature gradient, $\nabla_{\rm rad}$, is larger than the adiabatic temperature gradient, $\nabla_{\rm ad}$; hence, the region is convective according to the Schwarzschild criterion). The temperature and composition profiles between this convective fully-mixed core and the radiative He mantle that surrounds it are more uncertain, due to the complex interplay between physical processes (involving opacities, equations of state, mixing processes, etc.) at work in that specific region. At first, the fully convective core is expected to grow steadily as the carbon and oxygen processed in the core are pushed above the Schwarzschild boundary by even a slight overshoot. The then radiative layers receiving this C/O enrichment become more opaque and eventually turn themselves convective, thus adding to the core expansion. This ``mechanics'' smoothly proceeds during the early stages of the He-burning phase until a point is reached where the radiative gradient develop a minimum at the convective Schwarzschild boundary. At that stage, the core evolution becomes unclear. According to the Schwarzschild criterion, this situation would lead to the appearance of a second convective region (a ``splitting of the core in two'') above the central convective core with a thin radiative layer in between (see Fig.~\ref{fig:sc_pb}, and Fig.~1 of Blouin et al. 2024). This second convective region would imply an acceleration of the stellar material at the top of the core, at the transition to the radiative He mantle (red vertical line on Fig.~\ref{fig:sc_pb}), which is unphysical because the convective flux should be zero (neutral stability) at this transition. Several prescriptions have been proposed to solve this issue. The most classical one is the ``semiconvection treatment,'' where the composition is adjusted layer by layer to produce $\nabla_{\rm rad} = \nabla_{\rm ad}$, which results in a smooth abundance profile, or a partial mixing of elements \citep{1970QJRAS..11...12S,1970AcA....20..195P,1972ApJ...171..309R,1972A&A....20..445S}. In the  Modules for Experiments in Stellar Astrophysics (MESA) stellar evolution code, this prescription is called the convective premixing scheme \citep{2019ApJS..243...10P}. Another solution consists in stopping the growth of the convective core before the second convection region appears \citep{2015MNRAS.452..123C}. This is known as maximal overshoot or predictive mixing in MESA \citep{2018ApJS..234...34P}. Thus far, none of these treatments is considered better than the others. For sdB stars, however, when attempting to reproduce the core masses derived by asteroseismology \citep{2010ApJ...718L..97V,2010A&A...524A..63V,2011A&A...530A...3C,2019A&A...632A..90C}, the premixing scheme seems to be preferred \citep{2021MNRAS.503.4646O}. Yet, the cores produced are still lower in mass than the seismically-derived values. Beyond these two treatments, other possibilities exist in 1D stellar models such as resolving the interaction between convection, overshooting, and diffusion (gravitational settling). This approach is implemented in the  STELlar modeling from the Université de Montréal (STELUM) evolutionary models, which we detail in Sect.~\ref{tools}. 

Further insights into the thermal and chemical structure of He-burning cores has also been obtained from explicit 3D hydrodynamical simulations.  \citet{2024MNRAS.527.4847B} modeled the inner 0.45 $M_{\odot}$ of a 3 $M_{\odot}$ CHeB star, starting with initial states having semiconvection (corresponding to the two treatments available in MESA mentioned above). They found that the semiconvective layers are gradually (but much more rapidly than typical evolutionary timescales) homogenized by overshooting motions from the convective core, which ultimately erase these partially mixed layers completely. In other words, the 3D simulations of \citet{2024MNRAS.527.4847B} suggest that there should be no semiconvection, nor partial mixing of elements developing above convective cores in CHeB stars. Rather, there should be a convective penetration layer, and then the temperature gradient smoothly transitions from $\nabla_{\rm ad}$ to $\nabla_{\rm rad}$ very rapidly, over a fraction of the local pressure scale height $H_P$. A similar conclusion is reached from 3D simulations of \citet{2022ApJ...928L..10A} in the more general context of convective and adjacent radiative zones, including both thermal and compositional buoyancy forces. 
In the context of convective cores of massive stars on the main sequence, this conclusion was also reached from 2D simulations of \citet{2023MNRAS.519.5333B}, showing the formation of a near adiabatic layer above the Schwarzschild boundary of the convective core, and from 2.5D and 3D simulations of \citet{2024A&A...683A..97A}, where similarly, a nearly adiabatic penetration layer is found between the convective core and the radiative layer above it.

Despite these impressive breakthroughs in 3D simulations that provide most needed guidance, 1D stellar models remain necessary to cover long timescale processes, secular evolution, and to provide integral representations of stellar interiors allowing for more direct comparisons with observations of g-modes as well as seismic probing of actual stars. The aim of the present paper is to thoroughly compute theoretical g-mode spectra using our current 1D sdB models, both static and evolutionary. Section~\ref{tools} presents these models, with an emphasis on the chemical and thermal structures deriving from the implemented prescriptions. Sections \ref{static} and \ref{evol} present the g-mode spectra obtained from these models, with an emphasis on the link between the chemical and thermal structures with the g-mode distribution, in particular the appearance of trapped g-modes. We present our summary and conclusions in Sect.~\ref{cc}.

\section{STELUM: Static and evolutionary models, and the pulsation code}
\label{tools}
\subsection{Static models}

To carry out quantitative seismic modeling of sdB and white dwarf pulsators, our group has developed different flavors of static stellar models, which are defined from a set of parameters to directly compute the structure of the star. One of the main drivers behind this strategy is the increased flexibility that such models provide, allowing us to explore a wide range of structural configurations at a manageable computational cost for the identification of one (or several) optimal seismic solution(s) that best match the observed pulsation periods of a given star (for a full description of the seismic forward modeling method applied to sdB stars, see \citealt{2019A&A...632A..90C}; and to white dwarfs, see \citealt{2022FrASS...9.9045G}). A second, more subtle reason is the wish to capture the ``instantaneous'' structure of the star ``seen'' by the pulsation modes propagating in it (the very definition of what a ``seismic model'' is), as independent as possible of uncertainties of stellar evolution that accumulate over time, an especially sensitive and too often disregarded issue for evolved stars. In this framework, reproducing with evolutionary models the internal structure derived from the optimal seismic (static) model is considered as a separated, subsequent problem. Static models have been used successfully to derive global and structural parameters for a handful of white dwarf pulsators \citep[][and references therein]{2022FrASS...9.9045G}. As first claimed by \citet{2018Natur.554...73G}, it is systematically found with this method that white dwarf cores are larger and more oxygen-enriched than predicted by canonical evolutionary models. This is directly connected to the problem of convective boundary mixing in the CHeB phase evoked in Sect.~\ref{intro} and has partly triggered a renewed interest in this problem \citep[see, e.g.,][]{2017RSOS....470192S}. 

\begin{figure}[h!]
\centering
\includegraphics[width=0.49\textwidth]{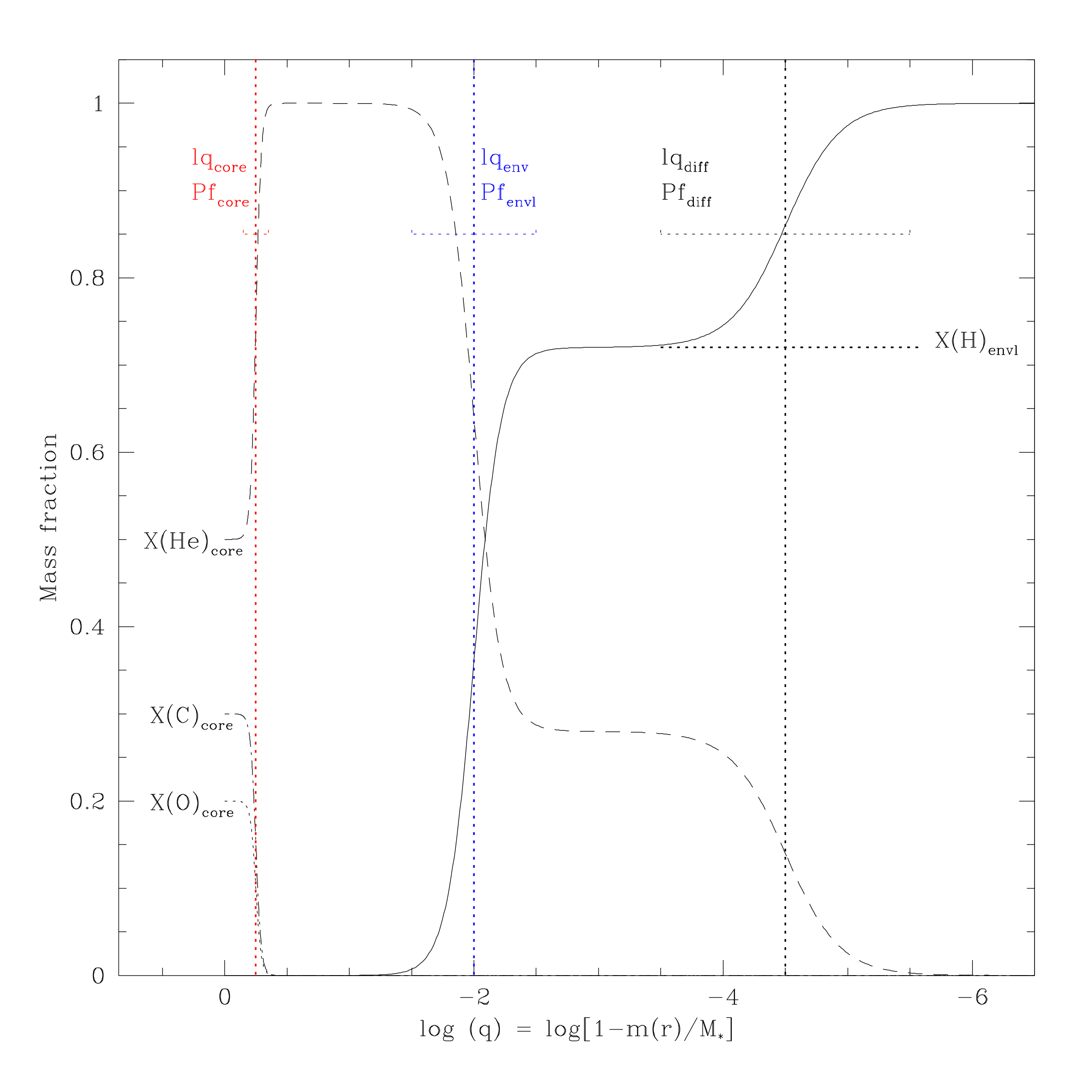}
\caption{Illustration of the model parameters specifying the chemical stratification of 4G/4G+ models (see text for a full description of these parameters). H, He, C, and O mass fractions are shown as a function of logarithmic fractional mass, defined as $\log (q)$, with plain, dashed, dot-dashed, and dotted curves, respectively.} 
\label{fig:model4Gparameters}
\end{figure}

For sdB stars, historical background and detailed description of the input physics implemented in static models are reported in \citet{2013A&A...553A..97V}. The flavor of static models that we consider in this paper are complete (from surface to center) stellar structures assumed to be in strict thermal and mechanical equilibrium (i.e., the luminosity is exactly balanced with the core He-burning energy production rate, plus the very small contribution of H-burning at the base of the H-rich envelope for the models with the most massive envelopes). They are the same as those used and described in \citet[][see their Sect.~3.2]{2019A&A...632A..90C}. We often refer to these models as the fourth-generation (4G) models, following a series of improvements that were implemented into them over the past two decades. For instance, although it is of secondary importance for the pulsation spectra of g-modes in retrospect, we introduced a double-layered envelope in our migration from 3G to 4G static models, in which a pure H-envelope is defined on top of a mixed He+H layer. The motivation at that time was to account for the fact that gravitational settling of helium does not have time to fully separate hydrogen from helium within the evolutionary lifetime for sdB stars with the thickest envelopes (that is the coolest ones, which are also the typical g-mode pulsators).   
We recall here the primary parameters necessary to define a static model (see Fig.~\ref{fig:model4Gparameters})\footnote{We extensively use the logarithmic fractional mass, defined as $\log (q) \equiv \log (1-m(r)/M_*$), as the radial coordinate to represent the models. We recall that $\log (q)=0$ is the center of the star and $\log (q)=-\infty$ is the surface. In practice, models are computed up to $\log (q)=-14.5$ for static models, and $\log (q)=-16.0$ for evolutionary models.}:
\begin{itemize}
\setlength\itemsep{0.5em}
\item the total mass of the star, $M_*$.
\item the mass contained in the core, $\log (M_{\rm core}/M_*)$ (hereafter ``lq\_core''). "Core" refers to the central region of He, C and O composition, up to the chemical transition to the pure He mantle. lq\_core represents then the core-mantle transition, including in figures that follow.
\item the chemical composition in the core (with X(He)$_{\rm core}$ + X(C)$_{\rm core}$ + X(O)$_{\rm core}$ + $Z$ = 1.0). We fixed $Z$ at 0.0134 in the present study.
\item the mass contained in the envelope (including the mixed He+H and pure H layers), $\log (M_{\rm env}/M_*)$ (hereafter, ``lq\_env''). lq\_env represents then the mantle-envelope transition, including in figures that follow. We note that for completeness that the stellar envelope incorporates a nonuniform iron distribution computed assuming equilibrium between radiative levitation and gravitational settling (see, e.g., \citealt{1997ApJ...483L.123C}).
\item  the mass of the pure H layer forming the upper envelope, $\log (M_{\rm H}/M_*)$ (hereafter ``lq\_diff''). We kept this parameter fixed at the value lq\_diff = $-$4.5 in what follows.
\item X(H)$_{\rm envl}$ specifies the abundance of H in the mixed H-He layer at the bottom of the envelope, here kept fixed at 0.715.
\item The shapes (or steepness, or extent) of chemical transitions (respectively from He-C-O core to He radiative mantle, He radiative mantle to He+H envelope, and He+H envelope to the pure H layer) are controlled by additional parameters called Pf$_{\rm core}$, Pf$_{\rm envl}$ and Pf$_{\rm diff}$. They have been kept fixed in this paper, calibrated at values derived from standard evolutionary models without diffusion or semiconvection.
\item Three additional parameters can also be used (not represented in Fig.~\ref{fig:model4Gparameters}). They are associated to the potential carbon pollution of the otherwise pure He radiative mantle resulting from the He-flash: ``flash\_c'', the amount of C in the He mantle (thought to be about a few percents), the mass of the polluted mantle (``lq\_flash''), and the profile factor Pf$_{\rm flash}$. We did not consider such a C pollution in this paper. 
\end{itemize}

With a model being fully defined from these parameters, all other secondary quantities, such as the effective temperature, surface gravity, or stellar radius, can be derived from the computation of the model in hydrostatic and thermal equilibrium. Finally, it is important to realize that static models do not explicitly implement overshooting or semiconvection phenomena as is done in evolution models. The convective or radiative character of a layer is determined according to the Schwarzschild criterion, given a chemical composition. As a consequence, the thermal and chemical stratification in 4G static models are decoupled. The chemical stratification is determined as described above, in which in particular the region below the specified (by parameter lq\_core) core-mantle transition is assumed to be fully mixed (He-C-O fully mixed core). The thermal stratification is determined by the Schwarzschild criterion (for convective region, the Mixing-Length Theory in its ML1/$\alpha=$1.0 version is used). Consequently, we can have for some configurations a radiative region below lq\_core, or the appearance of the ``splitting of the core in two,'' with a second convection zone above the central convective core and a narrow radiative inter layer (see the introduction to this paper and Fig.~\ref{fig:sc_pb}). To remedy this situation that can be unphysical, we have recently introduced a new flavor of static structures, the ``4G+'' models, in which the region below the core-mantle transition is treated as fully convective, by imposing $\nabla_T = \nabla_{\rm ad}$ in that region. We explore in Sect.~ \ref{static} the g-mode spectra from both the 4G and 4G+ models.

\subsection{Evolutionary models}
\label{description_evol}
Following the first seismic modeling of g-mode sdB pulsators using static models \citep{2010ApJ...718L..97V,2010A&A...524A..63V,2011A&A...530A...3C}, we felt the need to develop evolutionary sdB models as well to compare the structures and associate an age to the derived central He abundance X(He)$_{\rm core}$, among other applications. These evolutionary models were developed by one of us (P.B.), and are now part of a full package called STELUM, which is a modular and integrated package mostly written in Fortran 90 that provides codes to compute static and evolutionary models, adiabatic and non-adiabatic pulsations, and other utilities such as opacity tables \citep{2015ASPC..493..125B}. STELUM is (specifically but not exclusively) designed for modeling white dwarfs and hot subdwarfs, starting from the zero-age EHB. The STELUM evolutionary code is described in \citet{2022ApJ...927..128B}, with details about the numerical scheme and its implementation, input physics, modeling of atmospheric layers, and the transport of chemical species, among others. For the latter, STELUM implements several transport phenomena (that are considered simultaneously): chemical diffusion, gravitational settling, thermal diffusion, ordinary convection, convective overshoot, semiconvection, thermohaline convection, stellar winds, and external accretion (only the last three are not directly relevant in the context of the present paper). These mechanisms are treated as time-dependent diffusive processes (there is no instantaneous mixing involved) and the composition changes are fully coupled to the evolution. Importantly in the context of CHeB stars, the code is able to follow phenomena occurring on timescales much shorter than the evolutionary timescale, such as competing diffusion processes. The time-dependent diffusive coefficients used for convection and overshooting can be found in \citet{2022ApJ...927..128B}. In the evolutionary models presented in this paper, there is no explicit process for allowing semiconvection. Instead, a similar effect is obtained by resolving in time the interaction between convection, overshooting and gravitational settling (at the cost of significantly increased computation time). In the partial mixing zone obtained, the behavior is schematically the following: if locally, the matter is convective, overshooting and gravitational settling tend to dilute the local C/O enrichment in a way that the region becomes radiative. If, locally, the matter is radiative, adjacent overshoot and gravitational settling produce a local enrichment in C/O such as the region eventually becomes convectively unstable, as stated by the Schwarzschild criterion. Hence, small radiative/convective layers constantly alternate in that region, slowly pushing the C/O enriched material from the fully convective core upwards, and a partial mixing (semiconvective) zone develops over the course of evolution.  

\begin{figure}[h]
\centering
\includegraphics[width=0.49\textwidth]{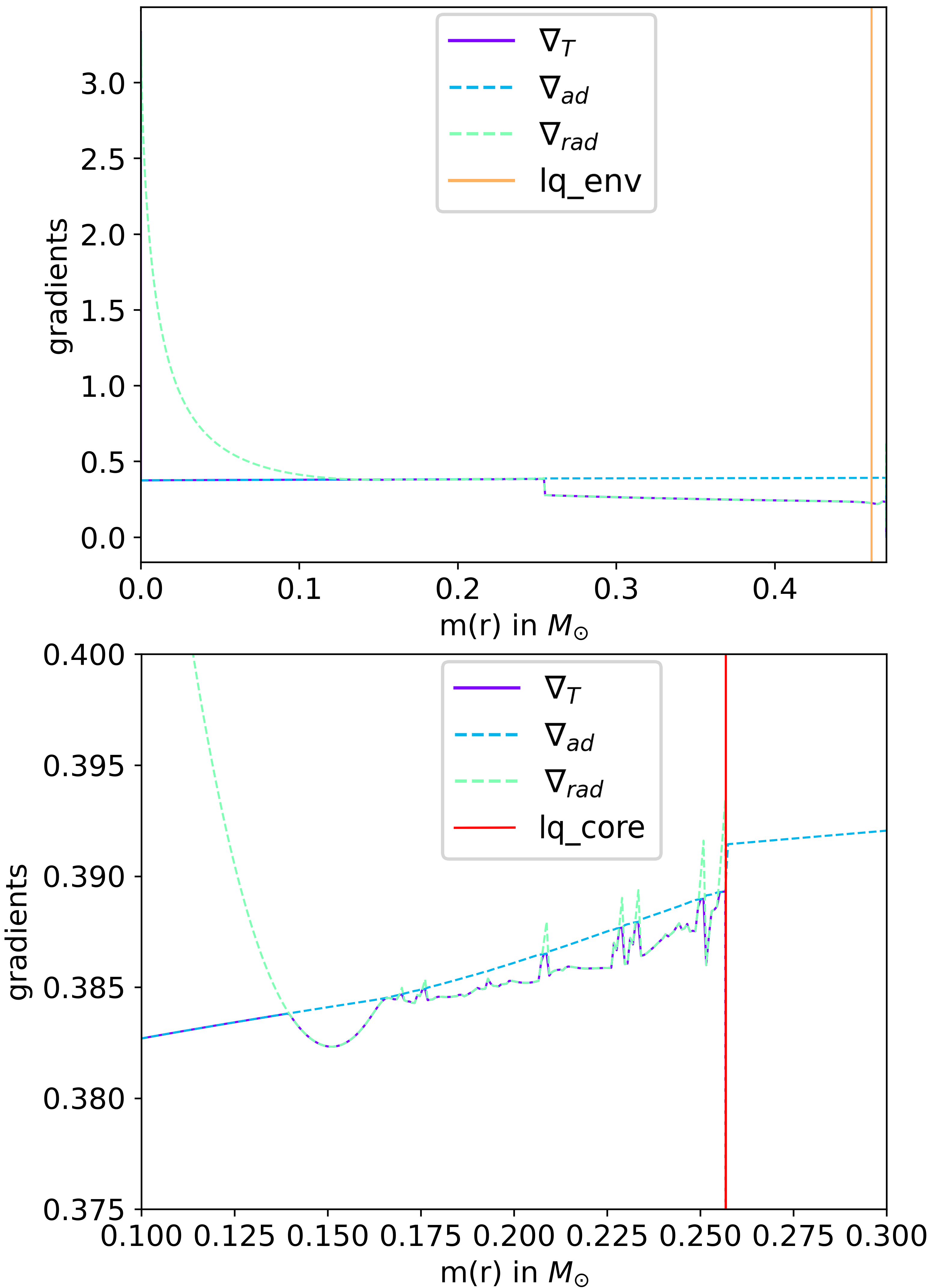}
\caption{Temperature gradients (real $\nabla_T$, adiabatic $\nabla_{\rm ad}$ and radiative $\nabla_{\rm rad}$) as a function of $m(r)$ in a sdB evolutionary model having $M_*=0.47 M_{\odot}$, lq\_env~=$-$2.0, and X(He)$_{\rm core}=0.2$. \textit{Top panel:} Throughout the whole star (the orange vertical line indicates the lq\_env transition). \textit{Bottom panel:}  Zoom on the overshooting and semiconvection zones.} 
\label{fig:grad_evol}
\end{figure}

\begin{figure}[ht!]
\centering
\includegraphics[width=0.49\textwidth]{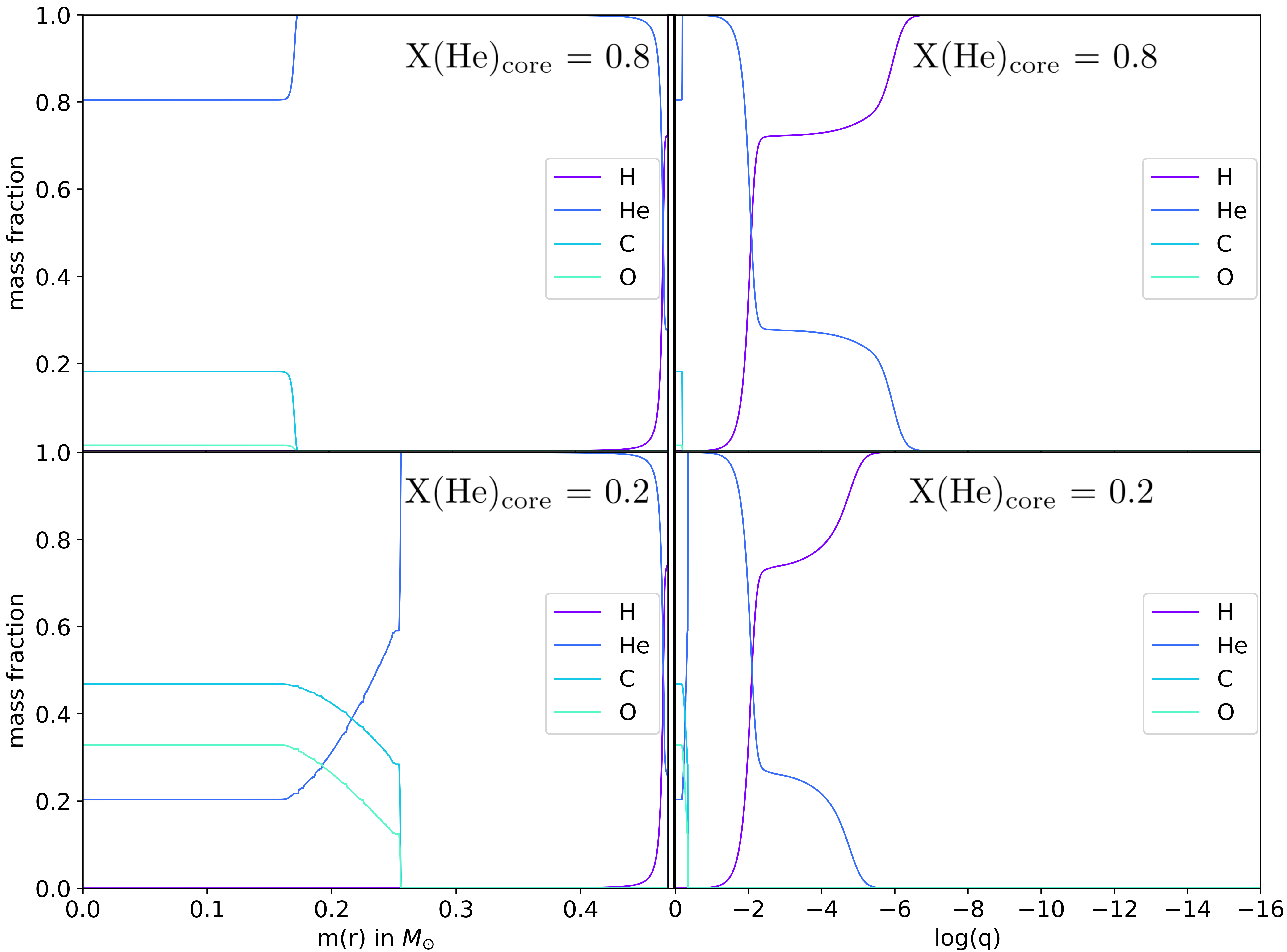}
\caption{Chemical abundances profiles for H, He, C and O as a function of $m(r)$ (left, emphasizing the core region) and $\log (q)$ (right, emphasizing the envelope region) for an evolutionary sdB model having $M_*=0.47 M_{\odot}$, lq\_env~=$-$2.0, and X(He)$_{\rm core}=0.8$ (top; before the onset of semiconvection) and X(He)$_{\rm core}=0.2$ (bottom, with semiconvection).} 
\label{fig:chem_evol}
\end{figure}

These prescriptions lead to the following internal structure, depicted in Fig.~\ref{fig:grad_evol} for the temperature gradients and in Fig.~\ref{fig:chem_evol} for the chemical profiles. We have a radiative overshooting zone around $m(r)\sim 0.15 M_{\odot}$, topped by an alternance of small convective and larger radiative layers that corresponds to the semiconvection zone, as obtained from the interaction convection/diffusion described above. We thus have localized mixing episodes in the semiconvection zone, leaving behind numerous small composition discontinuities, as can be seen in the bottom left panel of Fig.~\ref{fig:chem_evol}. These discontinuities are generally short-lived, subsequent localized convective instabilities eventually erasing them as the evolution proceeds, while new ones appear elsewhere in the semiconvective zone. This behavior may be linked to the possible onset of layering in double-diffusive (semiconvective) conditions \citep{Mirouh2012,Brown2013,Wood2013,Garaud2015}, but this connection remains to 
be explored further.

\begin{figure}[h!]
\centering
\includegraphics[width=0.49\textwidth]{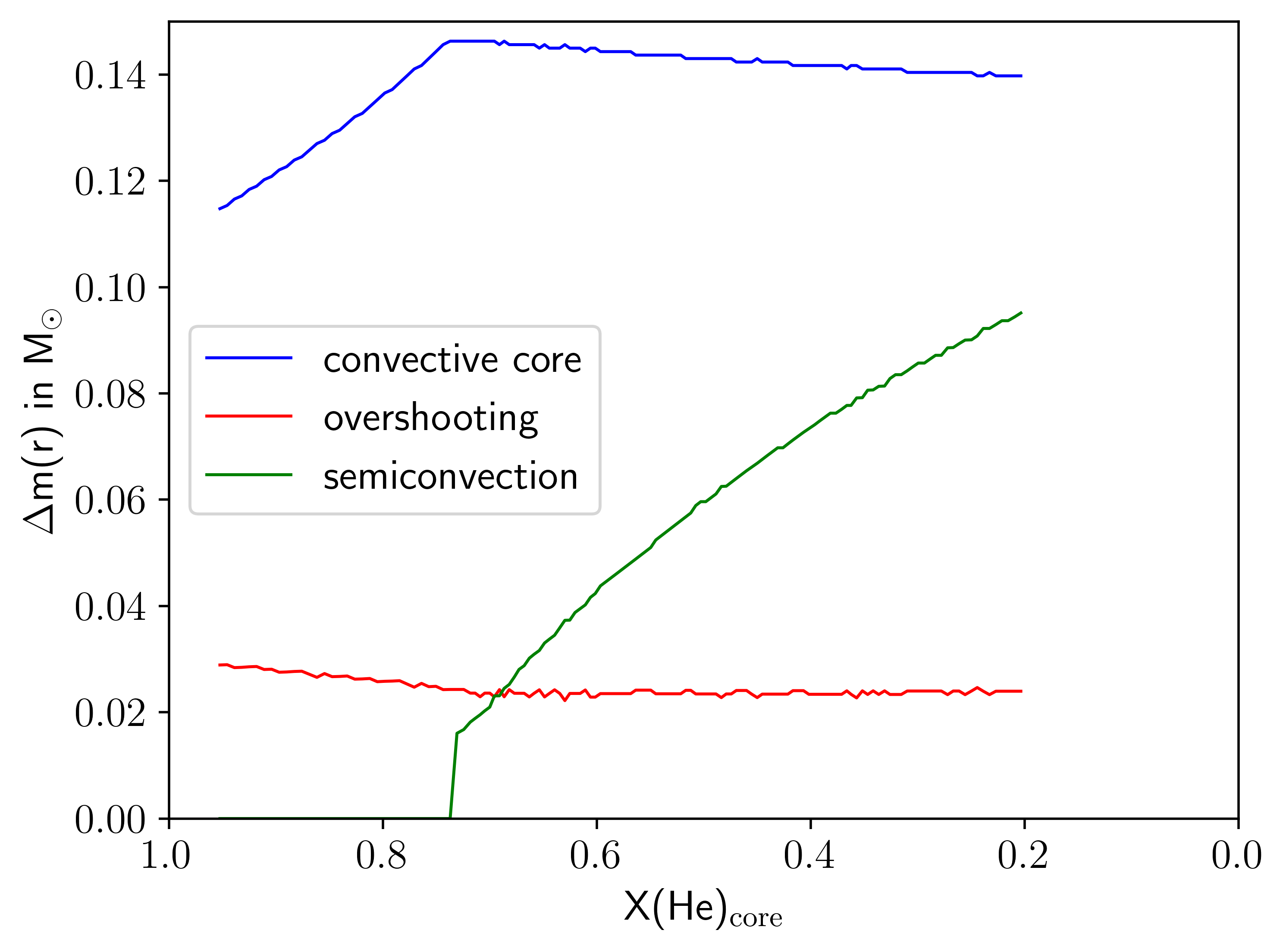}
\caption{Mass (in $M_{\odot}$) contained in the convective core, overshooting zone, and semiconvection zone for an evolutionary sdB sequence having $M_*=0.47 M_{\odot}$ and lq\_env~$=-$2.0, from X(He)$_{\rm core}=0.98$ to 0.20.}  
\label{fig:masses_evol}
\end{figure}

Finally, we show in Fig.~\ref{fig:masses_evol} the masses $\Delta m(r)$ (in $M_{\odot}$) contained within the convective core, the overshooting region, and the semiconvection region along a representative evolutionary sequence for a 0.47 $M_{\odot}$ sdB star. The convective core steadily grows with diminishing core helium abundance until X(He)$_{\rm core} \sim$ 0.7, reaching a maximum value of $\sim$ 0.15 $M_{\odot}$. The onset of semiconvection at X(He)$_{\rm core}\sim$ 0.7 halts the convective core growth, which instead shows a small decrease in mass for the rest of the evolution. Meanwhile, the semiconvection zone steadily grows with decreasing core helium abundance, reaching a value of $\sim$ 0.09 $M_{\odot}$ at X(He)$_{\rm core}=0.2$. The overshooting zone remains relatively constant in 
mass throughout the evolutionary sequence, showing only a slowly decreasing mass from the zero-age EHB to the onset of semiconvection, after which it stays of nearly constant mass (of about 0.025 $M_{\odot}$) for the rest of the evolution.

\subsection{Pulsation computation}
The pulsation properties of a given stellar model are computed using the Montr\'eal pulsation code PULSE \citep{1992ApJS...80..725B,2008Ap&SS.316..107B}, that is now included in the STELUM package. Calculations are carried out in the adiabatic approximation for the present purposes (as we are not interested in studying the stability of the modes in this paper). For static models (Sect.~\ref{static}) as well as evolutionary ones (Sect.~\ref{evol}), we computed theoretical spectra of g-mode pulsations for degrees $\ell=1$ to 4 and for periods between 1 000 s and 15 000 s, with the additional constraint to keep g-modes only up to $k=70$. This limit roughly corresponds to the period cutoff above which the modes are not expected to be observable, as they are no longer reflected back at the surface of the star, their energy leaking out through the atmosphere \citep[see the supplementary material provided in][]{2011Natur.480..496C}.
This range amply covers the g-modes observed in these stars.

\section{The g-mode spectra of static models}
\label{static}
In this section, we present the results from our study of 4G and 4G+ models, showing the pulsation spectra obtained by varying the main model primary parameters (the core helium abundance X(He)$_{\rm core}$, the core mass lq\_core, the envelope mass lq\_env, and the total stellar mass $M_*$). We highlight in particular the presence or not of trapped modes and investigate their trapping cavities.

\begin{figure}[t]
\centering
\includegraphics[width = 0.49\textwidth]{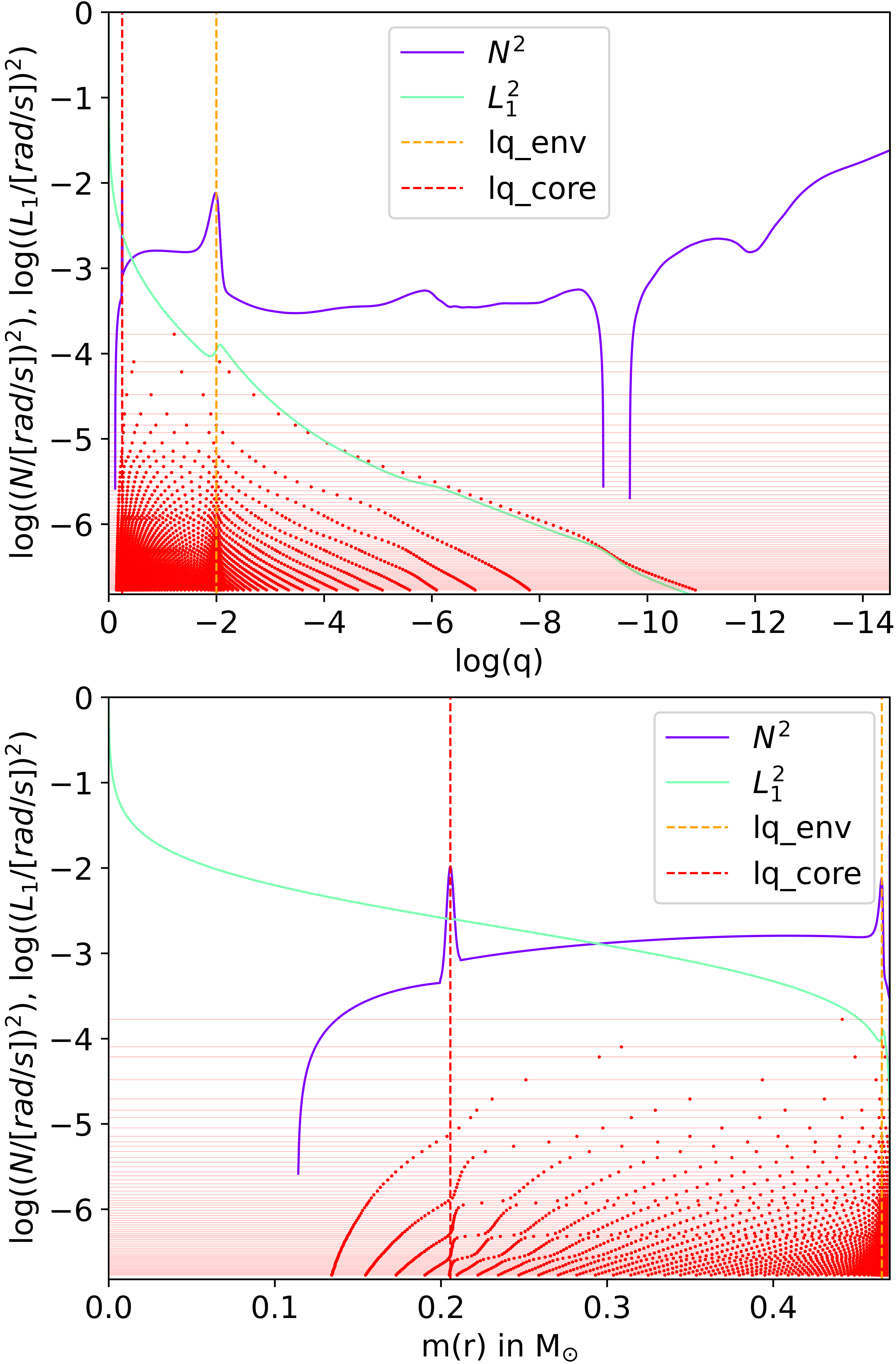}
\caption{Propagation diagram of a 4G model having $M_*=0.47 M_\odot$, lq\_core~$=-$0.25, lq\_env~$=-$2, X(He)$_{\rm core}=0.9$, and modes $\ell$=1 ranging from radial order $k=1$ to 70. The logarithm of the square of the Brunt-Väisälä frequency ($\log N^2$) is depicted in blue and the logarithm of the square of the Lamb frequency for $\ell=1$ ($\log L^2_1$) in green. The red dotted vertical line is the lq\_core (core-mantle) transition, and the orange dotted vertical line is the lq\_env (mantle-envelope) transition. Red horizontal lines are the $\log(\sigma^2)$ of each mode and red dots indicate the radial node positions in $\log(q)$.} 
\label{fig:propa_diag}
\end{figure}

As shown on Fig.~\ref{fig:chem_evol}, our sdB models are stratified, with several chemical transitions. These transitions influence the propagation of g-modes (which generally propagate in cavities with $\sigma^2 < L^2, N^2$, where $\sigma$ is the frequency of the mode, and $L$ and $N$ are the Lamb and Brunt-Väisälä frequencies, respectively), by confining them preferentially in some regions of the star. This behavior is well shown on so-called ``propagation diagrams'', such as the one displayed in Fig.~\ref{fig:propa_diag} for a representative 4G model having $M_*= 0.47 M_\odot$, lq\_core~$=-$0.25, lq\_env~$=-$2, and X(He)$_{\rm core}=0.9$ (top: $\log(q)$ scale, enlightening the envelope regions; bottom: $m(r)$ scale, enlightening the core regions). The Brunt-Väisälä (blue) and Lamb (green, for $\ell=1$) frequencies are represented, as well as the frequencies of each mode $\sigma^2$ from $k=1$ to $k=70$ (horizontal red lines). Each radial node of the radial displacement of a given mode is a red dot. We see that g-modes are mostly confined in the radiative mantle, delimited by the two chemical transitions lq\_core (red vertical line) and lq\_env (orange vertical line), as most nodes are contained within this region. Additionally, the chemical transitions at lq\_core and lq\_env display a ``node pinching'' phenomenon, with node density (and, hence, mode local wavelength) drastically increasing (decreasing) in their vicinity. 

\subsection{General influence of chemical transitions on g-mode spectra}
\label{gen_influence}
\begin{figure*}[t]
\centering
\includegraphics[width=0.70\textwidth]{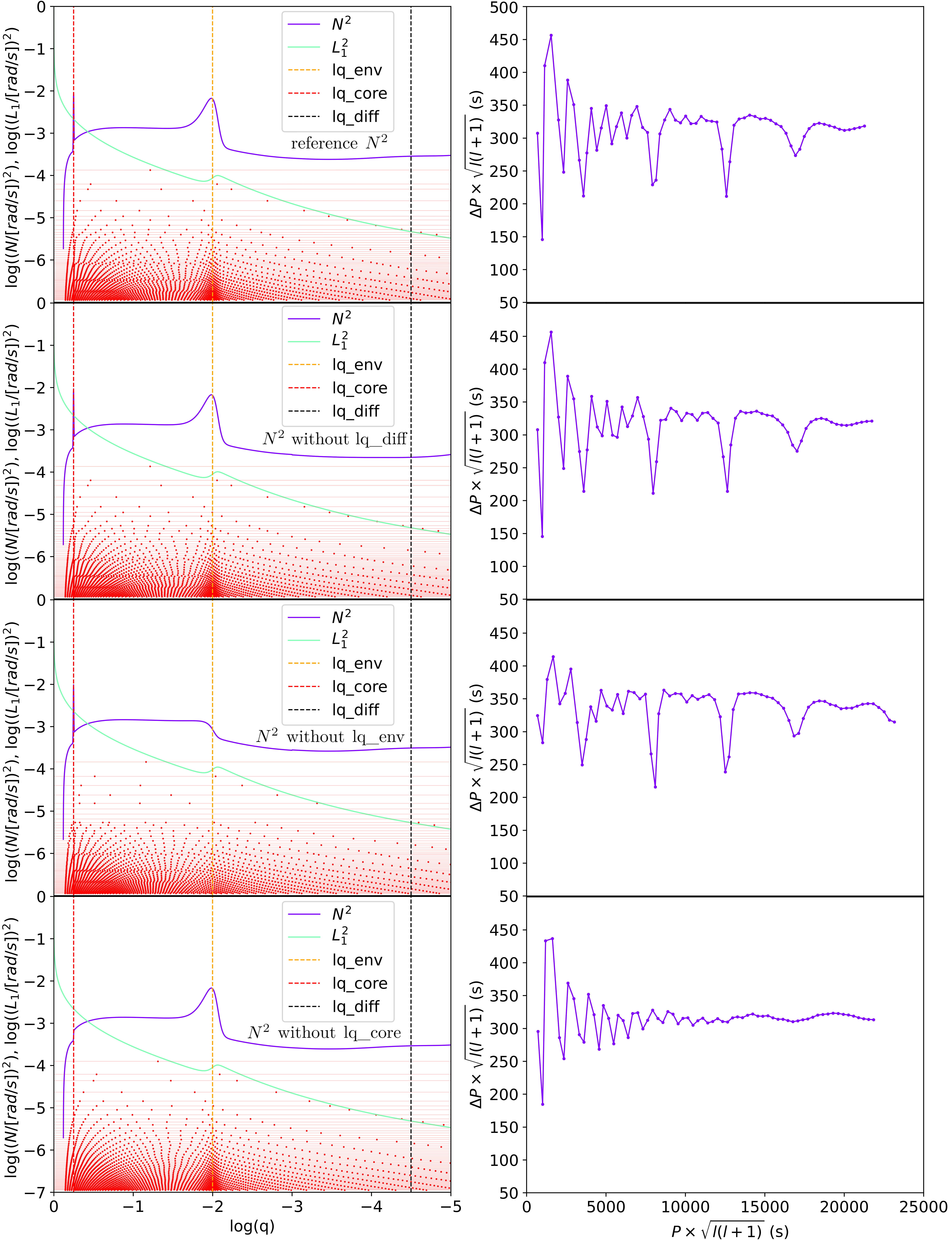}
\caption{Influence of removing chemical gradients at lq\_core, lq\_env, and lq\_diff on a 4G model having $M_*=0.47 M_\odot$, lq\_core~$=-$0.25, lq\_env~$=-$2, and X(He)$_{\rm core}=0.9$. Left panels show the logarithm of the square of the Brunt-Väisälä frequency ($\log N^2$, blue) and the logarithm of the square of the Lamb frequency for $\ell=1$ ($\log L^2_1$, green) between $\log(q)=0$ and $-5$, for the reference model (left panel, first row), when lq\_diff is removed (left panel, second row), when lq\_env is removed (left panel, third row), and when lq\_core is removed (left panel, fourth row). The red dotted vertical line is the lq\_core (core-mantle) transition, the orange dotted vertical line is the lq\_env (mantle-envelope) transition, and the black dotted line is the lq\_diff (envelope of H+He/envelope of pure H) transition. Red horizontal lines are the $\log(\sigma^2)$ of each mode and red dots indicate the radial node positions in $\log(q)$. Right panels show the pulsation spectrum associated to the models of the left panels, presented as the reduced period spacing ($\Delta \Pi = \Delta P \times \sqrt{\ell(\ell+1)}$) as a function of the reduced period ($P\times \sqrt{\ell(\ell+1)}$).}
\label{fig:chem_transi_influence}
\end{figure*}

First, let us examine the general influence of the chemical transitions core-mantle (C-O-He to He at lq\_core), mantle-envelope (He to He+H at lq\_env), and envelope-envelope (He+H to H at lq\_diff) on a representative pulsation spectrum of a 4G model having $M_*= 0.47 M_\odot$, lq\_core~$=-$0.25, lq\_env~$=-$2, and X(He)$_{\rm core}=0.9$. We recall the Brunt-Väisälä frequency as a function of the compositional gradient (also called the Ledoux term):

\begin{equation}
    N^2 = \frac{g^2\rho}{P}\frac{\chi_T}{\chi_\rho}\left(\nabla_{ad} - \nabla_T + B\right),
\;\;\;    \label{BN_freq}
\end{equation}
where $\nabla_T$ is the actual temperature gradient, $\nabla_{ad}$ the adiabatic temperature gradient, and $B$ is the Ledoux term, or compositional gradient, and in which:
\begin{gather}
    \chi_T = \left(\frac{\partial \ln P}{\partial \ln T}\right)_\rho,\\
    \chi_\rho = \left(\frac{\partial \ln P}{\partial \ln \rho}\right)_T,\\
    B = -\frac{1}{\chi_T}\left(\frac{\partial \ln P}{\partial \ln \mu}\right)_{\rho,T}\frac{d \ln \mu}{d \ln P}\;\;\;.
    \label{ledoux_expression}
\end{gather}

Studying the influence of chemical transitions can be done by directly modifying the structure of the static model, in particular by removing the chemical gradient associated to chemical transitions by setting to zero locally the Ledoux term of the Brunt-Väisälä frequency. In Fig.~\ref{fig:chem_transi_influence}, the left panels show propagation diagrams, zoomed on the inner part of the star between $\log (q)=0$ and $-5$ to highlight the regions where chemical transitions are located. From top to bottom panels, we find the reference model (first row), the model where the chemical gradient around lq\_diff~$=-$4.5 is removed (second row), the model where the chemical gradient around lq\_env~$=-$2 is removed (third row), and the model where the chemical gradient around lq\_core~$=-$0.25 is removed (fourth row). The impact on the Brunt-Väisälä frequencies can readily be seen (even though the chemical gradient at lq\_diff is noticeably very small). The right panels of Fig.~\ref{fig:chem_transi_influence} show the pulsation spectrum associated to each model, presented as the reduced period spacing ($\Delta \Pi = \Delta P \times \sqrt{\ell(\ell+1)}$) as a function of the reduced period ($P\times \sqrt{\ell(\ell+1)}$). Comparing first the reference model pulsation spectrum (first row, right panel) to the one without lq\_diff (second row, right panel), it is clear that removing the chemical transition around lq\_diff has practically no impact for all g-modes. This is primarily due to the low weight of these upper layers on the g-modes (in other words, the weight function, see Eq.(\ref{eq:wfi}) below, is small in such regions of low $\rho r^2$ values), and secondly due to the small value of the chemical gradient at lq\_diff (i.e. from He+H, where the X(H)$_{\rm diff}=$ 0.715, to pure H). Comparing now the reference model pulsation spectrum (first row, right panel) to the one without lq\_env (third row, right panel), we show that nullifying the chemical transition around lq\_env primarily affects low- to mid-radial-order modes (from $k=1$ to $k\sim 20$), specifically by strongly decreasing the variations of reduced period spacing observed for those modes in the reference pulsation spectrum. We note that this attenuation effect is stronger for low-order modes ($k=1$ to $k\sim$ 10) than mid-order ones ($k\sim10$ to $k\sim20$). Finally, comparing the reference model pulsation spectrum (first row, right panel) to the one without lq\_core (fourth row, right panel), we find that suppressing the chemical gradient around lq\_core strongly affects mid- to high-order modes ($k\sim 10$ to $k=70$). Indeed, while in the reference model, the pulsation spectrum of high order-modes alternates between local minima of reduced period spacing (this is due to mode trapping, as we  explain later in this section) and modes of more regular period spacing, the pulsation spectrum of the same model with no chemical transition at lq\_core shows a flattened spectrum for high-order modes. We note as well here that mid-order modes are less affected than high-order ones. Essentially, the pulsation spectra can be divided into three parts, where low-order ($k=1$ to $k=10$) modes are sensitive to changes around lq\_env (in line with results from \citealt{2000ApJS..131..223C}), high-order modes are affected by changes in lq\_core, and mid-order modes are affected by both chemical transitions at lq\_env and lq\_core, but to a lessened extent. We recall that observed g-modes in sdB stars correspond to mid- and high-order, low-degree g-modes.

The impact of chemical gradients on the nodes of the pulsation modes is also directly noticeable on the propagation diagrams in Fig.~\ref{fig:chem_transi_influence}. We see that removing the chemical gradient around lq\_env (third row, left panel) directly relaxes the ``node pinching'' seen in propagation diagrams where lq\_env is not removed (first, second and fourth row, left panels). In addition, while removing the chemical gradient around lq\_core only moderately relaxes node pinching around it, it however clearly modifies node behavior in its vicinity. Indeed, we observe smoother transitions between radial nodes of different orders around lq\_core when the chemical gradient around is nullified (fourth row, left panel), while we observe discontinuities between radial nodes for propagation diagrams when this same chemical gradient is kept (first, second and third row, left panels). Notably, those discontinuities in radial nodes directly correspond to minima of reduced period spacing at high orders (hence, as we  explain below, to trapped modes).

These results are in direct line with earlier results from \citet{2000ApJS..131..223C,2002ApJS..139..487C}, which identified the lq\_env transition as responsible for mode trapping (minima of period spacing) in low- to mid-order g-modes.

\subsection{Influence of X(He)$_{\rm core}$ on g-mode spectra}
\subsubsection{4G models}
\label{XHe_4G}

\begin{figure}[t]
\centering
\includegraphics[width=0.49\textwidth]{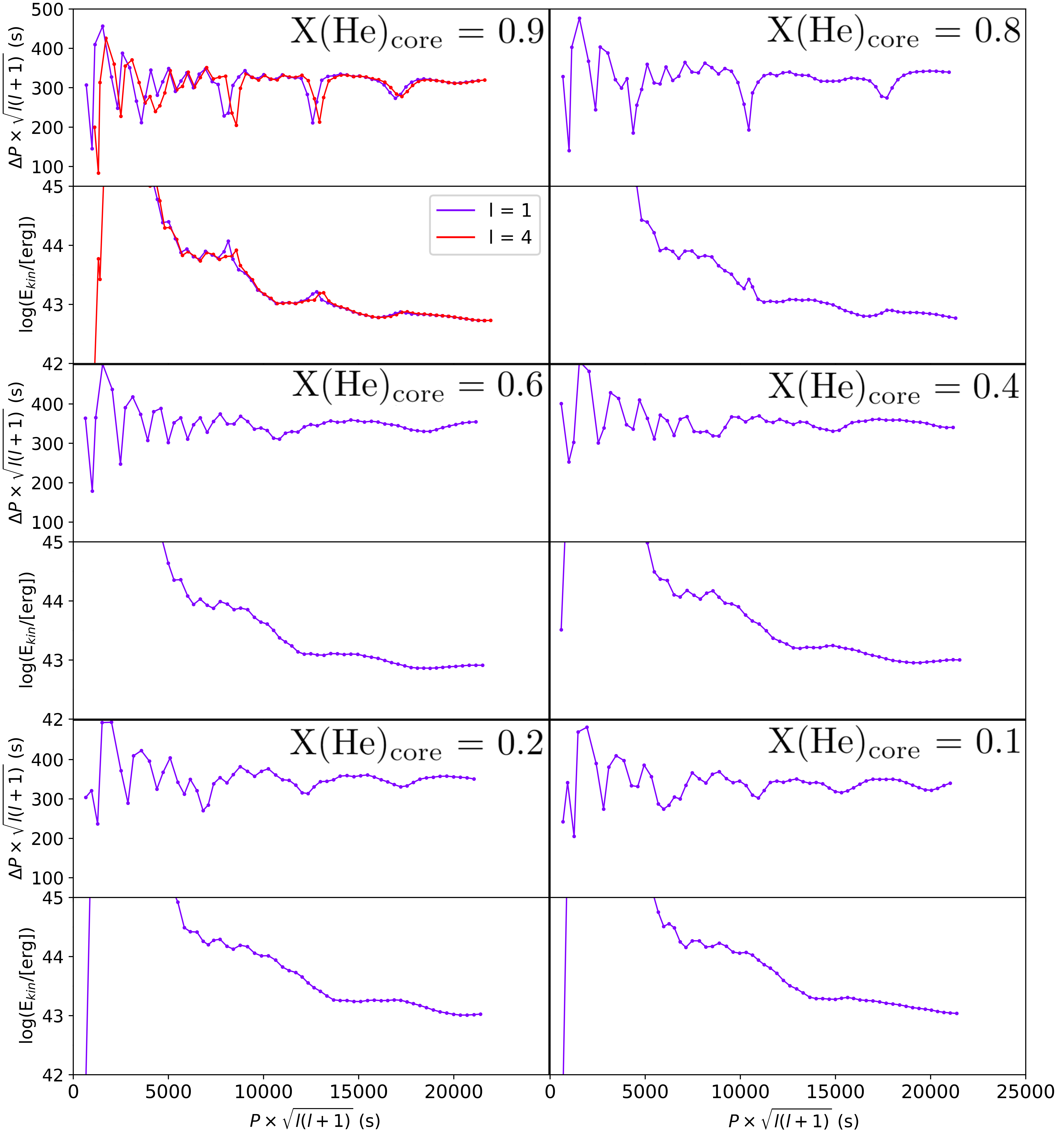}
\caption{Pulsation spectra (top panels) and associated kinetic energies (bottom panels) of a 4G model having $M_*=0.47 M_\odot$, lq\_core~$=-$0.25, lq\_env~$=-$2, at X(He)$_{\rm core}=0.9$, 0.8, 0.6, 0.4, 0.2, and 0.1.}  
\label{fig:4G_hecore}
\end{figure} 

Static models are not dependent of stellar evolution, but we can mimic an evolutionary track by computing various static models with decreasing core helium abundance X(He)$_{\rm core}$ (but, keeping lq\_core fixed). To this end, we defined a reference 4G model of $M_*= 0.47 M_\odot$, lq\_env $= -$2 and lq\_core $= -$0.25, and mimicked the CHeB phase by computing this reference model at X(He)$_{\rm core}$ ranging from 0.985 to 0.005, by steps of 0.005. Figure~\ref{fig:4G_hecore} displays six panels at X(He)$_{\rm core}=0.9$, 0.8, 0.6, 0.4, 0.2 and 0.1 for this reference model. Each panel is divided in two, with on top the pulsation spectrum presented as the reduced period spacing as a function of the reduced period. The bottom panels show the associated kinetic energy, also as a function of the reduced period. We recall here the definition of the kinetic energy of a mode:
\begin{equation}
E_{\rm kin}=\frac{\sigma^2}{2}\int_0^R[\xi_r(r)^2+\ell(\ell+1)\xi_h(r)^2]\rho r^2 dr
\label{eq:ekin}
\end{equation}
with $\xi_r$ and $\xi_h$ the radial and horizontal displacement eigenfunctions (see for example \citealt{2000ApJS..131..223C}), $\rho$ the stellar density and $R$ the stellar radius. As is well known from asymptotic pulsation theory, g-modes of different degree $\ell$ have similar reduced periods in the asymptotic limit, i.e., for high radial orders. We verified it is the case for X(He)$_{\rm core}=0.9$ (top left panel of Fig.~\ref{fig:4G_hecore}), showing $\ell=1$ (blue) and $\ell=4$ (red) pulsation spectra. For the other X(He)$_{\rm core}$ and the rest of the paper, we only show the g-modes of degree $\ell=1$, for the sake of clarity. 

For 4G models and high X(He)$_{\rm core}$, we observe in Fig.~\ref{fig:4G_hecore} a few minima of the reduced period spacing, interposing in the sequence of more regular spacing, in the limit of high radial orders. For X(He)$_{\rm core}\sim 0.6$ and below, the phenomenon attenuates to stabilize into shallower and less selective trapping, with rather a ``wavy'' period spacing, slowly oscillating around a mean value. The periodic signatures observed in Fig.~\ref{fig:4G_hecore} pulsation spectra are clear deviations from the constant period spacing at high radial orders expected from the asymptotic theory. They are supposed to originate from structural changes in the star, such as chemical and thermal transitions \citep{1992ApJS...80..369B}. To study in more detail this behavior and identify the structural changes responsible for it, we made use of the so-called weight functions $F$ (abbreviated ``wfi'' in figures that follow), as well as of the normalized buoyancy radius $\phi$. Weight functions give the contribution of the different regions of the star on the frequencies $\sigma^2$ of the modes, such as \citep{2000ApJS..131..223C}:
\begin{equation}
\sigma^2 \propto \int_0^R F(\xi_r,P',\Phi',r) dr
\label{eq:wfi}
,\end{equation}
with $F(\xi_r,P',\Phi',r) \propto \rho r^2$ (the full expression can be found on Equation (38) of \citealt{2000ApJS..131..223C}; $P'$ and $\Phi'$ being the Eulerian perturbations of, respectively, the pressure and the gravitational potential). These weight functions are arbitrarily normalized to unity at their maximum for each mode of a given model.
For its part, the buoyancy radius allows, when plotting the Brunt-Väisälä frequency as a function of it, to identify the structural changes in the star. The buoyancy radius at a position $r$ in the star is defined as (using the definition from \citealt{2003MNRAS.344..657M} and \citealt{2008MNRAS.386.1487M}):
\begin{equation}
  \Lambda_r^{-1} = \int^r_{r_0} \frac{\lvert N \rvert}{r^\prime}dr^\prime
,\end{equation}
\label{eq:local_Bradius}
where $N$ is the Brunt-Väisälä frequency and $r_0$ is the first radius at which $\log(N^2)$ is defined, which in our case is the top of the convective core. The total buoyancy radius can thus be written as:
\begin{equation}
  \Lambda_0^{-1} = \int^{R}_{r_0} \frac{\lvert N \rvert}{r^\prime}dr^\prime
\label{eq:total_Bradius}
,\end{equation}
with $R$ the stellar radius. The normalized buoyancy radius, $\phi$ = $\Lambda_0/\Lambda_r$, can be used as a new radial coordinate, varying from $0$ (at the top of the convective core) to $1$ (at the surface of the star), and is particularly efficient at highlighting regions of interests such as the lq\_core and lq\_env chemical transitions.

\begin{figure}[h!]
\centering
\includegraphics[width=0.49\textwidth]{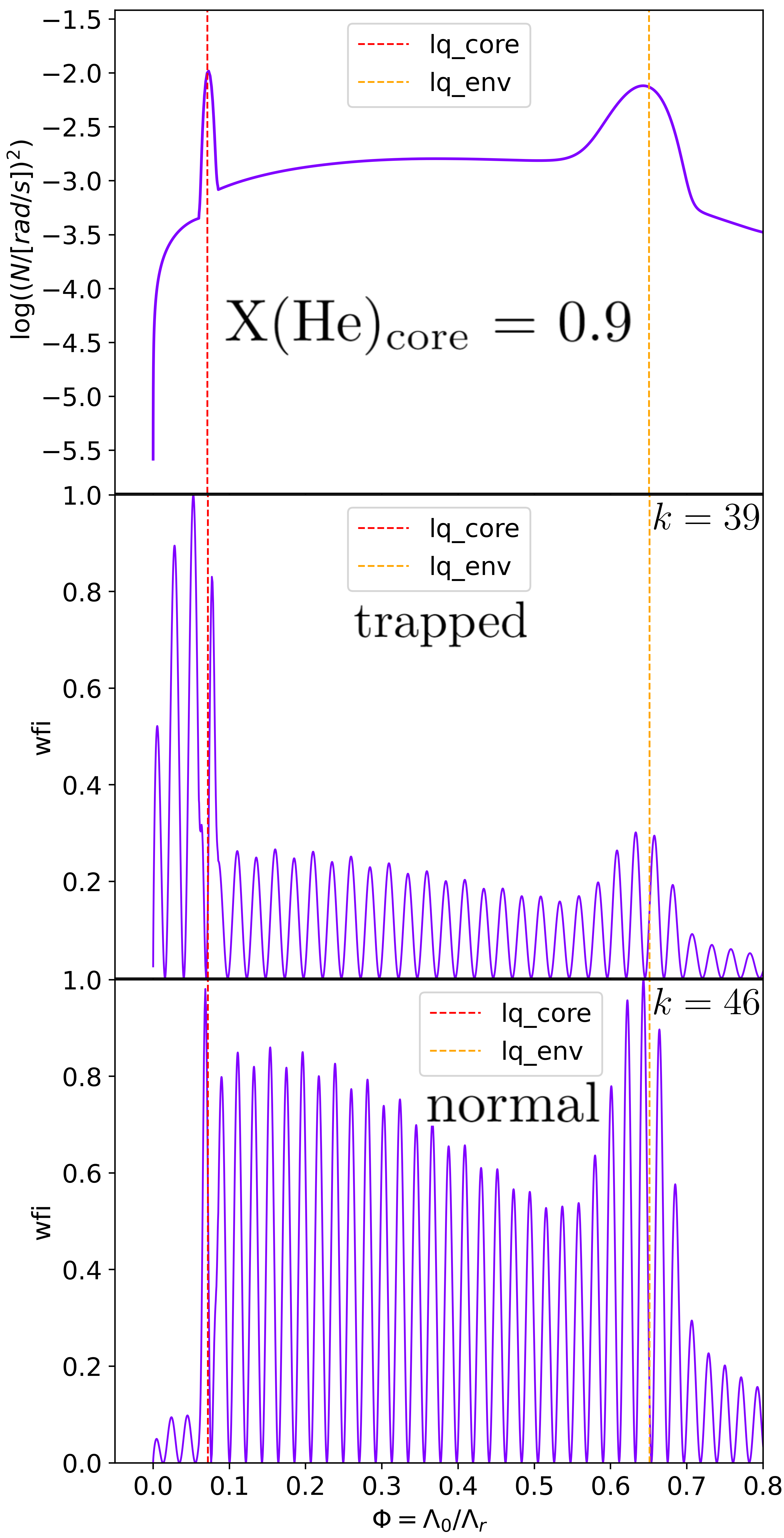}
\caption{\textit{Top panel}: $\log(N^2)$ as a function of the normalized buoyancy radius ($\phi = \Lambda_0/\Lambda_r$) at X(He)$_{\rm core}=0.9$. The red dotted vertical line is the lq\_core (core-mantle) transition, and the orange dotted vertical line is the lq\_env (mantle-envelope) transition. \textit{Middle and bottom panels:} Weight functions (wfi) of modes from a 4G model having $M_*=0.47 M_\odot$, lq\_core~$=-$0.25, lq\_env~$=-$2, and X(He)$_{\rm core}=0.9$. \textit{Middle panel}: Mode trapped in the radiative zone below lq\_core. \textit{Bottom panel}: Normal mode.} 
\label{fig:4G_hecore_wfi_09}
\end{figure}

Figure~\ref{fig:4G_hecore_wfi_09} depicts the Brunt-Väisälä frequency as a function of the normalized buoyancy radius (top panel) at X(He)$_{\rm core}=0.9$, as well as the weight functions of a trapped (a minimum of $\Delta \Pi$, here of radial order $k=39$, middle panel) and a normal mode (a mode in the regular period spacing sequence between two minima, here $k=46$, bottom panel).
It should be stressed that the width of a given transition in terms of $\phi$ is not the physical width of this transition in the star, but rather its contribution to the total buoyancy radius. The weight functions reveal that a trapped mode has a significant amplitude just below the core-mantle transition (red vertical line), while the amplitude of the normal mode there is much lower. On the contrary, in the He mantle (region between red and orange vertical lines), the amplitude of the weight function is lower for the trapped mode than for the normal mode. This allows us to identify the trapping cavity, in this particular 4G model and at X(He)$_{\rm core}=0.9$, to be a small radiative zone located below the core-mantle transition at lq\_core, down to the top of the convective core. This radiative zone can also be clearly identified on Fig.~\ref{fig:4G_hecore_wfi_09} top panel, as the region below the peak associated with the chemical transition at lq\_core. As the modes trapped in the radiative part of the core have a higher amplitude there than the normal modes, they also have a higher kinetic energy than the normal modes due to the $\rho r^2$ term in the $E_{\rm kin}$ definition (Eq.~(\ref{eq:ekin})). This is readily observed in the panels that show the kinetic energies in Fig.~\ref{fig:4G_hecore}. This shows a notable contrast  with the situation seen for white dwarfs, where some g-modes are trapped in the envelope and therefore correspond to local minima of kinetic energy (\citealt{1992ApJS...80..369B}, \citealt{2000ApJS..131..223C,2002ApJS..139..487C}). Let us note that the modes are not perfectly trapped in the radiative part of the core, since they have non-negligible amplitudes in the mantle and even in the envelope as well. The core-mantle transition is responsible for partial mode reflection only. This is true for any mode, actually, and the mode trapping phenomenon should be viewed as a continuum, being more or less selective depending on the mode. Modes that are almost perfectly trapped in a deep region of the star have little chance to be observed (see an example on Fig.~\ref{fig:radiative_arch} below), but modes partly trapped and having non-negligible amplitudes in the upper parts as well, such as the one presented in Fig.~\ref{fig:4G_hecore_wfi_09}, might be observable.

\begin{figure}[t]
  \centering
  \includegraphics[width=0.49\textwidth]{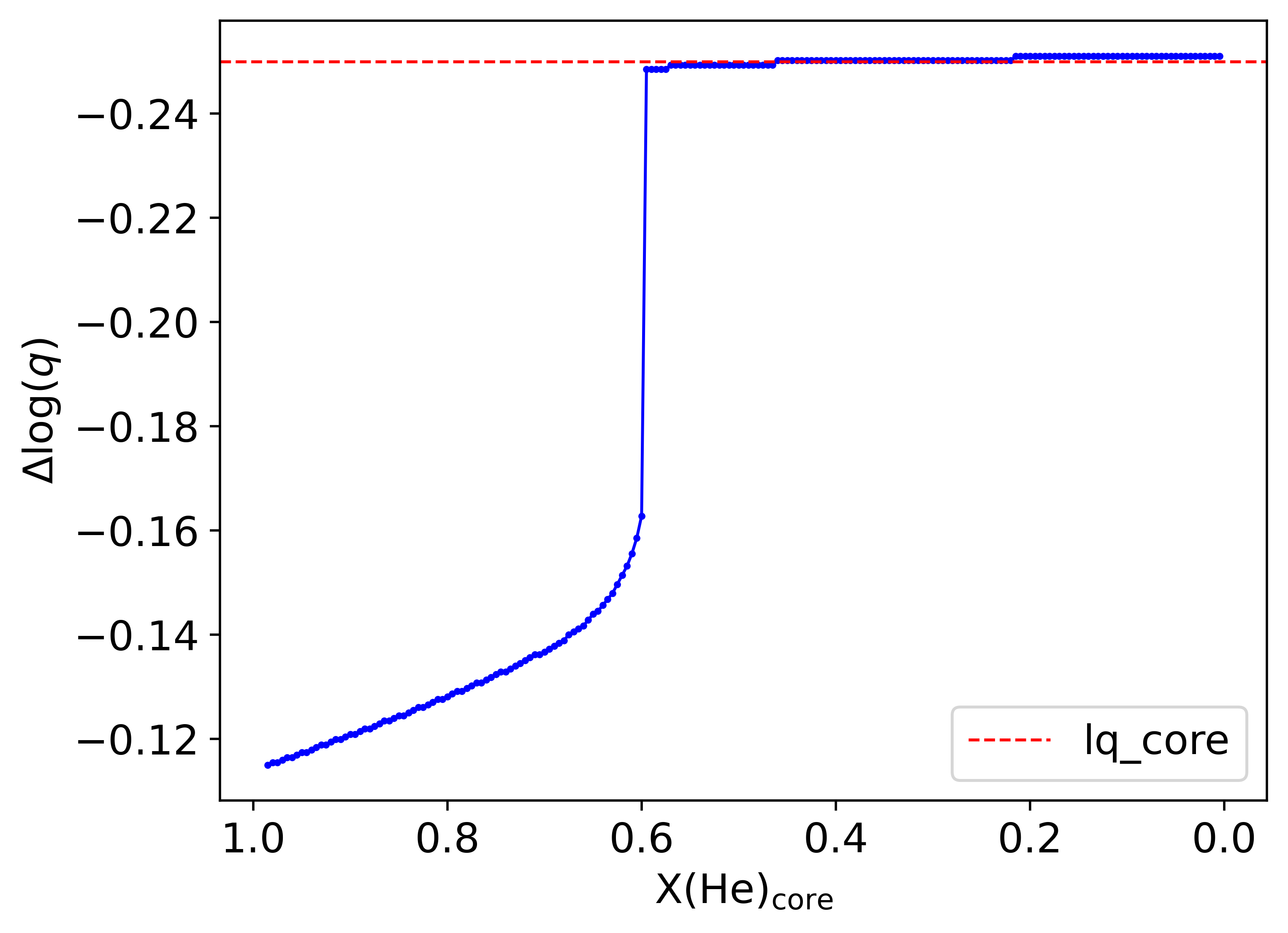}
  \caption{Convective core mass in $\log(q)$ scale as a function of X(He)$_{\rm core}$. The convective core grows until X(He)$_{\rm core}= 0.6$ where it reaches the lq\_core transition, fixed at lq\_core~$=-$0.25 (red dashed line). Each blue dot represents a 4G model having $M_*=0.47 M_\odot$, lq\_core~$=-$0.25, lq\_env~$=-$2, at corresponding X(He)$_{\rm core}$.}
  \label{fig:4G_core_growth}
\end{figure}

The existence of a radiative zone in the core where g-modes can propagate is a direct consequence of the decoupling of the thermal (set according to the Schwarzschild criterion) and chemical (fully mixed core up to lq\_core) structures in 4G models. We show in Fig.~\ref{fig:4G_core_growth} the mass in $\log(q)$ of the convective part of the core (that is, the mass of a given zone with respect to $M_\ast$, and of boundaries determined by the Schwarzschild criterion) as a function of X(He)$_{\rm core}$. The convective core grows from $\log(q) \sim -0.11$ at  X(He)$_{\rm core}\sim 0.985$ to $\log(q) =-0.25$ at X(He)$_{\rm core}\sim 0.6$, where it reached the fixed lq\_core transition. In other words, from X(He)$_{\rm core}\sim 0.985$ to $\sim0.6$, we have a growing convective part of the core, while conversely the radiative part shrinks. At X(He)$_{\rm core}\sim0.6$ (for lq\_core fixed to $-$0.25) and below, the core is fully convective, and g-modes cannot propagate in the core anymore. The presence, strength and number of trapped modes in our reference 4G model is directly linked to the mass (``size'' in $\log(q)$) of the radiative cavity. When the convective core reaches lq\_core, the radiative cavity does not exist anymore, and we observe flattened pulsation spectra showing rather a wavy pattern (Fig.~\ref{fig:4G_hecore}, X(He)$_{\rm core}=0.6$ to 0.1). 

\begin{figure}[t]
  \centering
  \includegraphics[width=0.49\textwidth]{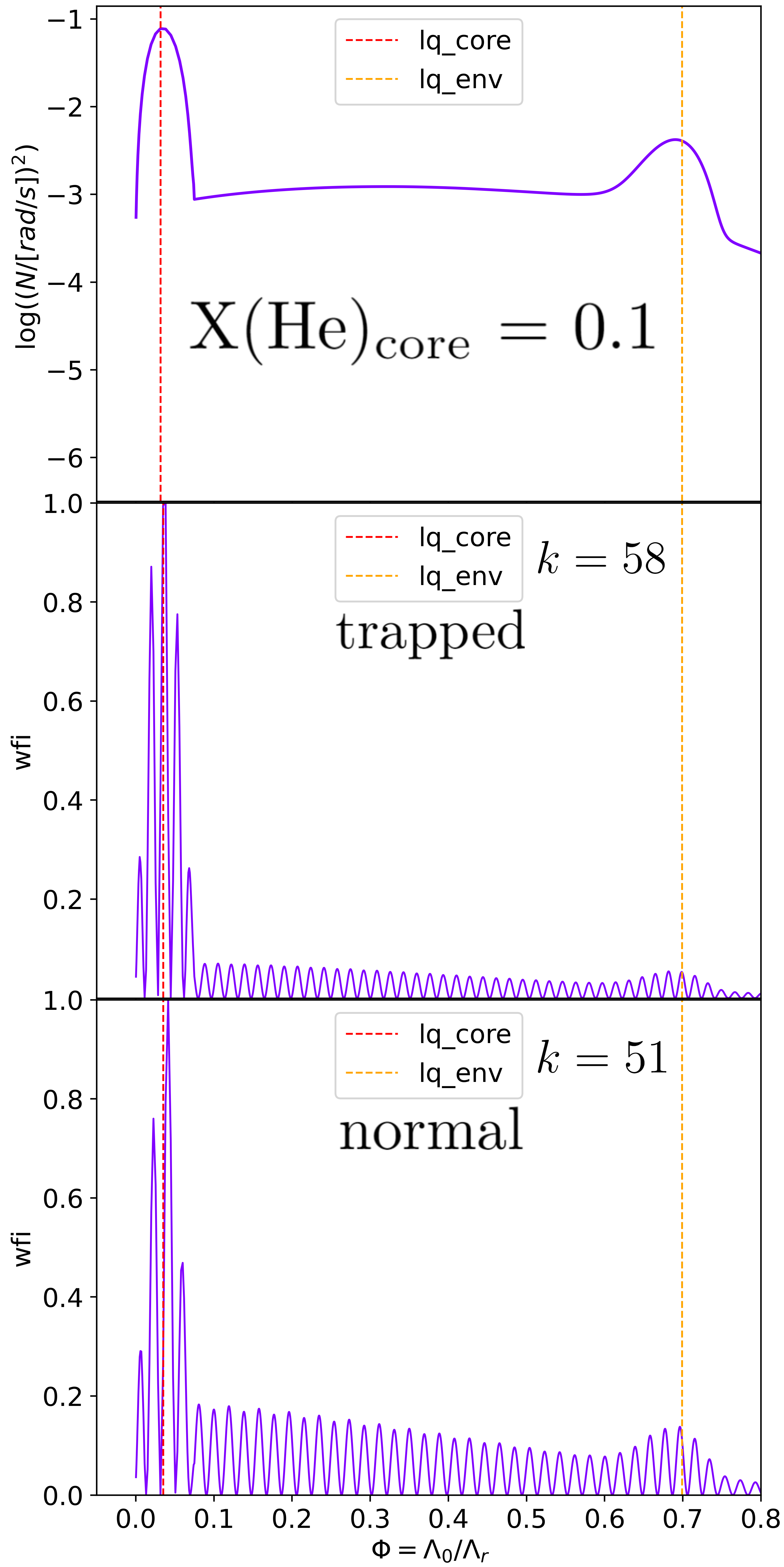}
  \caption{\textit{Top panel}: $\log(N^2)$ as a function of the normalized buoyancy radius ($\phi = \Lambda_0/\Lambda_r$) at X(He)$_{\rm core}=0.1$ for a 4G model having $M_*=0.47 M_\odot$, lq\_core~$=-$0.25, lq\_env~$=-$2, and X(He)$_{\rm core}=0.1$. The red dotted vertical line is the lq\_core (core-mantle) transition, and the orange dotted vertical line is the lq\_env (mantle-envelope) transition. \textit{Middle and bottom panels:} weight functions (wfi) of two modes: a mode trapped in the width of the chemical transition at lq\_core (middle) and a normal mode (bottom).}
  \label{fig:4G_hecore_wfi_01}
\end{figure}

We display on the top panel of Fig.~\ref{fig:4G_hecore_wfi_01} the Brunt-Väisälä frequency as a function of the normalized buoyancy radius at X(He)$_{\rm core}=0.1$. We observe a broader peak in terms of normalized buoyancy radius at the core-mantle transition compared to higher central helium abundances (e.g., with respect to the top panel of Fig.~\ref{fig:4G_hecore_wfi_09}, at X(He)$_{\rm core}=0.9$). Equally importantly, there is no radiative zone below the lq\_core transition any longer, which is consistent with Fig.~\ref{fig:4G_core_growth}. The weight functions associated to a trapped mode (aka a mode having a minimum of the period spacing, here of $k=58$; middle panel of Fig.~\ref{fig:4G_hecore_wfi_01}) and a normal mode (of $k=51$, bottom panel of Fig.~\ref{fig:4G_hecore_wfi_01}) show that, for X(He)$_{\rm core}=0.1$, none of them propagate in the core and both have a strong, large peak at lq\_core. It can be inferred that for minima of the ``wavy'' pulsation spectra (Fig.~\ref{fig:4G_hecore}, X(He)$_{\rm core}=0.6$ to $0.1$), the trapping region is the enlarged lq\_core chemical transition itself. Trapped and normal modes show different amplitudes for their weight functions in the radiative mantle (between lq\_core and lq\_env), with the normal mode having a higher amplitude in it than a trapped mode. This characteristic is consistent with the amplitude difference between a trapped and a normal mode seen at higher X(He)$_{\rm core}$ (Fig.~\ref{fig:4G_hecore_wfi_09}). By comparing this latter figure to Fig.~\ref{fig:4G_hecore_wfi_01}, it can be derived that the amplitude differences of weight functions in the radiative mantle between a trapped and a normal mode are related to the depth of the minima of reduced period spacing, as observed in Fig.~\ref{fig:4G_hecore}. The smaller the difference in amplitude in the radiative mantle between a trapped and normal mode, the shallower the minima in the reduced period spacing.

Now that the trapping cavities have been identified, we can compare the period spacings between the trapped modes found in our models with the ones predicted by theory. The theoretical asymptotic reduced period spacing between two modes propagating in a given cavity reads as (\citealt{1980ApJS...43..469T}, see also \citealt{1992ApJS...80..369B}):
\begin{equation}
  \overline{P_{k+1} - P_{k}} = \frac{2\pi^2}{\sqrt{\ell(\ell+1)}}\left(\int^{r_t}_{r_b} \frac{\lvert N\rvert}{r}dr\right)^{-1}\;\;\;,
  \label{eq_spacing}
\end{equation}
where $N$ is the Brunt-Väisälä frequency, $\ell$ is the degree of the mode, $r_t$ is the radius of the upper boundary of the cavity we integrate on, and $r_b$ the bottom of it. The boundaries of the integral can be shifted for multiple purposes: integrating over the whole propagation cavity of g-modes gives the mean period spacing between two consecutive modes of adjacent radial order (aka $k$ and $k+1$), the so-called $\Pi_{0,l}$ spacing. Integrating over a trapping cavity (if existing) instead gives the mean period spacing between two consecutive trapped modes, called $\Pi_{T,l}$. Multiplying $\Pi_{0,l}$ and $\Pi_{T,l}$ by $\sqrt{\ell(\ell+1)}$ gives the so-called asymptotic period spacing $\Pi_0$ (which is then related to the inverse of the total buoyancy radius, see Eq.~(\ref{eq:total_Bradius})), and the asymptotic period spacing between consecutive trapped modes $\Pi_T$, both independent of the degree $\ell$.

\begin{figure}[t]
  \centering
  \includegraphics[width=0.49\textwidth]{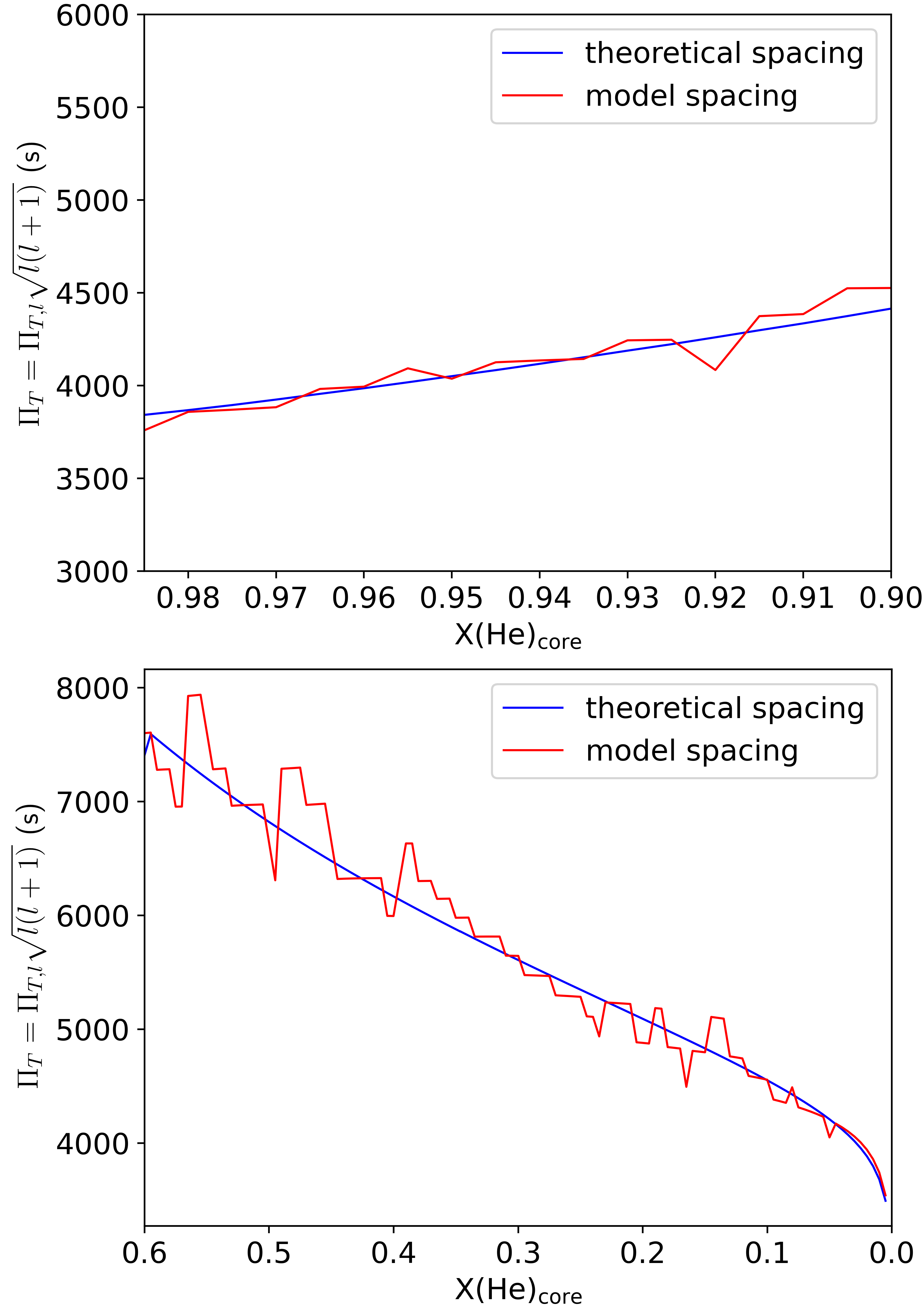}
  \caption{\textit{Top panel}: Model (red) and theoretical (blue) mean reduced period spacing between two consecutive modes trapped in the radiative zone below the lq\_core chemical transition, for X(He)$_{\rm core}$ $> 0.9$. \textit{Bottom panel}: Model (red) and theoretical (blue) mean reduced period spacing between two consecutive modes trapped in the width of the lq\_core transition, for X(He)$_{\rm core}<0.6$. Both panels are for a 4G model with $M_*=0.47 M_\odot$, lq\_core~$=-$0.25, lq\_env~$=-$2.}
  \label{fig:4G_spacing_th_obs}
\end{figure}

We show in Fig.~\ref{fig:4G_spacing_th_obs} the spacing between two consecutive trapped modes $\Pi_T$ computed by numerical integration of Eq.~(\ref{eq_spacing}) (blue curve), and the same spacings directly obtained from pulsation spectra of our models (red curve), such as from Fig.~\ref{fig:4G_hecore}. On the top panel, corresponding to models from X(He)$_{\rm core}=0.985$ to $0.9$, the theoretical spacing is computed from the top of the convective core ($r_b$ of Eq.~\ref{eq_spacing}) to the middle of the lq\_core transition ($r_t$ of Eq.~\ref{eq_spacing}). We find a very good agreement between the asymptotic theory and our models for this range of X(He)$_{\rm core}$, where the trapping cavity is the radiative part of the core. We observe a relative difference between both spacings of the order of 1.5\%, and always less than 4\%. On the bottom panel of Fig.~\ref{fig:4G_spacing_th_obs}, the period spacings are computed for models from X(He)$_{\rm core}=0.6$ to $0.005$, where modes are trapped in the width of the enlarged lq\_core transition. The theoretical spacing integral is computed over this trapping region, and we find again a good agreement between theory and models, with a relative difference ranging from 2 to 8\%. Such good agreements reinforce our identification of g-mode trapping cavities in our models.

There is a range from X(He)$_{\rm core}\sim0.9$ to $0.6$ where the period spacings between trapped modes are not constant in our 4G models. In this range of X(He)$_{\rm core}$, the influence of the radiative zone on top of the convective core and the one of the width of the core-mantle transition itself become of the same order (the radiative part of the core shrinks because the convective core grows, lq\_core being kept fixed, Fig.~\ref{fig:4G_core_growth}). We find in pulsation spectra some rather deep and some shallower trapped modes (see X(He)$_{\rm core}=0.8$ of Fig.~\ref{fig:4G_hecore} for example). This is expected during the transition from higher to lower values of X(He)$_{\rm core}$ since the trapping region progressively changes. The core splitting phenomenon also occurs during this range of X(He)$_{\rm core}$ (for the lq\_core transition being fixed at $-0.25$), and is believed to be non-physical (Fig.~\ref{fig:sc_pb}; this phenomenon is studied in more detail in Sect.~\ref{lqcore_4G}). The non-constant spacings could be a result of both effects occurring at the same time.

\subsubsection{4G+ models}
\label{XHe_4GP}

\begin{figure}[h!]
\centering
\includegraphics[width=0.49\textwidth]{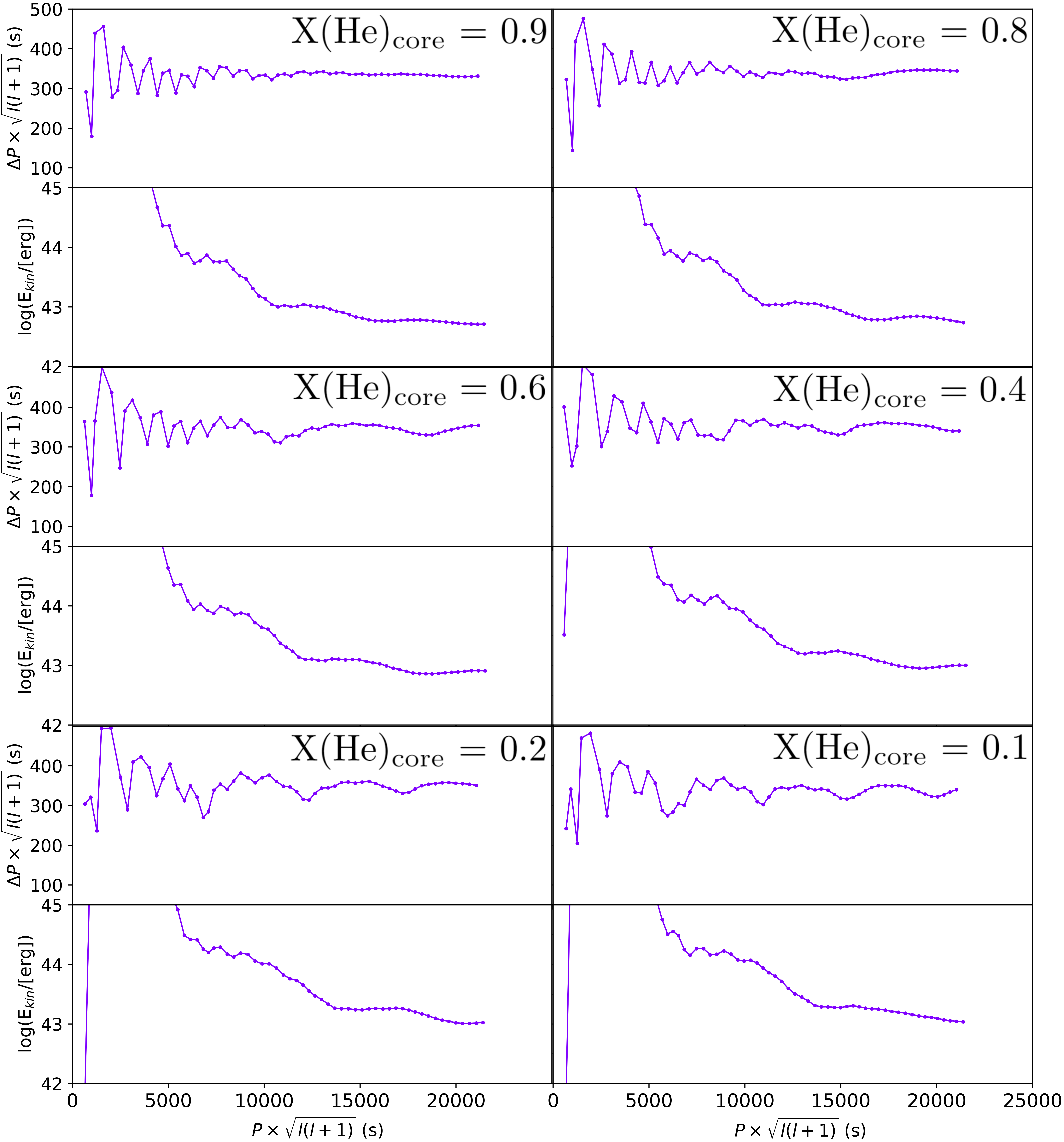}
\caption{Pulsation spectra (top panels) and associated kinetic energies (bottom panels) of a 4G+ model for $M_*=0.47M_\odot$, lq\_core~$=-$0.25, lq\_env~$=-$2, at X(He)$_{\rm core}=0.9$, 0.8, 0.6, 0.4, 0.2, and 0.1.}  
\label{fig:4GP_hecore}
\end{figure}

We go on to investigate the impact of core-He burning on 4G+ static models, in which the core (i.e., the region below lq\_core) is assumed to be fully convective. To this end, a reference 4G+ model was computed using the same parameters as the 4G reference model, having $M_* = 0.47 M_\odot$, lq\_core $= -$0.25, and lq\_env $= -$2, with X(He)$_{\rm core}$ ranging from 0.985 to 0.005, by steps of 0.005. Figure~\ref{fig:4GP_hecore} shows six panels at X(He)$_{\rm core}=0.9$, 0.8, 0.6, 0.4, 0.2, and 0.1, for the aforementioned 4G+ model. Each panel is divided in two similarly to Fig.~\ref{fig:4G_hecore}, displaying the pulsation spectrum on top, and the associated kinetic energy at the bottom, both as a function of the reduced period. For high X(He)$_{\rm core}$, Fig.~\ref{fig:4GP_hecore} shows pulsation spectra of almost constant period spacing for high radial order modes, which then progresses as long as core He abundance diminishes into more wavy pulsation spectra, slowly oscillating around a mean value. As the transition from flat to wavy pulsation spectra occurs smoothly over decreasing X(He)$_{\rm core}$, we deduced that the increasing ``waviness'' of the pulsation spectra follows the increase of the chemical gradient at core-mantle transition with core-He burning, from a more and more C- and O-enriched core to the pure He radiative mantle, as was the case with 4G models. Indeed, the stronger chemical gradient raises the corresponding peak in the Brunt-Väisälä frequency, which in turn increases the width of the lq\_core transition in terms of the normalized buoyancy radius (compare Fig.~\ref{fig:4G_hecore_wfi_09} and Fig.~\ref{fig:4G_hecore_wfi_01}).

As described in Sect.~\ref{tools}, 4G+ models are fully convective below lq\_core at any X(He)$_{\rm core}$. As a result, 4G and 4G+ models are only different for their thermal structures during X(He)$_{\rm core}$ values where the convective core grows in 4G models, which, in the specific case of lq\_core~$=-$0.25, is for X(He)$_{\rm core}>0.6$ (Fig.~\ref{fig:4G_core_growth}). This is verified when comparing 4G (Fig.~\ref{fig:4G_hecore}) and 4G+ (Fig.~\ref{fig:4GP_hecore}) pulsation spectra. At X(He)$_{\rm core}>0.6$, 4G models have modes trapped in the radiative part of the core below lq\_core (Sec. \ref{XHe_4G}), which cannot be the case for 4G+ models, as this same region is fully convective, leading to a flat spectrum. However, at X(He)$_{\rm core}<0.6$, 4G and 4G+ models are equivalent with a fully convective core, leading to exactly the same pulsation spectra. It is therefore direct to verify that for 4G+ models, high-order modes corresponding to minima of period spacing are modes trapped in the width of the core-mantle transition at lq\_core. As pulsation spectra of 4G and 4G+ models are equivalent for X(He)$_{\rm core}<0.6$, we refer to Fig.~\ref{fig:4G_hecore_wfi_01} for an example of the Brunt-Väisälä frequency and weight functions associated to those models.

Additionally, turning  our attention to low-order modes ($k=1$ to $k\sim10$), the pulsation spectra of 4G and 4G+ models at X(He)$_{\rm core}>0.6$ (specifically X(He)$_{\rm core}=0.9$, 0.8), shows little to no differences, indicating that the presence or not of a radiative region below lq\_core is of minimal impact on low-order modes. This is understood by coming back to the propagation diagram in $m(r)$ scale of Fig.~\ref{fig:propa_diag} (bottom panel): low-order modes, which are the few top-most red horizontal lines in frequency, show no nodes below lq\_core and are in consequence unaffected by core conditions.

\subsection{Influence of core mass on g-mode spectra}

We present in this subsection the results obtained by varying the lq\_core parameter between $-$0.1 and $-$0.4. We  consider the 4G and 4G+ models separately. 

\subsubsection{4G models}
\label{lqcore_4G}
\begin{figure}[t]
\centering
\includegraphics[width=0.49\textwidth]{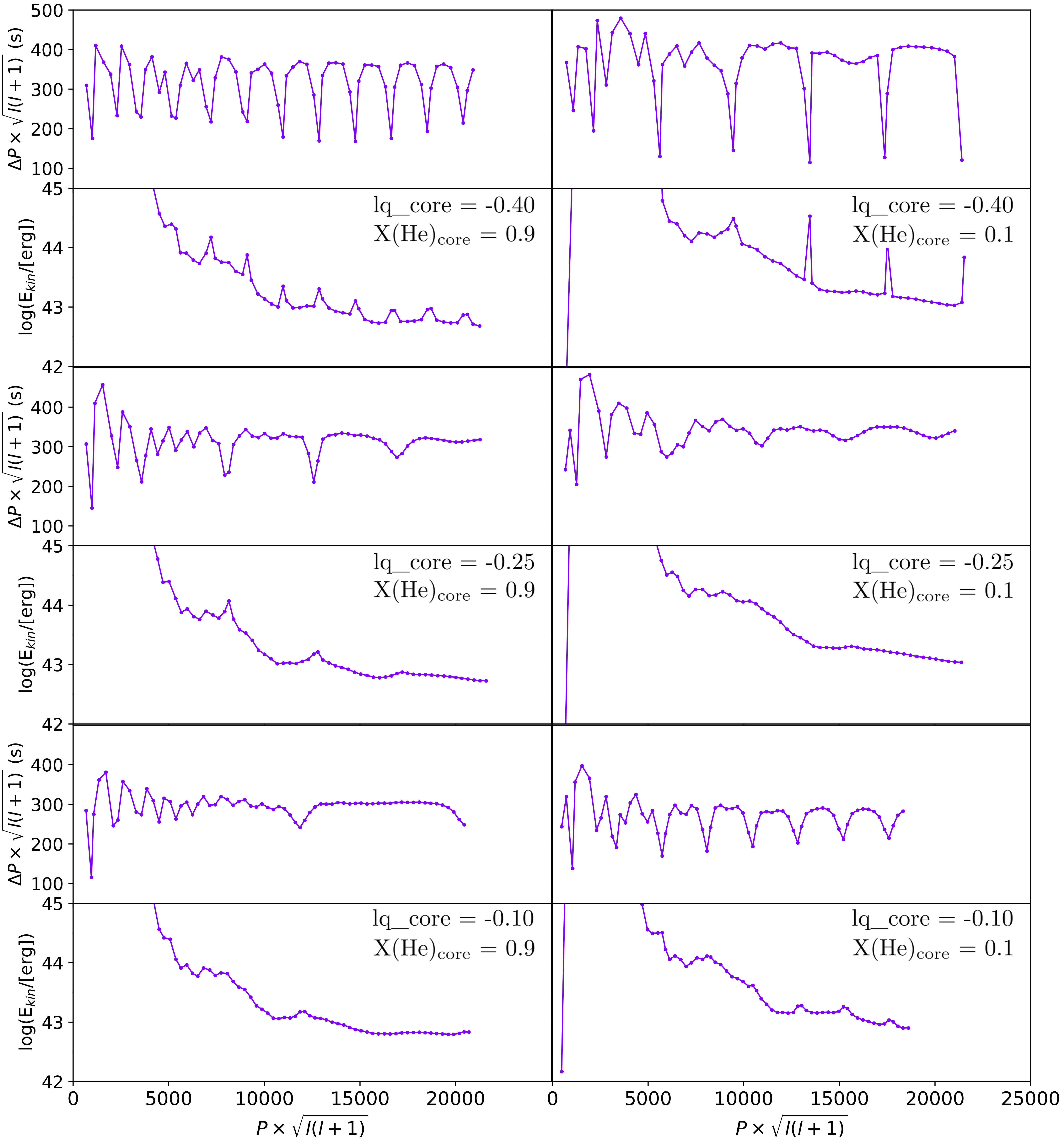}
\caption{Pulsation spectra and associated kinetic energies, for 4G models of $M_*=0.47M_\odot$ and lq\_env $=-$2, and lq\_core $=-$0.40 (top panels), $-$0.25 (middle panels) and $-$0.10 (bottom panels). Two extremes of core-He burning are represented: X(He)$_{\rm core}=0.9$ (left) and 0.1 (right).}
\label{fig:4G_lqcore}
\end{figure}

 We show in Fig.~\ref{fig:4G_lqcore} six split panels, with on top the pulsation spectra with their associated kinetic energy on the bottom, obtained at X(He)$_{\rm core}=0.9$ (left) and X(He)$_{\rm core}=0.1$ (right), for three cases: lq\_core $=-$0.40 (top panels), $-$0.25 (middle panels), and $-$0.10 (bottom panels). Each model presented has $M_*=0.47M_\odot$ and lq\_env $=-$2. A core-mantle transition higher in the star (lq\_core $=-$0.40) implies a higher core mass than a smaller one (lq\_core $=-$0.10).

\begin{figure}[t]
\centering
\includegraphics[width=0.49\textwidth]{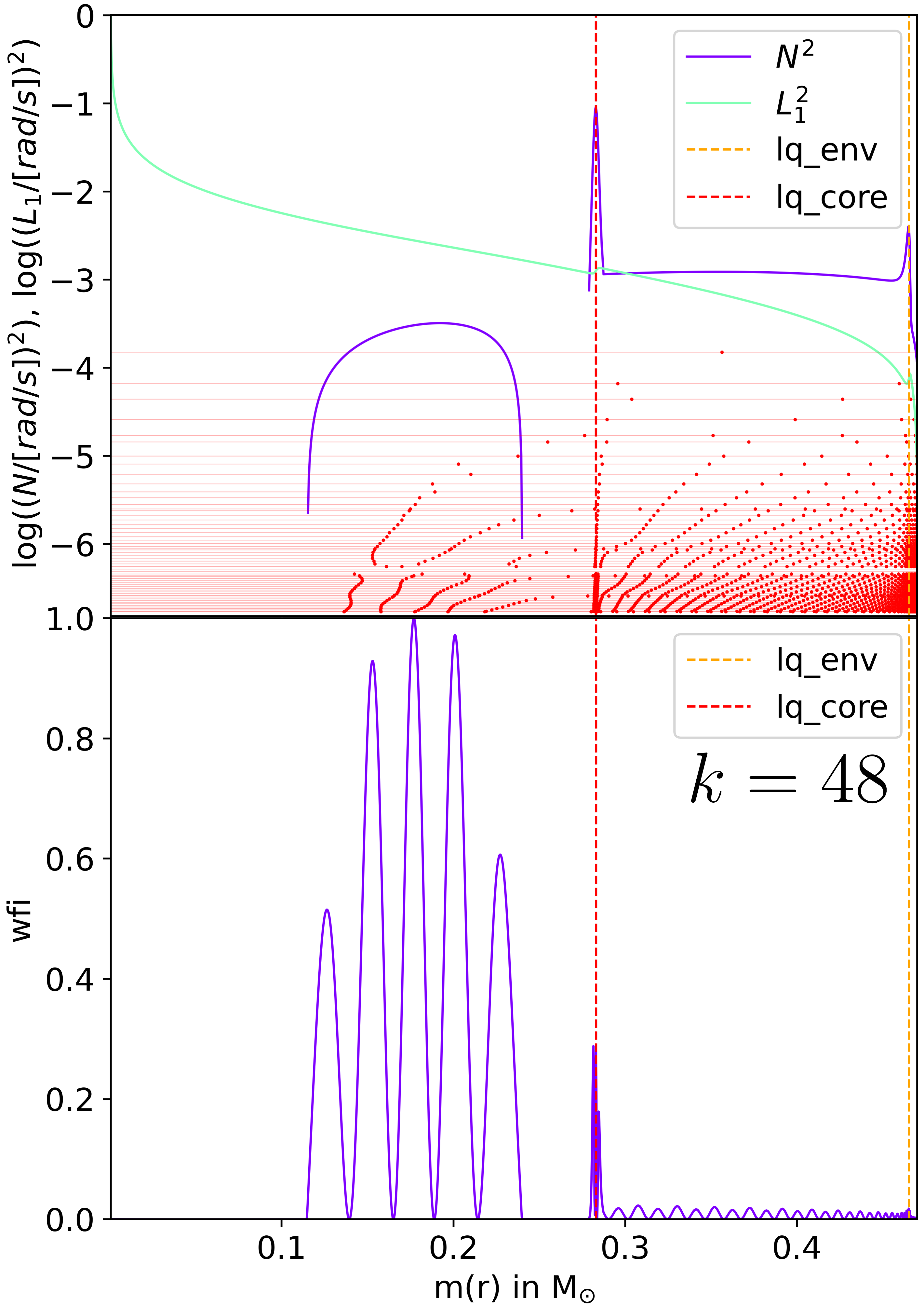}
\caption{\textit{Top panel}: Propagation diagram of a 4G model showing the ``radiative arch'' due to core splitting for $M_*= 0.47 M_\odot$, lq\_core $=-$0.4, lq\_env $=-$2, X(He)$_{\rm core}=0.1$. $N^2$ is the Brunt-Väisälä frequency (blue), $L_1$ is the Lamb frequency at $\ell=1$ (green). Red horizontal lines are the $\log(\sigma^2)$ of each mode of radial order $k=1$ to 70, and red dots indicate the radial node positions in $\log(q)$. \textit{Bottom panel}: Weight function of a mode trapped in the radiative arch, for the same model parameters as the top panel.}  
\label{fig:radiative_arch}
\end{figure}

For lq\_core~$=-$0.4, at high X(He)$_{\rm core}=0.9$ (top-left panel), we observe deep and numerous trapped modes. Those modes are trapped in the radiative cavity below lq\_core, as explained in Sect.~\ref{XHe_4G}. As the thermal and chemical structure of static models are decoupled, modifying lq\_core does not hold any consequences on the convective core growth, and this growth always occurs from an initial convective core at $\log(q)\sim-0.11$ at X(He)$_{\rm core}=0.985$. As a consequence, instead of a radiative cavity ranging from $\log(q)\sim-0.11$ to $-0.25$ (at X(He)$_{\rm core}=0.985$) for lq\_core~$=-$0.25 models, we now have a much larger cavity from $\log(q)\sim-0.11$ to $-0.40$ at the same X(He)$_{\rm core}$. We showed in Sect.~\ref{XHe_4G} that the strength (depth) and number of trapped modes at high X(He)$_{\rm core}$ for 4G models is linked to the size in $\log(q)$ of the radiative cavity below lq\_core. This is further verified here, and explains the deep and numerous trapped modes at high X(He)$_{\rm core}$. As X(He)$_{\rm core}$ decreases, the radiative cavity shrinks (amounts to a lower amount of mass with respect to $M_\ast$), and the depth of the trapped modes decreases while the period spacing between two consecutive trapped modes increases. However, due to the evolution of the radiative temperature gradient, the core is eventually split into two convective parts separated by a radiative cavity (see Introduction and Fig.~\ref{fig:sc_pb}). This gives the pulsation spectrum found at X(He)$_{\rm core}=0.1$ for lq\_core~$=-$0.40 (top-right panel, Fig.~\ref{fig:4G_lqcore}), which shows the strongest trapping phenomenon found in 4G models. The top panel of Fig.~\ref{fig:radiative_arch} shows the propagation diagram associated to the lq\_core~$=-$0.40 model at X(He)$_{\rm core}=0.1$. This propagation diagram displays what we called the ``radiative arch'' (due to its shape) below lq\_core, effectively showing that the core is split into two convective parts separated by a radiative region. The analysis of the weight function of a trapped mode at lq\_core~$=-$0.40 for X(He)$_{\rm core}=0.1$ given on the bottom panel of Fig.~\ref{fig:radiative_arch}, shows a very high amplitude in the radiative arch, and a much lower amplitude in the radiative mantle: the trapping cavity of those trapped modes is the radiative arch, namely, the radiative part encapsulated by two convection zones in the core. 

\begin{figure}[t]
\centering
\includegraphics[width=0.49\textwidth]{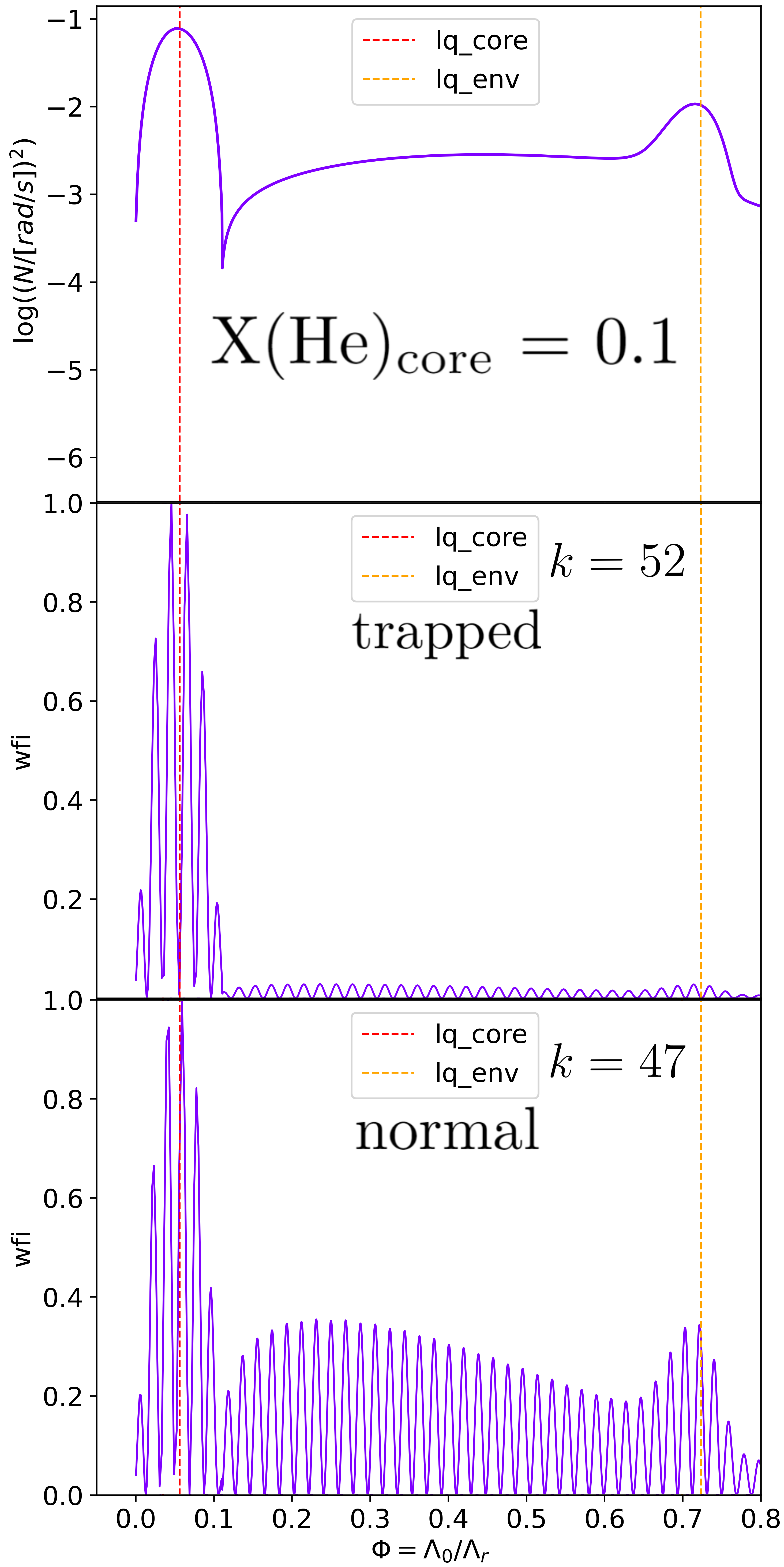}
\caption{\textit{Top panel}: $\log(N^2)$ as a function of the normalized buoyancy radius ($\phi = \Lambda_0/\Lambda_r$) at X(He)$_{\rm core}=0.1$ for a 4G model having $M_*=0.47 M_\odot$, lq\_core~$=-$0.1, lq\_env~$=-$2, and X(He)$_{\rm core}=0.1$. The red dotted vertical line is the lq\_core (core-mantle) transition, and the orange dotted vertical line is the lq\_env (mantle-envelope) transition. \textit{Middle and bottom panels:} Weight functions (wfi) of two modes: a mode trapped in the width of the chemical transition at lq\_core (middle) and a normal mode (bottom).}
\label{fig:4G_lqcore_wfi_01}
\end{figure}

\begin{figure}[t]
\centering
\includegraphics[width=0.49\textwidth]{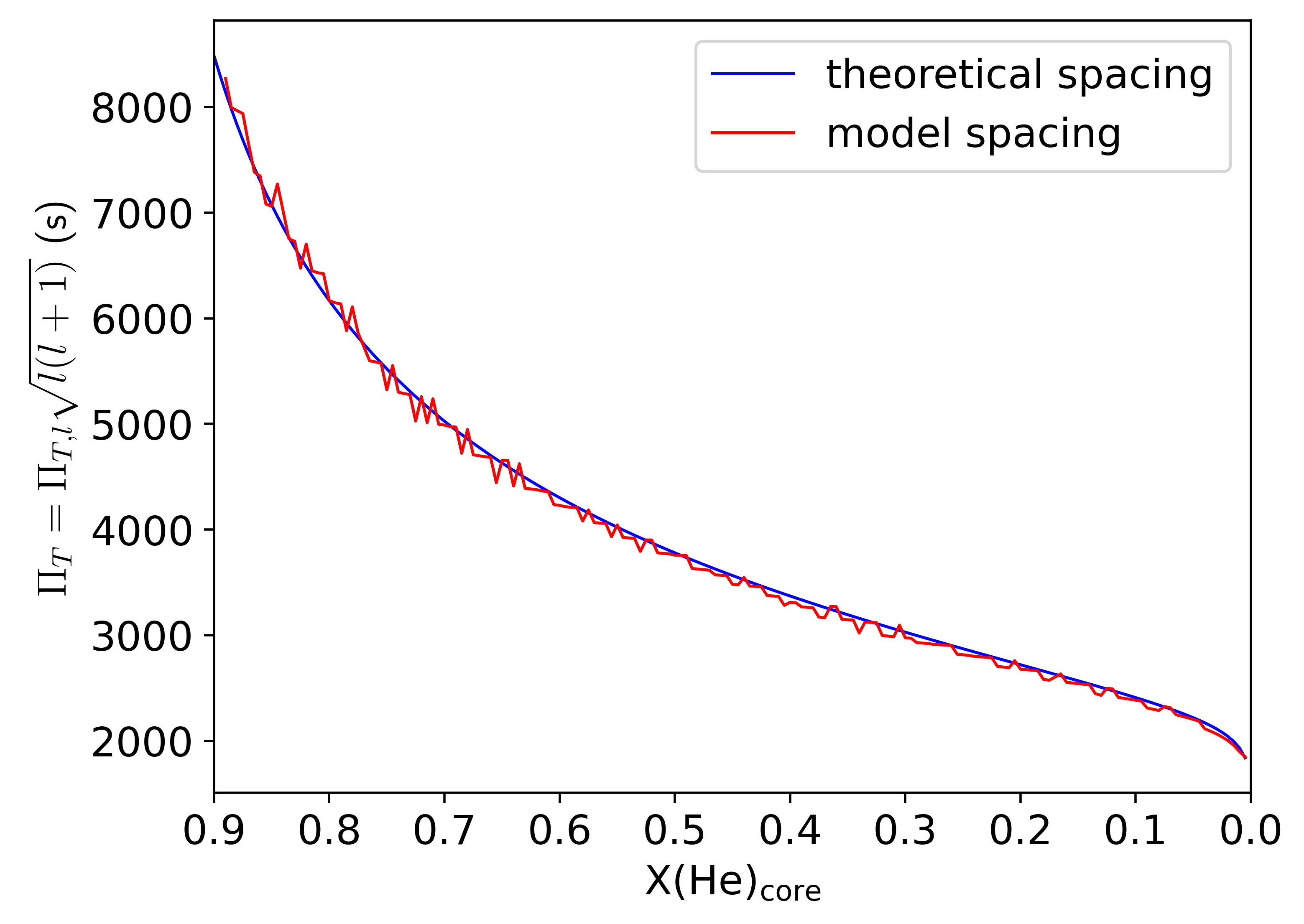}
\caption{Model (red) and theoretical (blue) mean reduced period spacing between two consecutive modes trapped in the lq\_core transition width, for X(He)$_{\rm core}$ $< 0.9$, for a 4G model having $M_*=0.47 M_\odot$, lq\_core~$=-$0.1, lq\_env~$=-$2.}
\label{fig:low_lqcore_spacing_th_obs}
\end{figure}

The lq\_core~$=-$0.1 case displays the other extreme case, where lq\_core is lower than the initial limit in $\log(q)$ of the convective core ($\log(q)\sim-0.11$ at X(He)$_{\rm core}=0.985$). Hence, no convective core growth occurs along decreasing X(He)$_{\rm core}$ and the region below lq\_core~$=-$0.1 (the core) is always fully convective. The pulsation spectra and kinetic energies at lq\_core~$=-$0.1 are shown on the bottom panels of Fig.~\ref{fig:4G_lqcore}, at X(He)$_{\rm core}=0.9$ (left) and X(He)$_{\rm core}=0.1$ (right). On one hand, the pulsation spectrum at X(He)$_{\rm core}=0.9$ shows for high-radial-order modes a single minimum of period spacing, and hints to another one right after the mode of highest radial order. On the other hand, the pulsation spectra at X(He)$_{\rm core}=0.1$ instead shows numerous trapped modes, shallower than the lq\_core~$=-$0.4 case, and a small period spacing between consecutive trapped modes. Following the evolution with decreasing X(He)$_{\rm core}$, we observe an associated smooth decrease in the period spacing between consecutive trapped modes, as well as a smooth increase of the depth of those modes, both partially attributed to the strengthening of the chemical gradient located around lq\_core with core-He burning. The comparison of trapped and normal modes at lq\_core~$=-$0.1 and X(He)$_{\rm core}=0.1$ is shown on Fig.~\ref{fig:4G_lqcore_wfi_01} (middle and bottom panels, respectively), with the Brunt-Väisälä frequency at X(He)$_{\rm core}=0.1$ (top panel), all shown as a function of the normalized buoyancy radius. The same behavior as for the weight functions of modes of ``wavy'' pulsation spectra (Fig.~\ref{fig:4G_hecore_wfi_01}) is observed, and shows that in the extreme case of lq\_core~$=-$0.1, the modes with deep minima at X(He)$_{\rm core}=0.1$ are trapped in the width of the lq\_core chemical transition as well. The comparison of theoretical period spacing (blue line of  Fig.~\ref{fig:low_lqcore_spacing_th_obs}), computed by the numerical integral of Eq.~(\ref{eq_spacing}) with integration boundaries being the width of the lq\_core transition, and the same spacing (in red) derived from the model pulsation spectra, reinforces this conclusion. There is again a clear agreement between the asymptotic theory and our models, and we find a relative difference between spacings of the order of 2\%, and always less than 5\%.

\begin{figure}[t]
\centering
\includegraphics[width=0.49\textwidth]{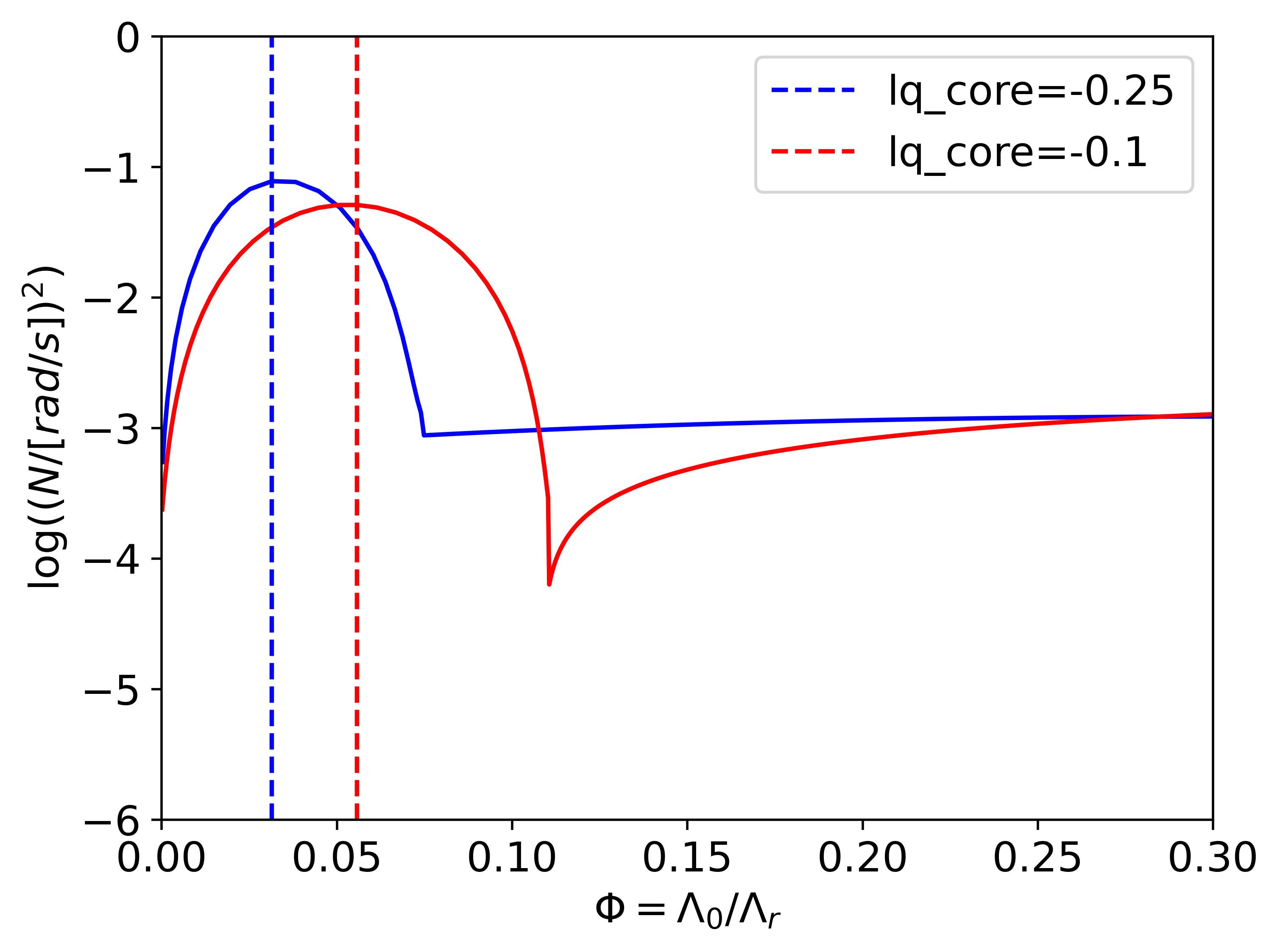}
\caption{$\log(N^2)$ as a function of the normalized buoyancy radius ($\phi = \Lambda_0/\Lambda_r$) at X(He)$_{\rm core}=0.1$ for 4G models having lq\_core~$=-$0.25 (blue) and $=-$0.10 (red), and fixed $M_*=0.47 M_\odot$, lq\_env~$=-$2, and X(He)$_{\rm core}=0.1$.}
\label{fig:buoyancy_radius_comparison}
\end{figure}

The waviness found at low X(He)$_{\rm core}$ and lq\_core~$=-$0.25 in both 4G and 4G+ pulsation spectra (Fig.~\ref{fig:4G_hecore} and Fig.~\ref{fig:4GP_hecore}, middle and bottom panels) is therefore explained by the mode trapping phenomenon taking place in the width of the lq\_core (core-mantle) chemical transition. While the origin of the mode trapping is the same for both lq\_core~$=-$0.25 and lq\_core~$=-0.10$, the latter shows more pronounced mode trapping. 
As shown in \citet{Cunha2019, Cunha2024}, a glitch in the Brunt-Väisälä frequency (in our case, the chemical transition lq\_core) induces a phase shift in the eigenfunction of a mode, which directly depends on the shape of the glitch itself. By imposing the continuity of the eigenfunction in the g-mode cavity containing a glitch, \citet{Cunha2019, Cunha2024} were able to derive an analytical expression of the impact on the period spacing of both step-like and Gaussian glitches (Eq.~(14) and (15) respectively of \citealt{Cunha2019}). In both cases, it is shown that the minima of period spacings, and so the trapping induced by a given glitch, are deeper with the increase of the amplitude of the glitch in Brunt-Väisälä frequency, and its position in normalized buoyancy radius. In Fig.~\ref{fig:buoyancy_radius_comparison}, we give a direct comparison of the Brunt-Väisälä frequency associated to the lq\_core transition at X(He)$_{\rm core}=0.1$, for lq\_core~$=-$0.10 (in red) and lq\_core~$=-$0.25 (in blue), as a function of the normalized buoyancy radius. It is readily apparent that the outer edge of the lq\_core transition is strongly shifted towards higher values of normalized buoyancy radius for lq\_core~$=-$0.10, compared to the lq\_core~$=-$0.25 case. While the lq\_core chemical transition in our case is more complex than a step-like or a Gaussian glitch, we believe the analytical results derived for both of those shapes might be extended to our case, and thus that the shift in normalized buoyancy radius of the outer edge of the lq\_core transition for the lq\_core~$=-$0.10 case is, at least partly, responsible for the stronger trapping observed in this case compared to the lq\_core~$=-$0.25 case (Fig.~\ref{fig:4G_lqcore}, middle right panel for lq\_core~$=-$0.25 and bottom right panel for lq\_core~$=-$0.10.)

Finally, for completeness, let us come back a moment to the intermediate case with lq\_core~$=-$0.25. It has been thoroughly studied in Sect.~\ref{XHe_4G}, and is displayed again in the middle panels of Fig.~\ref{fig:4G_lqcore}, at X(He)$_{\rm core}=0.9$ (left) and X(He)$_{\rm core}=0.1$ (right). In such cases, which include any lq\_core higher than $\log(q)= -$0.11 (the initial limit in $\log(q)$ of the convective core) so that convective core growth occurs, but low enough to allow the convective core reaching lq\_core before depletion of He in the core, we find that the core splitting phenomenon also occurs, albeit over a very narrow range of X(He)$_{\rm core}$. In such a case, the minimum of radiative gradient seen on Fig.~\ref{fig:sc_pb} associated with core splitting eventually overtakes the adiabatic gradient, thus lifting the core splitting and transitioning into a fully convective core instead. This is exactly the reason of the jump in convective core mass seen on Fig.~\ref{fig:4G_core_growth} at X(He)$_{\rm core}=0.6$: before core splitting is lifted, the convective core ranges from the center of the star to the lower boundary of the radiative arch (see Fig.~\ref{fig:radiative_arch}, top panel and Fig.~\ref{fig:sc_pb}), while after core splitting is lifted, it suddenly ranges from the center of the star to lq\_core, thus inducing the jump in convective core mass seen in Fig.~\ref{fig:4G_core_growth}.

\subsubsection{4G+ models}

\begin{figure}[t]
\centering
\includegraphics[width=0.49\textwidth]{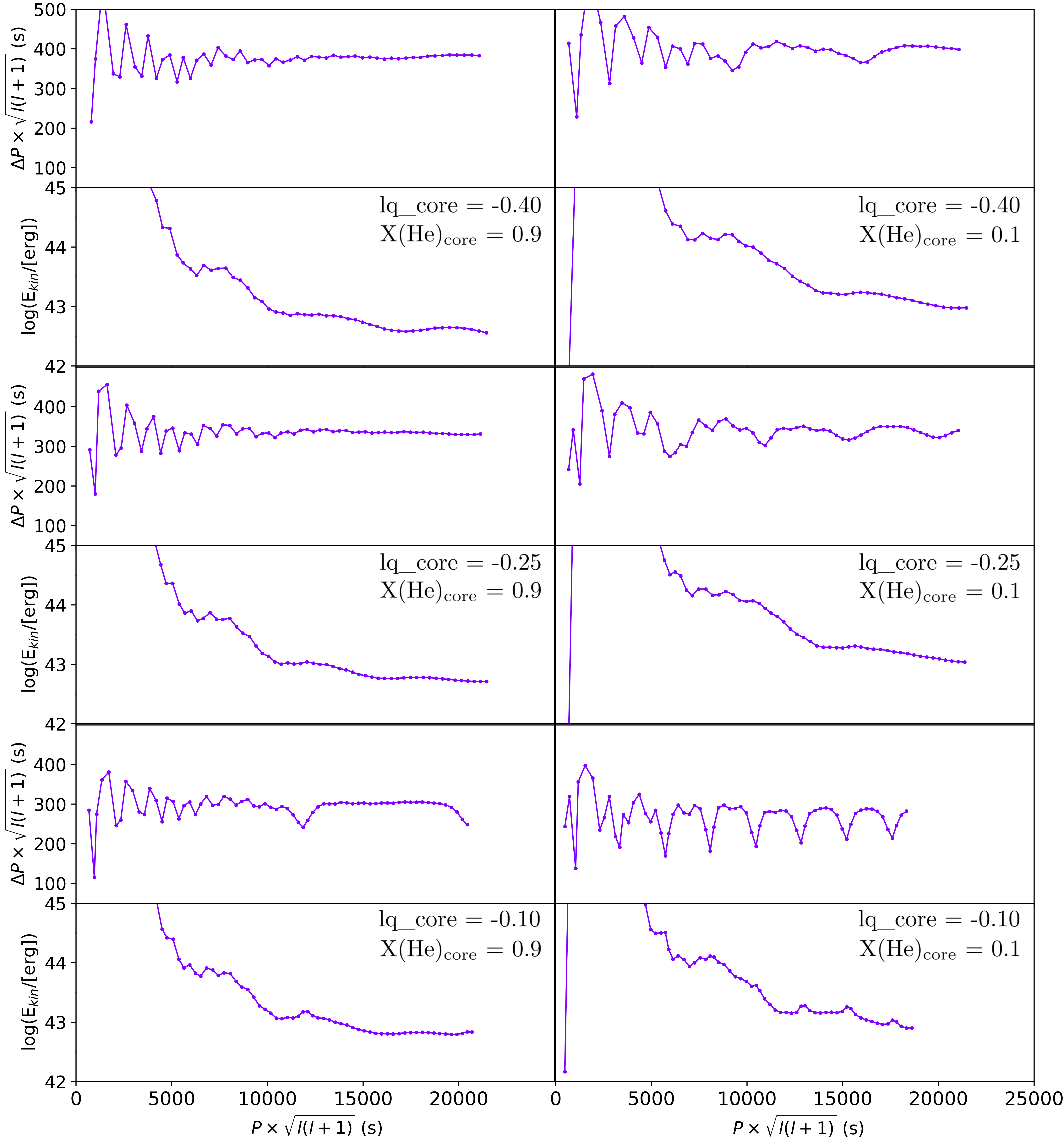}
\caption{Pulsation spectra and associated kinetic energies, for 4G+ models of fixed $M_*= 0.47 M_\odot$ and lq\_env~$=-$2, and varying lq\_core~$=-$0.40 (top panels), $-$0.25 (middle panels) and $-$0.10 (bottom panels). We represent both extremes of core-He burning at X(He)$_{\rm core}=0.9$ (left) and 0.1 (right).}
\label{fig:4GP_lqcore}
\end{figure}

In this subsection, we reproduce on 4G+ models the study of the influence of core mass done in Sect.~\ref{lqcore_4G}. In Fig.~\ref{fig:4GP_lqcore} we display six split panels, with on top the pulsation spectra, and on the bottom their associated kinetic energy, obtained at X(He)$_{\rm core}=0.9$ (left) and X(He)$_{\rm core}=0.1$ (right), for lq\_core$=-$0.40 (top panels), $-$0.25 (middle panels), and $-$0.10 (bottom panels).

Starting from lq\_core~$=-$0.40, at high X(He)$_{\rm core}=0.9$ (top-left panel), we observe a completely flat spectrum (constant reduced period spacing) for high order modes, while at lower X(He)$_{\rm core}=0.1$, we find a wavy pulsation spectrum. This heavily contrasts with the previous pulsation spectra of 4G models at lq\_core~$=-$0.40 found in Fig.~\ref{fig:4G_lqcore}, both at X(He)$_{\rm core}=0.1$ (top-right panel) and X(He)$_{\rm core}=0.9$ (top-left panel). Indeed, as seen in Sect.~\ref{XHe_4GP}, the direct consequence of having a fully convective core for any X(He)$_{\rm core}$ is preventing convective core growth and phenomena associated with it. With respect to mode trapping, compared to 4G models, we do not have a radiative cavity below lq\_core anymore, for any lq\_core and X(He)$_{\rm core}$. As this radiative cavity is responsible for mode trapping in high X(He)$_{\rm core}$ pulsation spectra of 4G models (the trapping then taking place in a radiative cavity below lq\_core in 4G models), and since at those X(He)$_{\rm core}$, the chemical gradient at the core-mantle transition is low, we now observe a constant period spacing in high-order modes in 4G+ models. Additionally, at lower X(He)$_{\rm core}$, while we previously found deeply trapped modes in the so-called radiative arch (see Fig.~\ref{fig:radiative_arch} and Fig.~\ref{fig:4G_lqcore}, top-left panel), we now have fully convective cores in 4G+ models. As a consequence, we instead observe a wavy pulsation spectrum, reminiscent of 4G models at lq\_core~$=-$0.25 and X(He)$_{\rm core}=0.1$ (Fig.~\ref{fig:4G_lqcore}, middle-right panel), where minima of period spacing are then modes trapped in the width of the lq\_core transition.

The pulsation spectra associated to lq\_core~$=-$0.25 in Fig.~\ref{fig:4GP_lqcore} show for X(He)$_{\rm core}=0.9$ (middle-left panel) a pulsation spectrum of constant period spacing for high order modes, as for lq\_core~$=-$0.40 (Fig.~\ref{fig:4GP_lqcore}, top-left panel), and a shallower mode trapping at X(He)$_{\rm core}=0.1$ (middle-right panel), again, similarly to the lq\_core~$=-$0.40 case (Fig.~\ref{fig:4GP_lqcore}, top-right panel). Indeed, both lq\_core~$=-$0.25 and lq\_core~$=-$0.40 4G+ models are much alike for any X(He)$_{\rm core}$, in the sense that their chemical and thermal structure solely differ through the positioning of lq\_core. This gives rise to the higher values of the period spacing seen in the lq\_core~$=-$0.40 case compared to lq\_core~$=-$0.25 ($\sim$ 380 s vs 300 s), but the general pattern of the pulsation spectra are very similar. 

Finally, we find for lq\_core~$=-$0.10 at X(He)$_{\rm core}=0.9$ (Fig.~\ref{fig:4GP_lqcore}, bottom-left panel) and X(He)$_{\rm core}=0.1$ (Fig.~\ref{fig:4GP_lqcore}, bottom-right panel), the exact same pulsation spectra as for 4G models of the same lq\_core (Fig.~\ref{fig:4G_lqcore}, bottom panels). This is the case because the core is already fully convective for any X(He)$_{\rm core}$ in 4G models at lq\_core~$=-$0.10, which directly implies that it is equivalent to a 4G+ model. The remarks for the pulsation spectrum of 4G models at lq\_core~$=-$0.10 then hold for 4G+ models at lq\_core~$=-$0.10 as well (Sect.~\ref{lqcore_4G}).

\subsection{Influence of the total stellar mass and envelope mass on g-mode spectra}

We now analyze the behavior of the pulsation spectra relative to the envelope mass lq\_env and the total stellar mass $M_*$ parameters. Their respective impact being the same for 4G and 4G+ models, we have grouped the explanations for their behaviors.

\subsubsection{Stellar mass}

\begin{figure}[h!]
\centering
\includegraphics[width=0.49\textwidth]{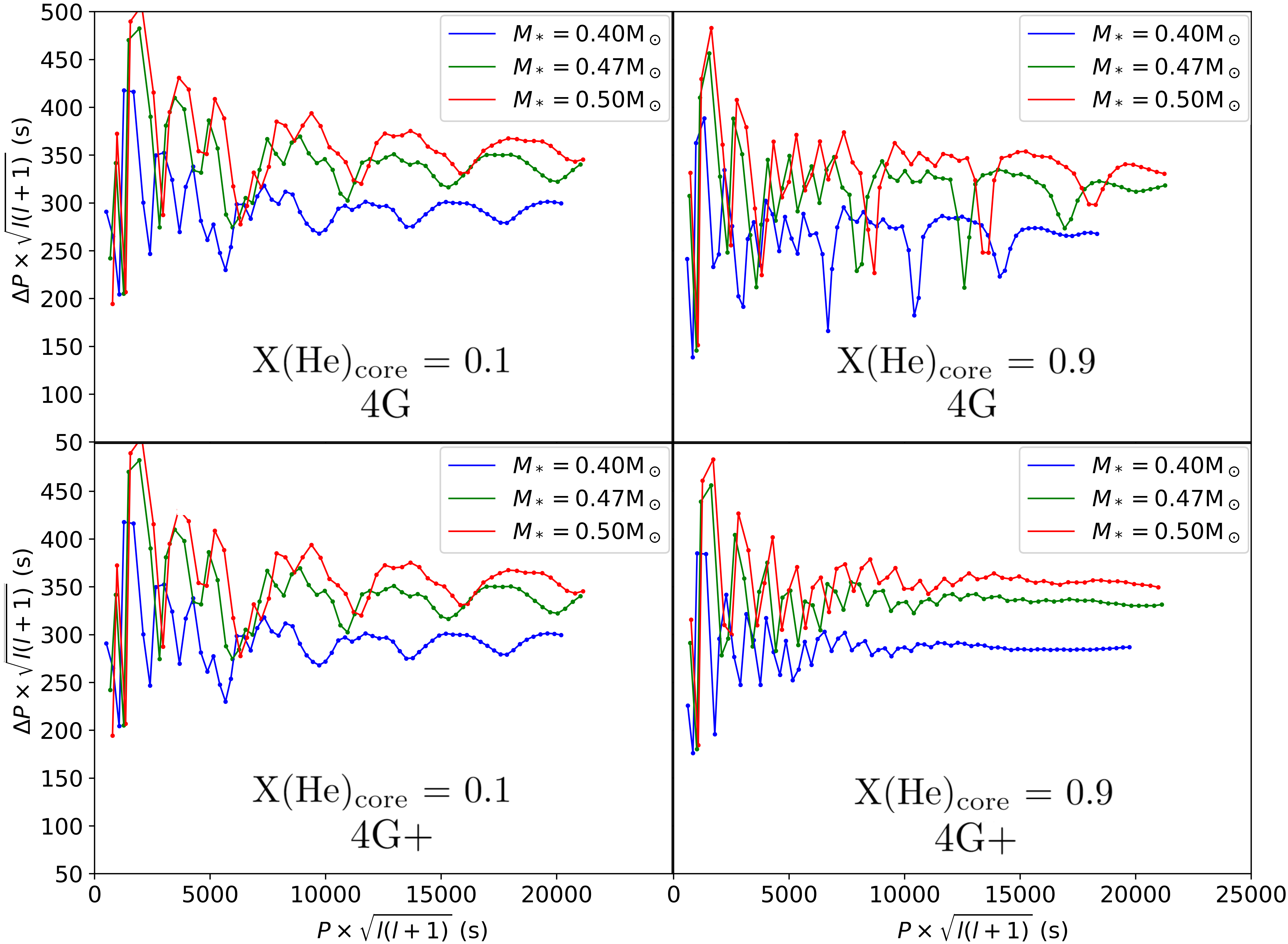}
\caption{Influence of the total stellar mass on the pulsation spectra of 4G and 4G+ models. All models share fixed lq\_env~$=-$2 and lq\_core~$=-$0.25. \textit{Top panels}: 4G models with $M_*=0.40$ (blue), 0.47 (green), 0.50 $M_\odot$ (red) at X(He)$_{\rm core}=0.1$ (left) and 0.9 (right). \textit{Bottom panels}: Same as top panels, but for 4G+ models.}
\label{fig:4G_4GP_mass}
\end{figure}

In order to study the influence of the stellar mass, we computed three 4G and 4G+ models, sharing lq\_env~$=-$2, lq\_core~$=-$0.25, and with varying mass $M_*=0.40, 0.47$, and 0.50 $M_\odot$. Figure~\ref{fig:4G_4GP_mass} shows the results of such computations, with top panels displaying the superposed pulsation spectra of 4G models for $M_*=0.40$ (blue), 0.47 (green), and 0.50 $M_\odot$ (red), at X(He)$_{\rm core}=0.1$ (left) and X(He)$_{\rm core}=0.9$ (right). Bottom panels show results from 4G+ models similarly. Increasing the total mass of the star has the same effects at any X(He)$_{\rm core}$, for both 4G and 4G+ models, which is to increase both the period and period spacing of the modes.

To show why this is the case, it is important to understand how quantities used to compute the mean period spacing (Eq.~(\ref{eq_spacing})) change in our models. To compute a static model at a given X(He)$_{\rm core}$, STELUM is given a number of layers, which are then discretized in $\log(q)$ scale. The number of layers is constant across static models (4800 layers), as well as the $\log(q)$ attributed to a given layer. With this discretization in mind, we always numerically integrate Eq.~(\ref{eq_spacing}) between two values of $\log(q)$ to find the period spacing associated to a given cavity. The value of this period spacing therefore varies if one or both of the following conditions are met:
\begin{itemize}
\setlength\itemsep{0.5em}
    \item [--] The cavity over which we integrate contracts or dilates in $\log(q)$ size;
    \item [--] The integrand $\lvert N \rvert/r$ varies in value.
\end{itemize}

An example of the first condition is illustrated on Fig.~\ref{fig:4G_lqcore} at X(He)$_{\rm core}$=0.9 for lq\_core~$=-$0.4 and $-$0.25. The radiative region in which modes are trapped is larger for lq\_core~$=-$0.4, and the period spacing between consecutive trapped modes is lower as a result. An example of the second condition is on the bottom panels of Fig.~\ref{fig:4G_hecore}: the width of the lq\_core transition in which modes are trapped is constant in $\log(q)$ size, but $\lvert N \rvert/r$ increases with decreasing X(He)$_{\rm core}$, and thus the period spacing between consecutive trapped modes decreases (see Fig. Fig.~\ref{fig:4G_spacing_th_obs}, bottom panel, for the numerical values of this spacing).

\begin{figure}[h!]
\centering
\includegraphics[width=0.49\textwidth]{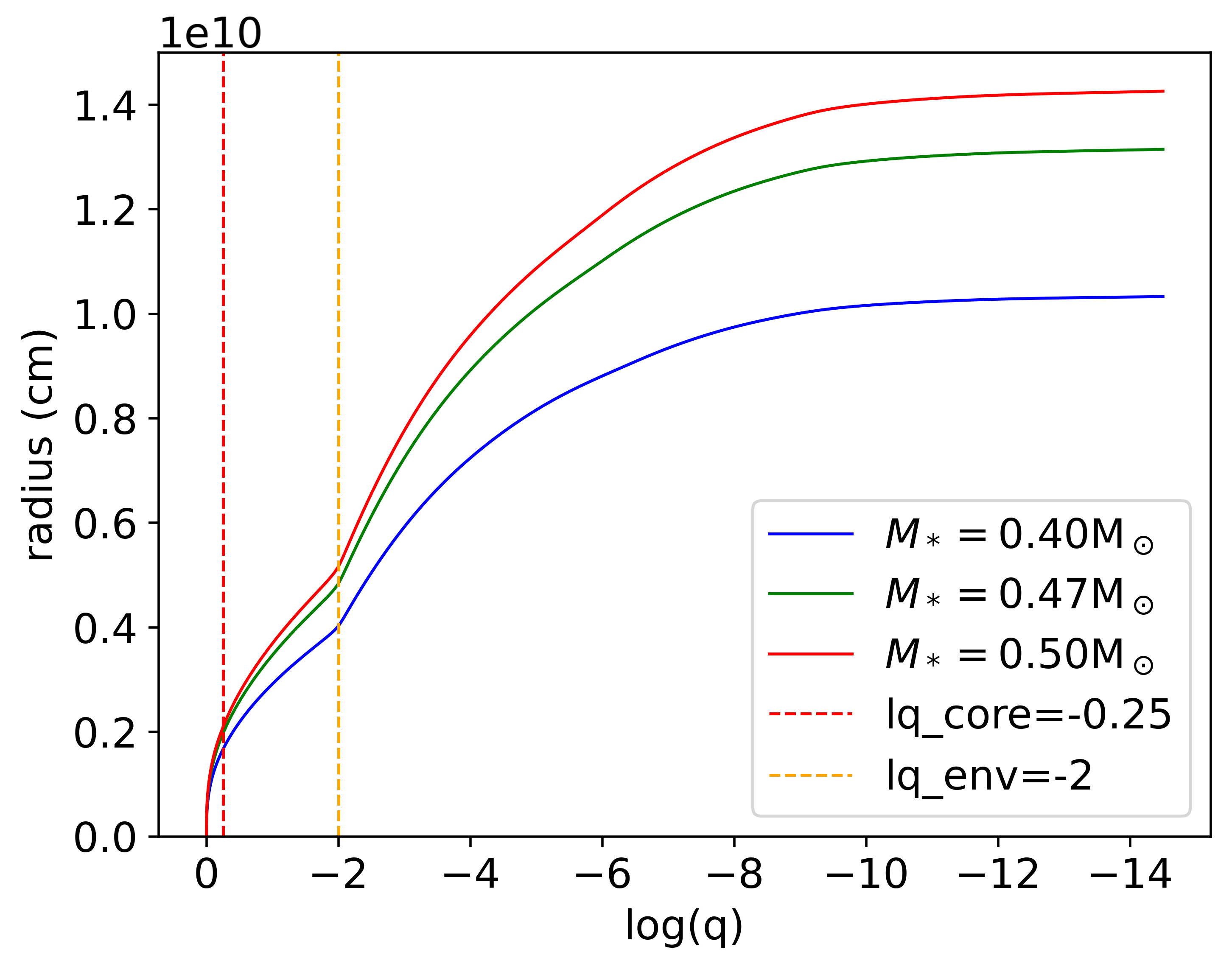}
\caption{Radius (from the center of the star) as a function of $\log(q)$, highlighting that a change in the total stellar mass modifies the radius of each region of the star. Three 4G models of lq\_core~$=-$0.25, lq\_env~$=-$2, and X(He)$_{\rm core}$ = 0.9 are represented, with masses 0.40 $M_\odot$ (blue), 0.47 $M_\odot$ (green), and 0.50 $M_\odot$ (red).}
\label{fig:4G_radius}
\end{figure}

Coming back to Fig.~\ref{fig:4G_4GP_mass}, the mean period spacing between modes of adjacent radial order $\Pi_0$ increases with the stellar mass. The boundaries of the integral in $\log(q)$ do not change (we integrate over the whole star for this period spacing), and $\lvert N \rvert$ remains similar inside the star for all models of different total mass, at a given X(He)$_{\rm core}$ (especially over the core and radiative mantle parts of the star, which hold the majority of the weight with respect to the mean period spacing). However, as displayed in Fig.~\ref{fig:4G_radius}, for 4G models of $M_*=0.40$ (blue), 0.47 (green), and 0.50 $M_\odot$ (red) as a function of $\log(q)$ (4G models are arbitrarily chosen, 4G+ models could have been used for equivalent results), increasing the total mass directly increases the radius $r$ at every layer of a given model. The $\lvert N \rvert/r$ quantity then decreases, so the mean period spacing increases, which is indeed observed on Fig.~\ref{fig:4G_4GP_mass}.

\subsubsection{Envelope mass}
\label{subssec:4G_lqenv}
\begin{figure}[h!]
\centering
\includegraphics[width=0.49\textwidth]{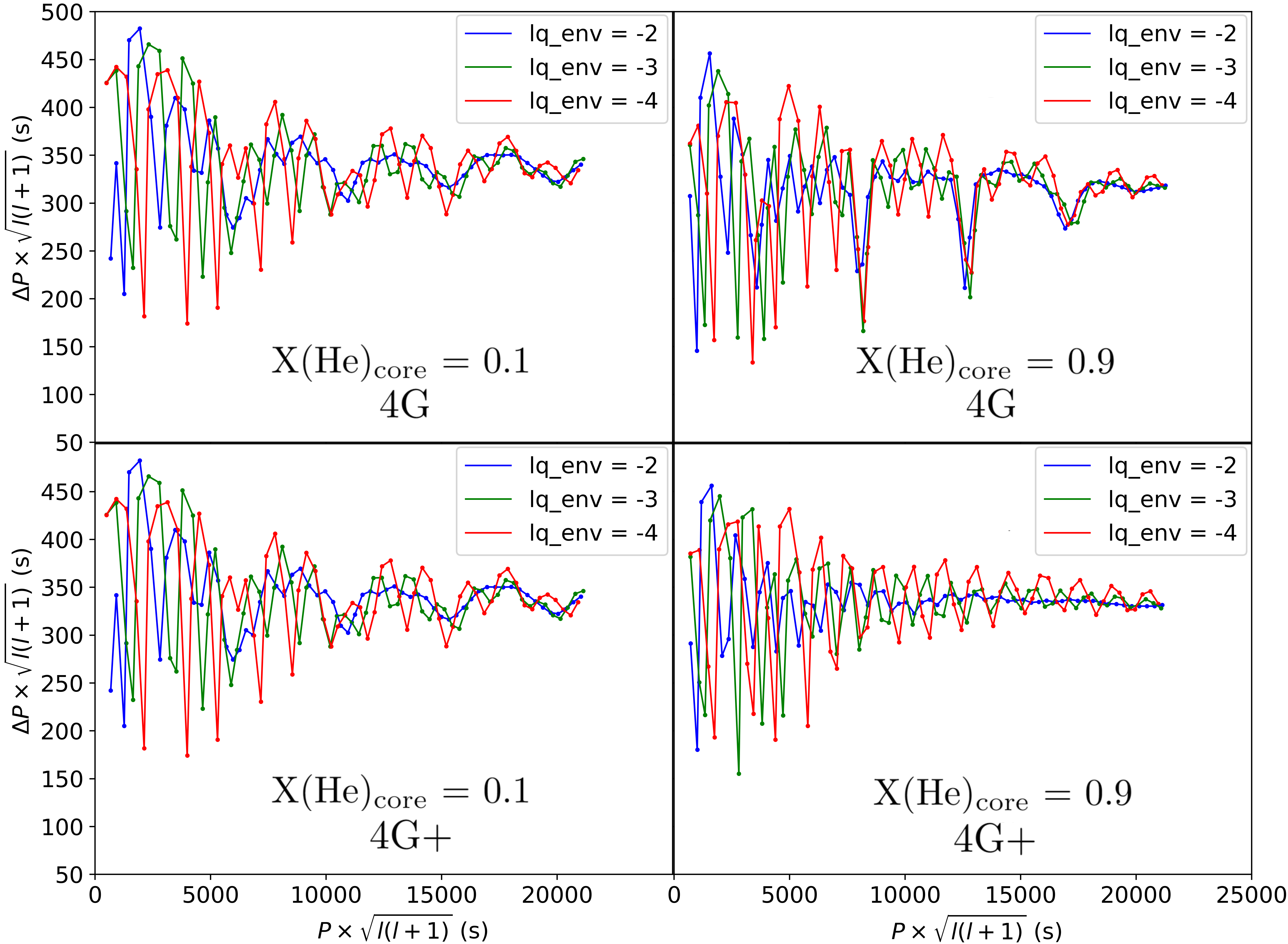}
\caption{Influence of the envelope mass on the pulsation spectra of 4G and 4G+ models. All models share fixed $M_*=0.47 M_\odot$ and lq\_core~ $=-$0.25. \textit{Top panels}: 4G models with lq\_env~$=-2$ (blue), $-$3 (green), $-$4 (red) at X(He)$_{\rm core}=0.1$ (left) and 0.9 (right). \textit{Bottom panels}: Same as top panels, but for 4G+ models.}  
\label{fig:4G_4GP_lqenv}
\end{figure}

We now analyze the impact of varying the envelope mass on the pulsation spectra by computing three 4G and 4G+ models, sharing lq\_core$=-$0.25 and $M_*=0.47 M_\odot$ but with varying lq\_env~$=-$2, $-$3, and $-$4 (note that lq\_env~$=-$2 is a more massive envelope than lq\_env~$=-$4). In Fig.~\ref{fig:4G_4GP_lqenv}, we show on top panels the superposed pulsation spectra from 4G models of lq\_env~$=-$2 (blue), $-$3 (green) and $-$4 (red), at X(He)$_{\rm core}=0.9$ (right) and X(He)$_{\rm core}=0.1$ (left). Bottom panels show the same quantities for 4G+ models. 

\begin{figure}[h!]
\centering
\includegraphics[width=0.49\textwidth]{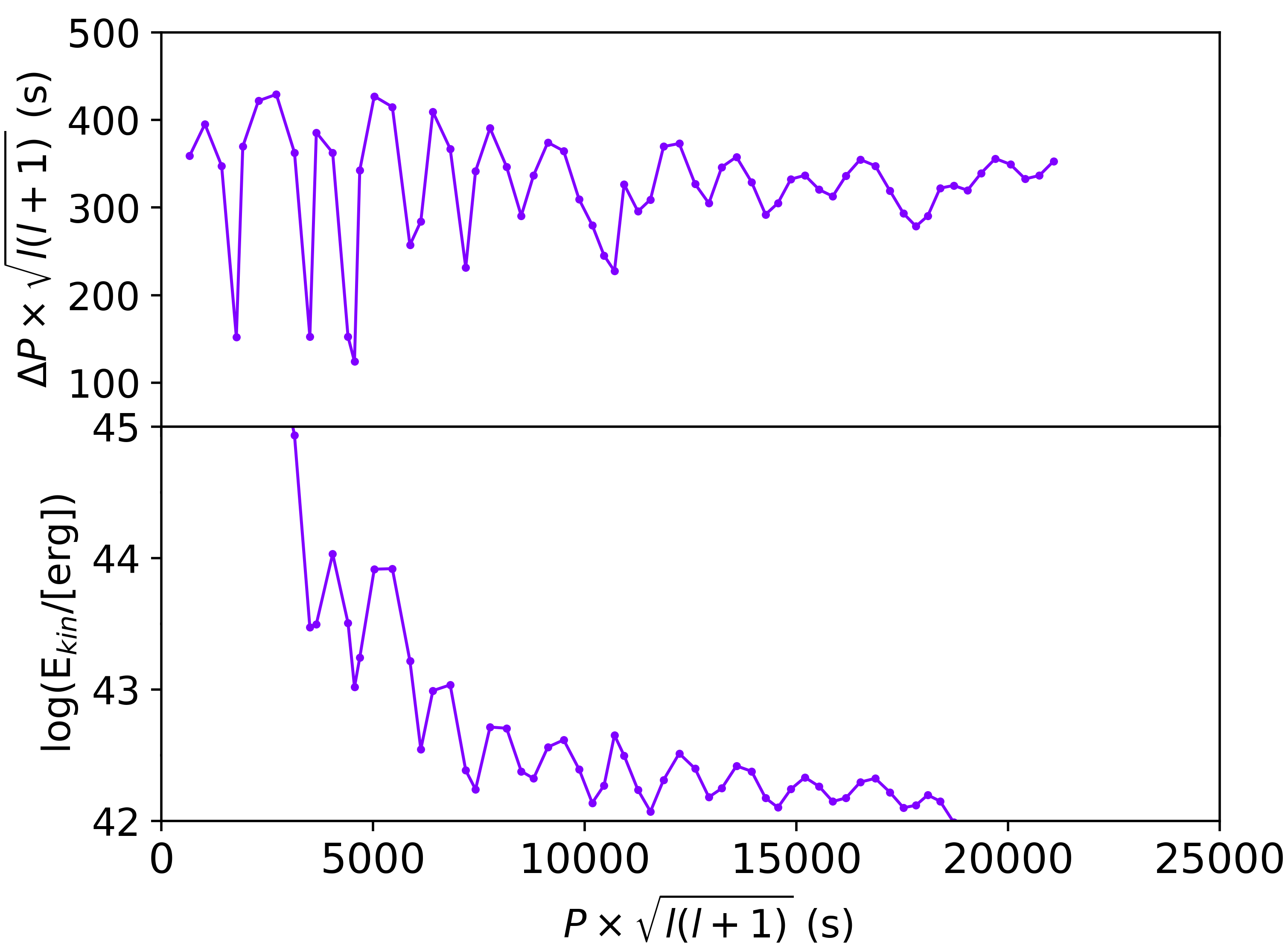}
\caption{Pulsation spectrum and its associated kinetic energy at X(He)$_{\rm core}$ = 0.9 from a 4G+ model for $M_*=0.47 M_\odot$, lq\_env~$=-$4 and lq\_core~$= -0.25$.}  
\label{fig:4GP_chaotic_Ekin}
\end{figure}

Whether for 4G or 4G+ models, and for any X(He)$_{\rm core}$, pulsation spectra at lower envelope mass lq\_env~$=-$3 and $-$4 show more rapid variations of the period spacing from mode to mode, compared to the lq\_env~$=-$2 model. While the spacing between those variations decrease with the envelope mass, their amplitude is increasing, as observed in the lq\_env~$=-$3 and $-$4 models. In addition, we give on Fig.~\ref{fig:4GP_chaotic_Ekin} a split panel with on top the pulsation spectrum from a 4G+ model at $M_*=0.47 M_\odot$, lq\_env~$=-$4, lq\_core~$=-$0.25, at X(He)$_{\rm core}=0.9$; and on the bottom its associated kinetic energy, showing that the latter directly follows the variations of period spacing found at higher lq\_env. Such pulsation spectra and kinetic energy point towards a possible secondary trapping phenomenon in the envelope of the star, as minima of period spacing are now corresponding to minima of kinetic energy (however, trapped modes associated to lq\_core (core-mantle transition) are still maxima of the kinetic energy). Indeed, note that while this phenomenon affects modes of all radial orders, it does not prevent other forms of trapping observed up to now, for example the trapping in the radiative cavity below lq\_core found in 4G models at X(He)$_{\rm core}=0.9$, or the wavy pulsation spectra found at X(He)$_{\rm core}=0.1$ for both 4G and 4G+ models. Both behaviors remain for each lq\_env values shown in Fig.~\ref{fig:4G_4GP_lqenv} (top-right panel for 4G models radiative cavity trapping at X(He)$_{\rm core}=0.9$, left panels for the wavy pulsation spectra at X(He)$_{\rm core}=0.1$).

\begin{figure}[t]
\centering
\includegraphics[width=0.49\textwidth]{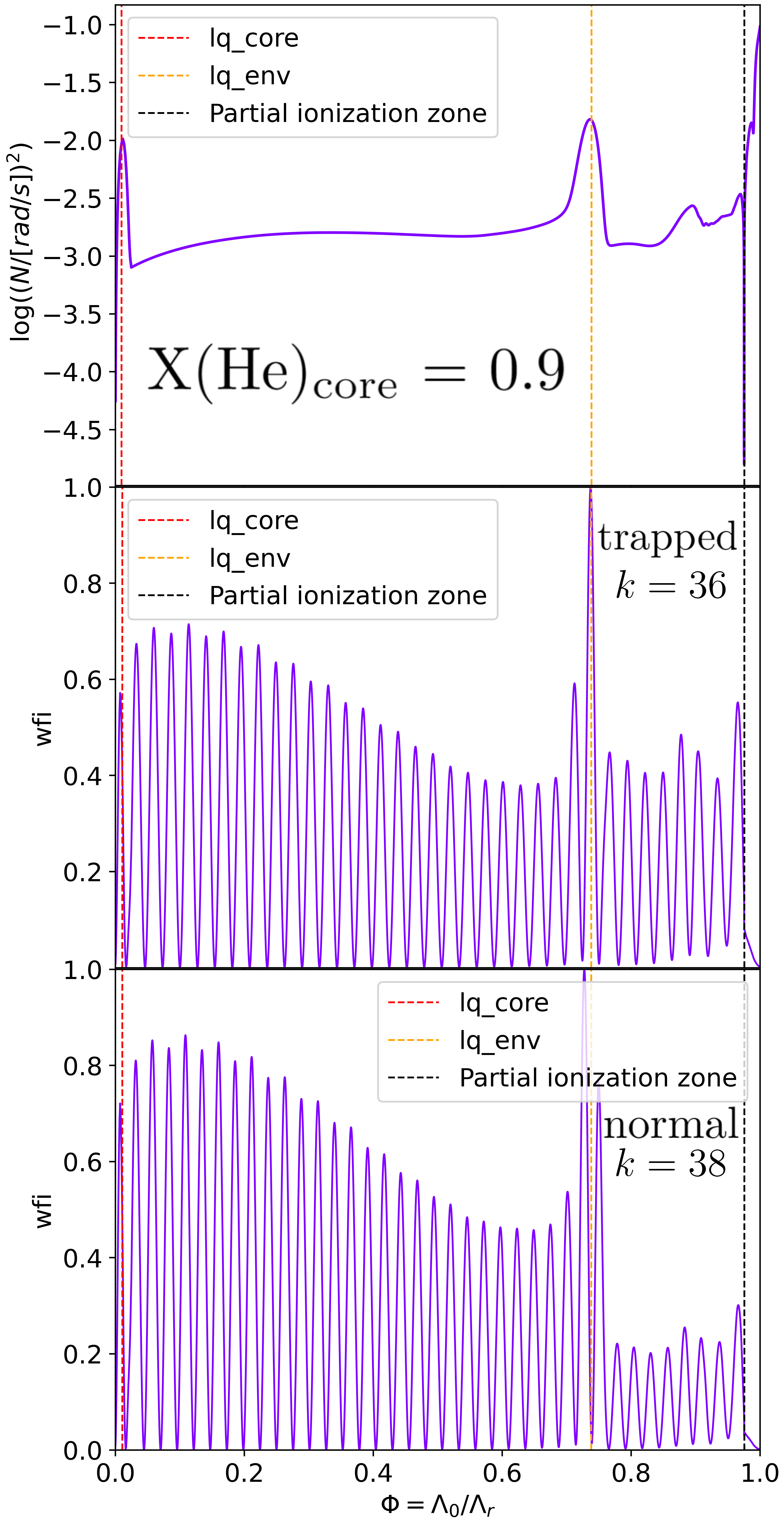}
\caption{\textit{Top panel}: $\log(N^2)$ as a function of the normalized buoyancy radius ($\phi = \Lambda_0/\Lambda_r$) at X(He)$_{\rm core}=0.9$ for a 4G+ model having $M_*=0.47 M_\odot$, lq\_core~$=-$0.25, lq\_env~$=-$4, and X(He)$_{\rm core}=0.9$. The red dotted vertical line is the lq\_core (core-mantle) transition, the orange dotted vertical line is the lq\_env (mantle-envelope) transition, and the black dotted vertical line is the bottom of the partial ionization zone of iron. \textit{Middle and bottom panels:} Weight functions (wfi) of two modes: a mode trapped between lq\_env and the bottom of the partial ionization zone (middle) and a normal mode (bottom).}
\label{fig:4GP_lqenv_wfi_09}
\end{figure}

It is important to point out that g-mode sdB pulsators seemingly have an envelope mass around lq\_env~$=-2$ \citep{2010ApJ...718L..97V,2010A&A...524A..63V,2011A&A...530A...3C,2019A&A...632A..90C}.  Therefore, studying the effect of smaller envelope masses at lq\_env~$=-$3 and $-$4 on pulsation spectra is more of a theoretical exercise. Still, it is possible to identify in which cavity are trapped the modes corresponding to these additional rapid variations in period spacing seen in Fig.~\ref{fig:4G_4GP_lqenv}. The top panel of Fig.~\ref{fig:4GP_lqenv_wfi_09} shows, for a 4G+ model\footnote{As 4G+ models do not show mode trapping from the lq\_core transition at high X(He) (see Sect.~\ref{XHe_4GP}), they are ideal to study this secondary trapping phenomenon in the envelope.} having $M_*=0.47M_\odot$, lq\_core~$=-$0.25, and lq\_env~$=-$4, the Brunt-Väisälä frequency as a function of the normalized buoyancy radius at X(He)$_{\rm core}=0.9$. The vertical black dotted line on the far right references the bottom of the convective zone seen at $\log(q)\sim-$9 in the top panel of Fig.~\ref{fig:propa_diag}, which is a convective zone due to partial ionization of iron. The middle and bottom panels of Fig.~\ref{fig:4GP_lqenv_wfi_09} show the weight functions of a trapped ($k=36$; middle panel) and a normal mode ($k=38$; bottom panel) with respect to the rapid variations of the period spacing, as a function of the normalized buoyancy radius. Between the mantle-envelope chemical transition at lq\_env (orange dotted vertical line) and the bottom of the partial ionization zone (black dotted vertical line), the amplitude of the weight function of a trapped mode is significantly higher than for a normal mode. 
This allows us to identify the trapping cavity of this secondary trapping phenomenon as the region between mantle-envelope transition at lq\_env and the bottom of the partial ionization zone.

\begin{figure}[t]
\centering
\includegraphics[width=0.49\textwidth]{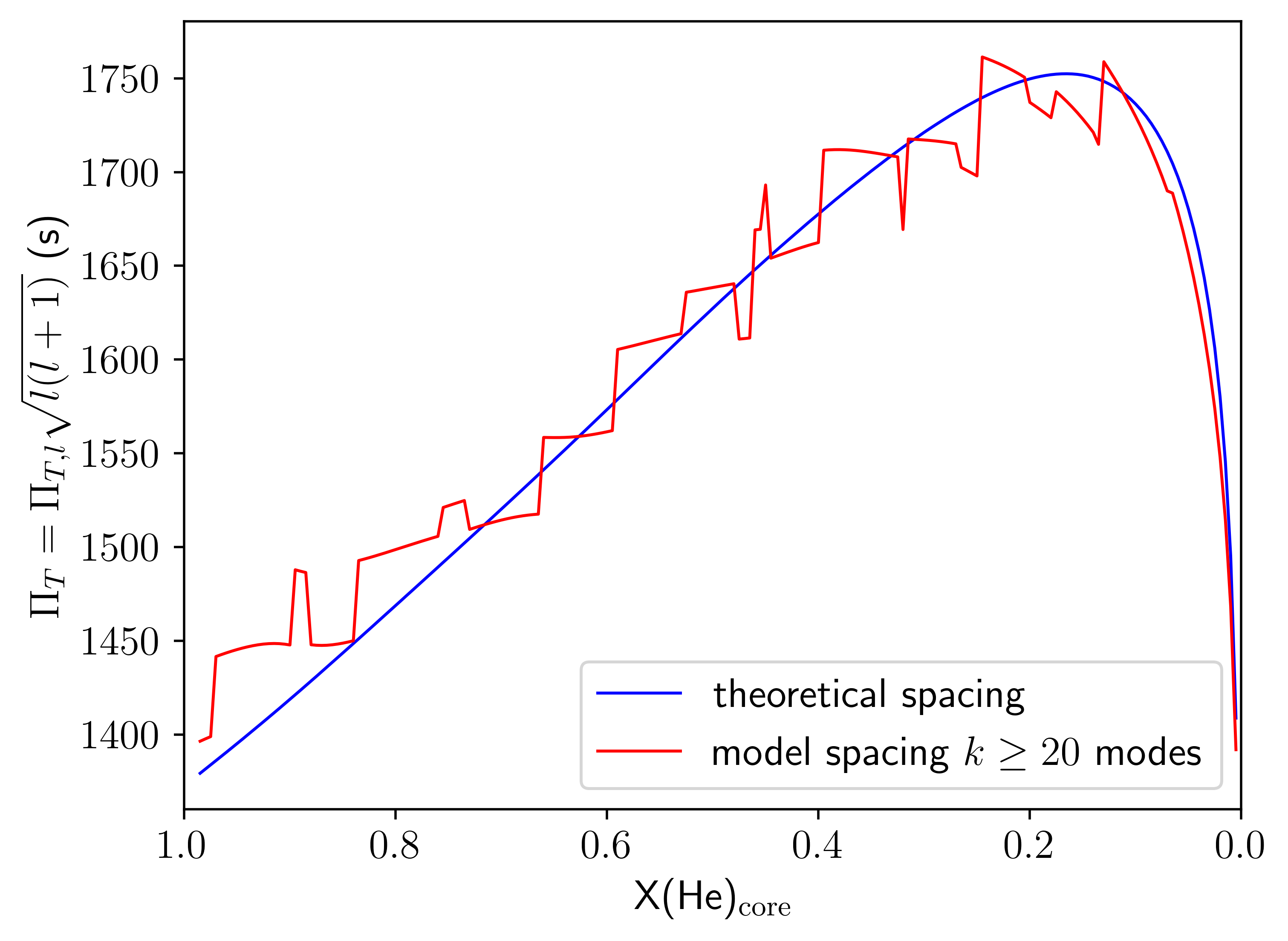}
\caption{Theoretical mean reduced period spacing between two consecutive modes trapped between the lq\_env chemical transition and the bottom of the partial ionization zone (blue), the same spacing from models (red), for modes of $k\geq 20$ (i.e., not including low-order modes). Computed from a 4G+ model with $M_*=0.47 M_\odot$, lq\_core~$=-$0.25 and lq\_env~$=-$4.}
\label{fig:4GP_spacing_lq_env}
\end{figure}

In Fig.~\ref{fig:4GP_spacing_lq_env}, as done previously, we further compare the theoretical period spacing between consecutive trapped modes in this region from asymptotic theory (blue curve), to the same period spacing derived directly from pulsation spectra for modes of radial order $k\geq 20$ (red curve). The numerical integral for the theoretical spacing is computed from Eq.~(\ref{eq_spacing}) between the middle of the lq\_env transition ($r_b$) and the bottom of the partial ionization zone ($r_t$). We find again an excellent agreement, with a relative difference   of the order of 1.5\% and always less than 4\%.

Finally, we can observe that the amplitude of the rapid variations of period spacing on Fig.~\ref{fig:4G_4GP_lqenv} clearly decreases with increasing period. \citet{2008MNRAS.386.1487M} showed that regions of sharp chemical gradient, inducing sharp variation of the Brunt-Väisälä frequency (in their case, a step-like chemical transition), should produce a periodic signature in pulsation spectra that does not decrease in amplitude with the period, while regions of smoother chemical gradient (in their case, a ramp function) should produce a periodic signature that decreases in amplitude with increasing period. In our models, the chemical gradient at lq\_core is much sharper than the one at lq\_env (see the Brunt-Väisälä frequency in Figs.~\ref{fig:propa_diag} and \ref{fig:chem_transi_influence}, for example). The observed decrease in amplitude of the rapid variations of the period spacing with the period found in our pulsation spectra could therefore be due to the smoothness of the mantle-envelope lq\_env chemical transition. This is further supported by \citet{Charpinet2014a} showing that above a radial order threshold, the local wavelength of a g-mode (the separation between two consecutive nodes of a given radial order) becomes small enough with respect to the width of lq\_env chemical transition, reducing its ability to trap modes.

\section{The g-mode spectra of evolutionary models}
\label{evol}
We finally turn our attention to evolutionary STELUM models. They are by essence distinct from the static ones, as a model at a given X(He)$_{\rm core}$ depends on those previously computed. Additionally, overshooting and semiconvection are implemented in those models (see Sect.~\ref{tools}). As those phenomena occur at the boundary between the convective core and the radiative mantle, we expect the pattern of the pulsation spectra of g-modes to be directly impacted. We computed 6 evolutionary tracks, of a same total mass 0.47 $M_\odot$, and of varying envelope mass lq\_env $= -2, -2.5, -3, -4, -4.5$, and $-$5. Unlike static models, the boundary corresponding to the lq\_core parameter is not fixed anymore (and, of course, is not specified as an input parameter), as the C-O-He/He chemical transition shifts with the growth of the convective core and with the mixing occurring above it. Let us mention that we stopped the evolutionary sequences at X(He)$_{\rm core} \sim 0.2$, before the onset of an instability called Breathing Pulses that is triggered in those models. Discussing the reality and physicality of such Breathing Pulses is not within the scope of this paper. 

\subsection{Impact of core-He burning in evolutionary models}

\begin{figure}[h!]
\centering
\includegraphics[width=0.49\textwidth]{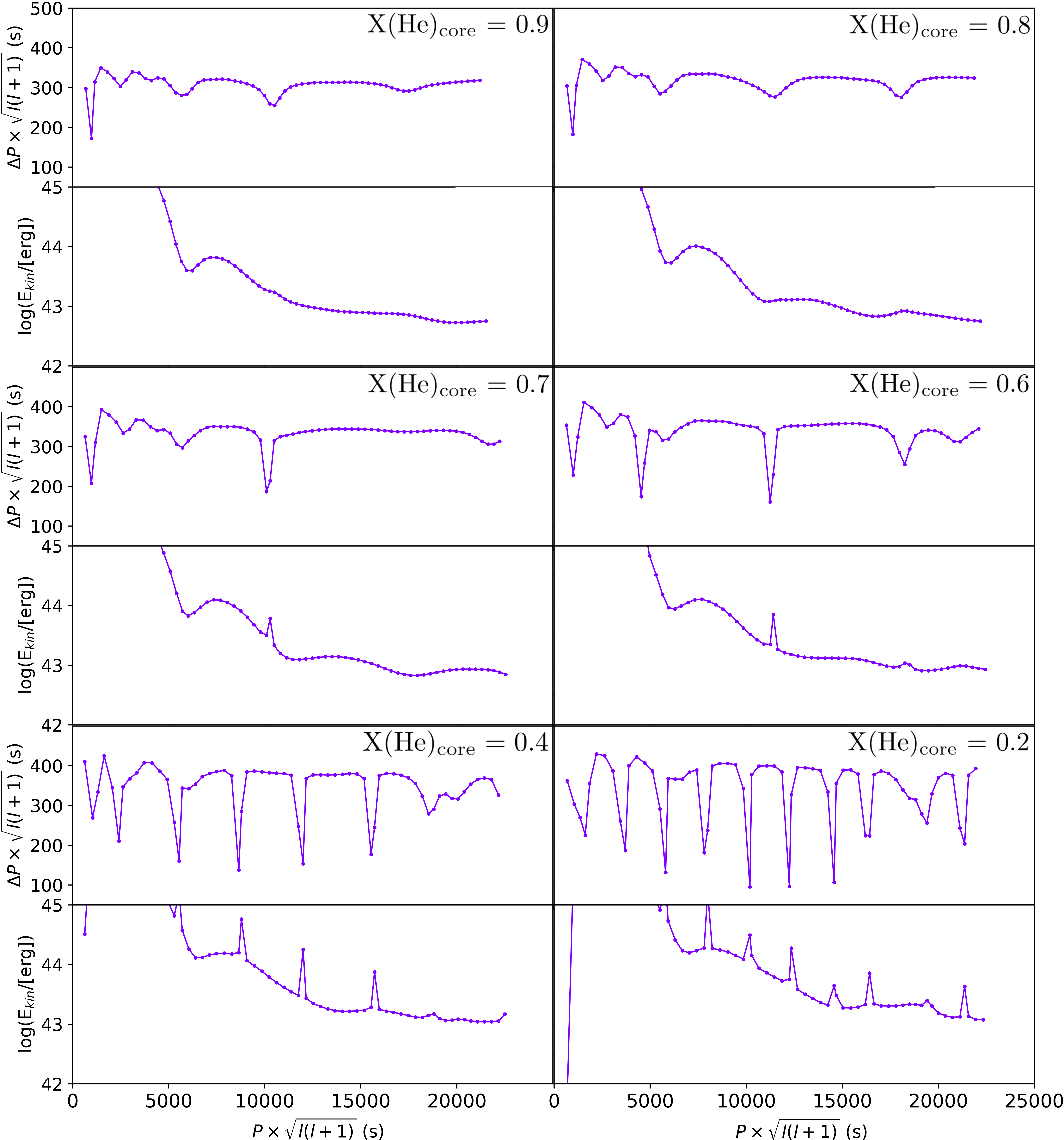}
\caption{Pulsation spectra (top panels) and associated kinetic energies (bottom panels) of an evolutionary sequence for $M_*=0.47M_\odot$ and lq\_env~$=-$2, at different X(He)$_{\rm core}$.}
\label{fig:Evol_hecore}
\end{figure}

Similarly to static models, we first study the impact of core-He burning on the pulsation spectra. Pulsation spectra and kinetic energies for the evolutionary sequence having $M_*=0.47M_\odot$ and lq\_env~$=-$2 are presented in Fig.~\ref{fig:Evol_hecore}, for X(He)$_{\rm core}=0.9$, 0.8, 0.7, 0.6, 0.4, and 0.2. This figure can be directly compared to Fig.~\ref{fig:4G_hecore} (reference 4G model) and Fig.~\ref{fig:4GP_hecore} (reference 4G+ model) for a given X(He)$_{\rm core}$. In evolutionary models, we distinguish two main different behaviors of the pulsation spectra: before and after the onset of semiconvection, which occurs at X(He)$_{\rm core}\sim 0.7$ (see Fig.~\ref{fig:masses_evol}, green curve).
Before the onset of semiconvection, the pulsation spectra of evolutionary models (top left and right panels of Fig.~\ref{fig:Evol_hecore}, for X(He)$_{\rm core}=0.9$ and 0.8, respectively) show a few minima in the period spacing. 

\begin{figure}[h!]
\centering
\includegraphics[width=0.40\textwidth]{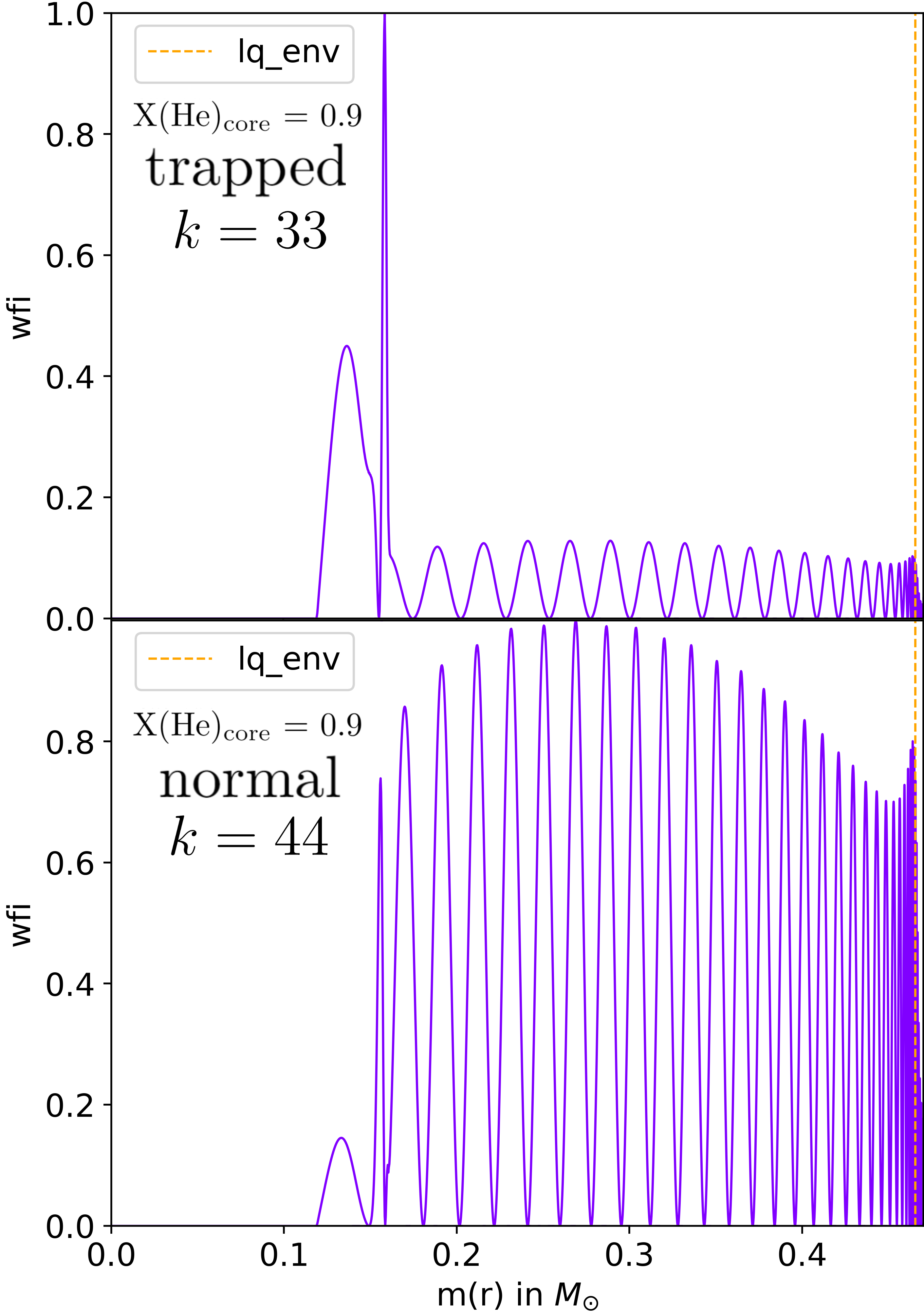}
\caption{Weight functions (wfi) of pulsation modes from an evolutionary model for $M_*=0.47 M_\odot$ and lq\_env~$=-$2, at X(He)$_{\rm core}=0.9$, highlighting the overshooting region as the trapping cavity. \textit{Top panel}: Trapped mode ($k=33$). \textit{Bottom panel}: Normal mode ($k=44$).}  
\label{fig:Evol_hecore_wfi_09}
\end{figure}

At high X(He)$_{\rm core}$, the core part of the evolutionary models is composed of a convective core topped by a small overshooting region (which is radiative), before the C-O-He/He transition from the core to the radiative mantle (see Fig.~\ref{fig:chem_evol} and Fig.~\ref{fig:masses_evol}). This chemical structure is similar to the one found for mid lq\_core values (around lq\_core~$=-$0.25) of 4G models before the convective core has reached lq\_core (Fig.~\ref{fig:4G_core_growth}), in which we find a convective core also topped by a radiative cavity below lq\_core. As such, there is a direct analogy between modes trapped in the radiative cavity below lq\_core of 4G models (Fig.~\ref{fig:4G_hecore}, top panels), and trapped modes at X(He)$_{\rm core}=0.9$ and 0.8 for evolutionary models, which are then trapped in the radiative overshooting region below the C-O-He/He transition. By analyzing the weight functions of a normal and a trapped mode in an evolutionary model at X(He)$_{\rm core}=0.9$, which are shown on Fig.~\ref{fig:Evol_hecore_wfi_09}, we see that the trapped mode ($k=33$; top panel) displays higher weight function values in the overshooting region than in the radiative mantle, while the normal mode ($k=44$; bottom panel) shows the opposite, thus anchoring the overshooting region as the trapping cavity. This situation is similar to Fig.~\ref{fig:4G_hecore_wfi_09}, showing a trapped and normal mode in the radiative cavity of a 4G model at X(He)$_{\rm core}=0.9$ and lq\_core~$=-$0.25. Let us note however that trapped modes at high X(He)$_{\rm core}$ for evolutionary models are shallower and less frequent than those of the same nature observed in the 4G models at similar X(He)$_{\rm core}$ (compare Fig.~\ref{fig:Evol_hecore} and Fig.~\ref{fig:4G_hecore}). This is a direct consequence of the overshooting region in evolutionary models being generally of smaller size in $\log(q)$ than the trapping radiative region under lq\_core of 4G models. While the former is relatively constant throughout core-He burning, with a $\Delta \log(q) \sim 0.0298$ (corresponding to $\Delta m(r) = 0.03 M_\odot$ in Fig.~\ref{fig:masses_evol}, red curve), the latter is much larger in static models, with a width of $\Delta \log(q) \sim 0.3$ at equivalent X(He)$_{\rm core}$ (before shrinking to becoming null when the convective core reaches lq\_core).

\begin{figure}[h!]
\centering
\includegraphics[width=0.40\textwidth]{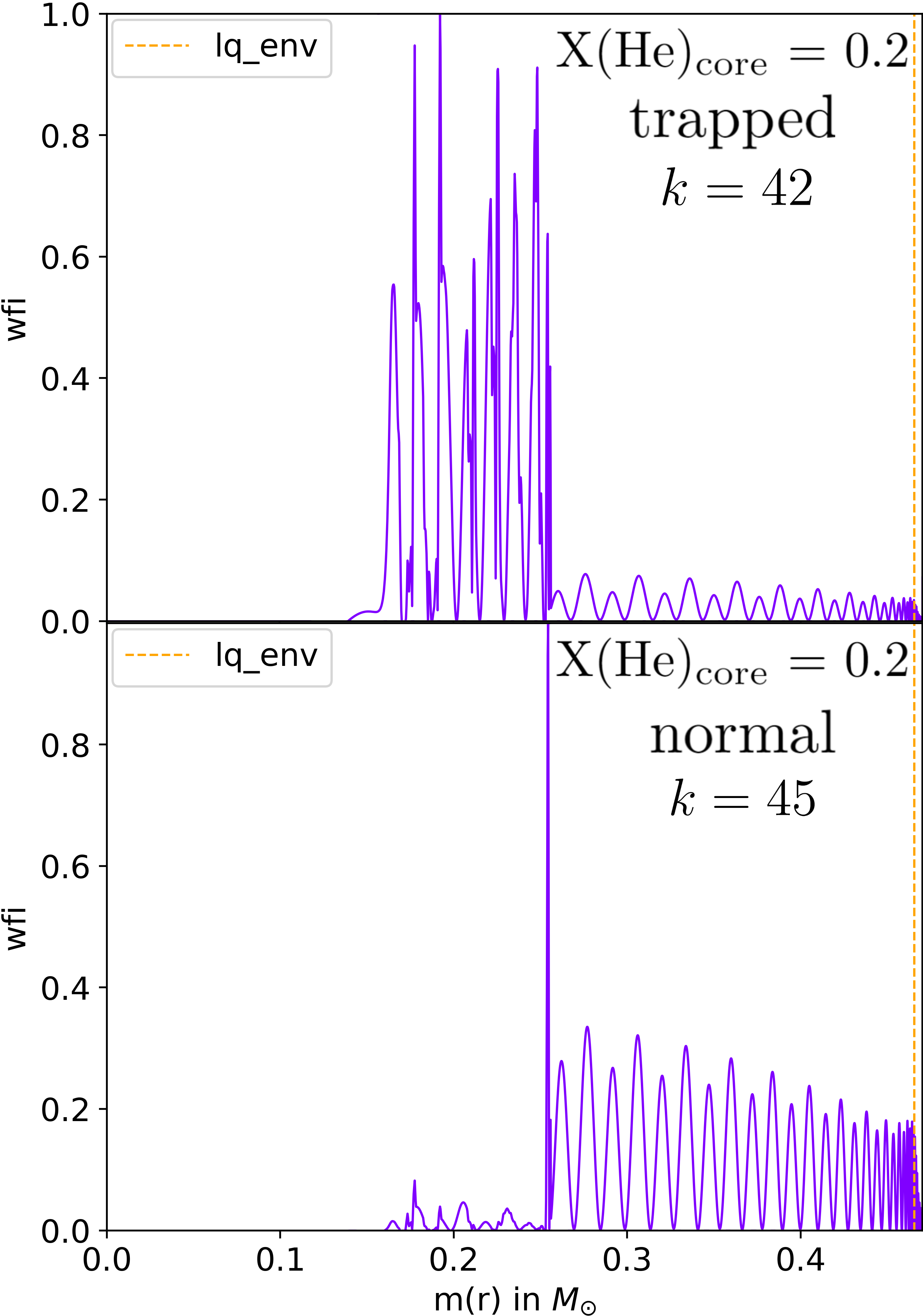}
\caption{Weight functions (wfi) of pulsation modes from an evolutionary model for $M_*=0.47 M_\odot$ and lq\_env~$=-$2, at X(He)$_{\rm core}=0.2$, highlighting the semiconvection region as the trapping cavity. \textit{Top panel}: Trapped mode ($k=42$). \textit{Bottom panel}: Normal mode ($k=45$).}   
\label{fig:Evol_hecore_wfi_02}
\end{figure}

After the onset of semiconvection at X(He)$_{\rm core}\sim 0.7$, the structure of an evolutionary model drastically changes. Instead of a growing convective core topped by an overshooting zone, we have for X(He)$_{\rm core}<0.7$ a convective core more or less fixed in mass (slightly decreasing with evolution), topped by an overshooting zone also of more or less constant mass, itself topped by a growing semiconvective region before the C-O-He/He chemical transition (Fig.~\ref{fig:masses_evol}). As shown in Fig.~\ref{fig:Evol_hecore} (middle and bottom panels), we find strongly trapped modes for X(He)$_{\rm core}<0.7$, with increasing depths for the minima and decreasing period spacing between two minima as long as  X(He)$_{\rm core}$ decreases. In Fig.~\ref{fig:Evol_hecore_wfi_02}, we show for X(He)$_{\rm core}=0.2$ the weight function of a trapped mode ($k=42$; top panel), displaying high amplitudes in the semiconvective region while being almost null in the radiative mantle, while conversely a normal mode ($k=45$; bottom panel) shows instead a high weight function in the radiative mantle and low one in the semiconvection zone. Therefore, we directly identify the semiconvection region as the trapping cavity for trapped modes at X(He)$_{\rm core}<0.7$ in evolutionary models. The ``strength'' of mode trapping increasing with decreasing X(He)$_{\rm core}$ is explained by the increase in chemical gradient at the core-mantle transition, and the growth of the semiconvection region, hence of the trapping cavity. Additionally, we note the atypical ``fuzzy'' shape of the weight functions of Fig.~\ref{fig:Evol_hecore_wfi_02} compared to previous ones (see Fig.~\ref{fig:4G_hecore_wfi_09}, Fig.~\ref{fig:4G_hecore_wfi_01}, or Fig.~\ref{fig:4G_lqcore_wfi_01}), which originates from the approach taken in STELUM to emulate semiconvection (see Sect~\ref{description_evol} and Fig.~\ref{fig:grad_evol}, bottom, in particular). Finally, while the overshooting zone is still present for X(He)$_{\rm core}<0.7$, the modes have 0, 1 or 2 radial nodes in this region. We observe that for both normal and trapped modes, the weight function in the overshooting region is near null, and  therefore the impact of the overshooting region is negligible on the pulsation spectra after the onset of semiconvection.

\subsection{Influence of envelope mass}

\begin{figure}[h!]
\centering
\includegraphics[width=0.49\textwidth]{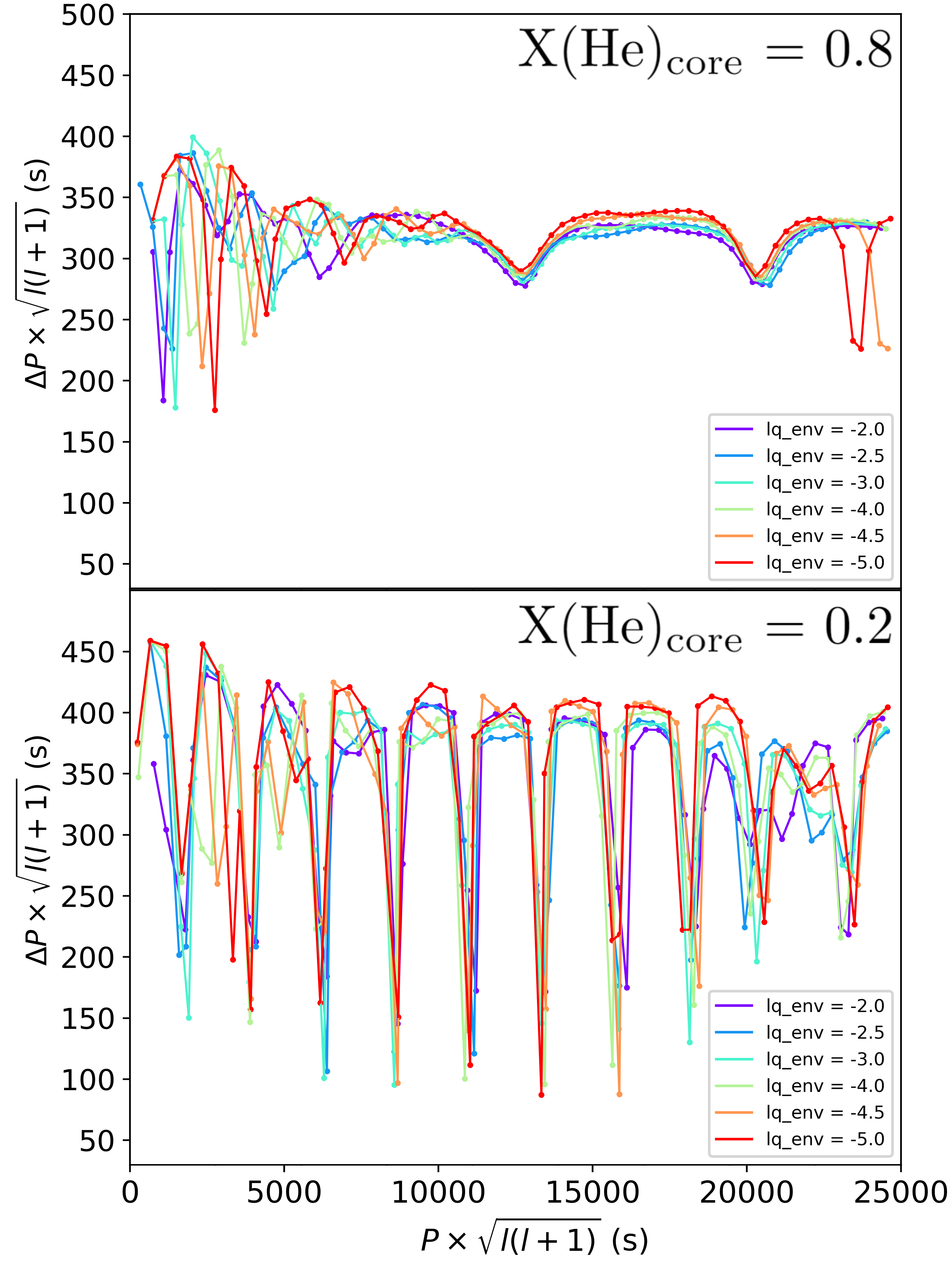}
\caption{Influence of the mass of the envelope on the pulsation spectra of evolutionary models sharing fixed $M_*=0.47 M_\odot$ and varying lq\_env~$=-2.0, -2.5, 3.0, -4.0, -4.5$ and $-$5.0 at X(He)$_{\rm core}=0.8$ (top), and X(He)$_{\rm core}=0.2$ (bottom).}  
\label{fig:Evol_lqenv}
\end{figure}

\begin{figure}[h!]
\centering
\includegraphics[width=0.49\textwidth]{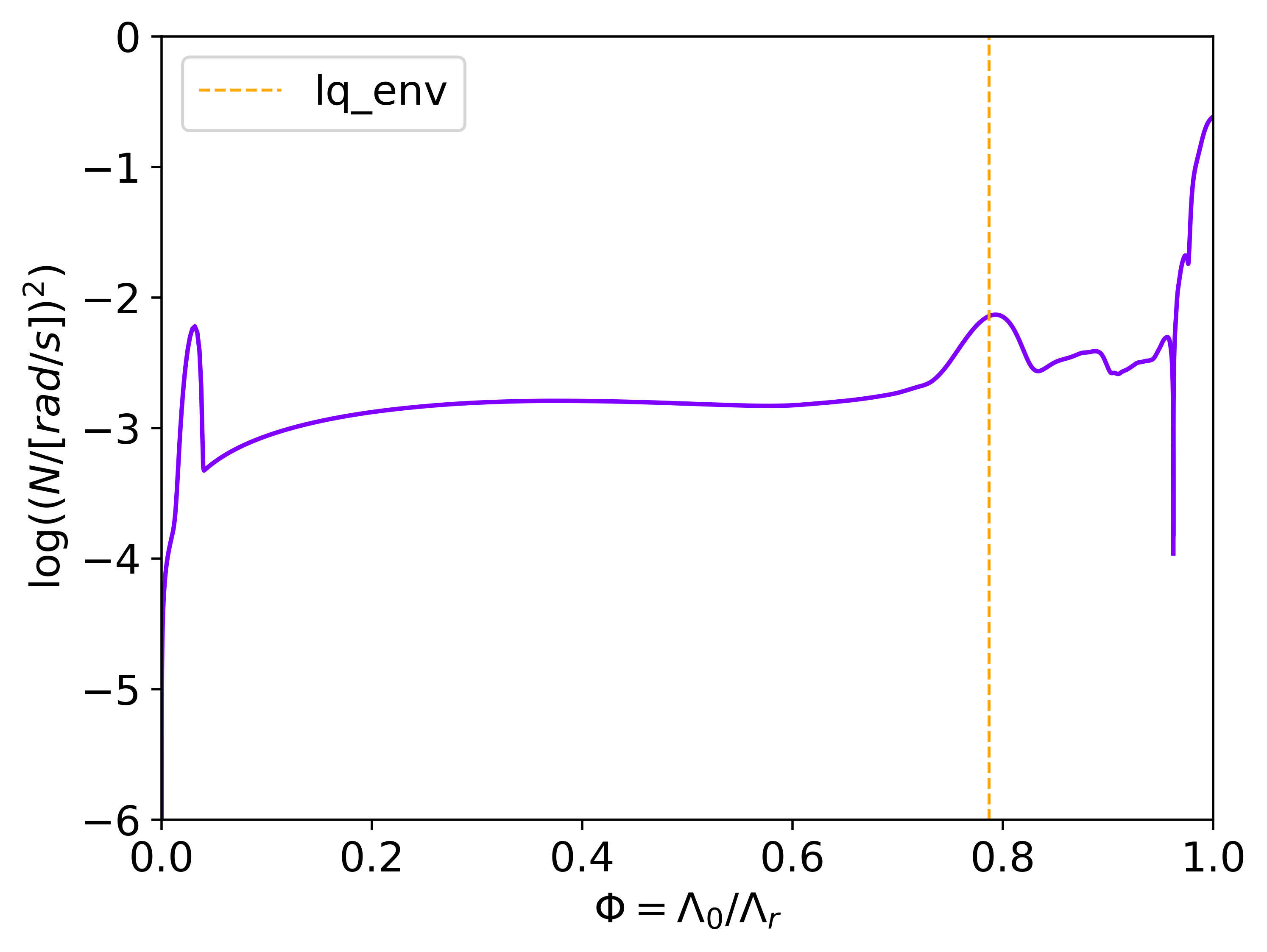}
\caption{$\log(N^2)$ as a function of the normalized buoyancy radius at X(He)$_{\rm core}=0.9$ for an evolutionary model with lq\_env~$=-5.0$. The orange dotted vertical line is the lq\_env (mantle-envelope) transition.}  
\label{fig:BN_Evol_dh500}
\end{figure}

In this subsection, we analyze the impact of changing the mass of the envelope in evolutionary models. We display in Fig.~\ref{fig:Evol_lqenv} the pulsation spectra of evolutionary models with lq\_env~$=-2, -2.5, -3, -4, -4.5$, and $-$5, at X(He)$_{\rm core}=0.8$ (top panel) and X(He)$_{\rm core}=0.2$ (bottom panel). While low radial-order modes notably differ, especially at higher X(He)$_{\rm core}$, high radial-order modes are left mostly unaffected by changing the envelope mass, and we do not observe the strong variations of the reduced period spacing previously found for lq\_env~$=-$3 and $-$4 in 4G and 4G+ models (Fig.~\ref{fig:4G_4GP_lqenv}). Figure~\ref{fig:BN_Evol_dh500} shows the  Brunt-Väisälä frequency of a lq\_env~$=-5.0$ evolutionary model at X(He)$_{\rm core}=0.9$, as a function of the normalized buoyancy radius: we observe that the lq\_env (mantle-envelope) transition of evolutionary models is much smoother than the lq\_env transition in 4G and 4G+ static models due to He diffusion present in evolutionary models (Fig.~\ref{fig:4GP_lqenv_wfi_09} for lq\_env~$=-4.0$, top panel, and Fig.~\ref{fig:4G_hecore_wfi_09} or Fig.~\ref{fig:4G_hecore_wfi_01} for lq\_env~$=-2.0$, top panels). Following the conclusion of Sect.~\ref{subssec:4G_lqenv}, we suggest to attribute the absence of rapid variations of the period spacing in evolutionary models at lq\_env~$=-3.0$ to $-5.0$ to the smoothness of the mantle-envelope lq\_env chemical transition (\citealt{Charpinet2014a, 2008MNRAS.386.1487M}).
Additionally, the separation between consecutive trapped modes is unaffected by changes in lq\_env, as was already the case for static models (looking beyond the variations of period spacing from mode to mode; see top right panel of Fig.~\ref{fig:4G_4GP_lqenv} for example). We note, however, that (1) the asymptotic period spacing do depend slightly on the mass of the envelope, with a higher value for the smaller envelopes; and (2) the depth of the minima of period spacing do also depends on lq\_env, especially for small X(He)$_{\rm core}$. 

\subsection{Asymptotic period spacings}
\label{asympto_per_spa}
We conclude our study of evolutionary models by comparing the theoretical and model values of the mean period spacing between modes of adjacent radial orders, and of the mean period spacing between consecutive trapped modes. We computed the theoretical period spacings $\Pi_0$ and $\Pi_T$ in evolutionary models by numerical integration over their respective regions using Eq.~(\ref{eq_spacing}): $r_b$, the lower integral boundary, is the top of the convective core (aka the bottom of the overshooting region), while $r_t$, the upper integral boundary, is the stellar radius $R$ for the $\Pi_0$ computation, and the top of the semiconvection zone for the computation of $\Pi_T$ (for X(He)$_{\rm core}<0.7$). 

\begin{figure}[h!]
\centering
\includegraphics[width=0.49\textwidth]{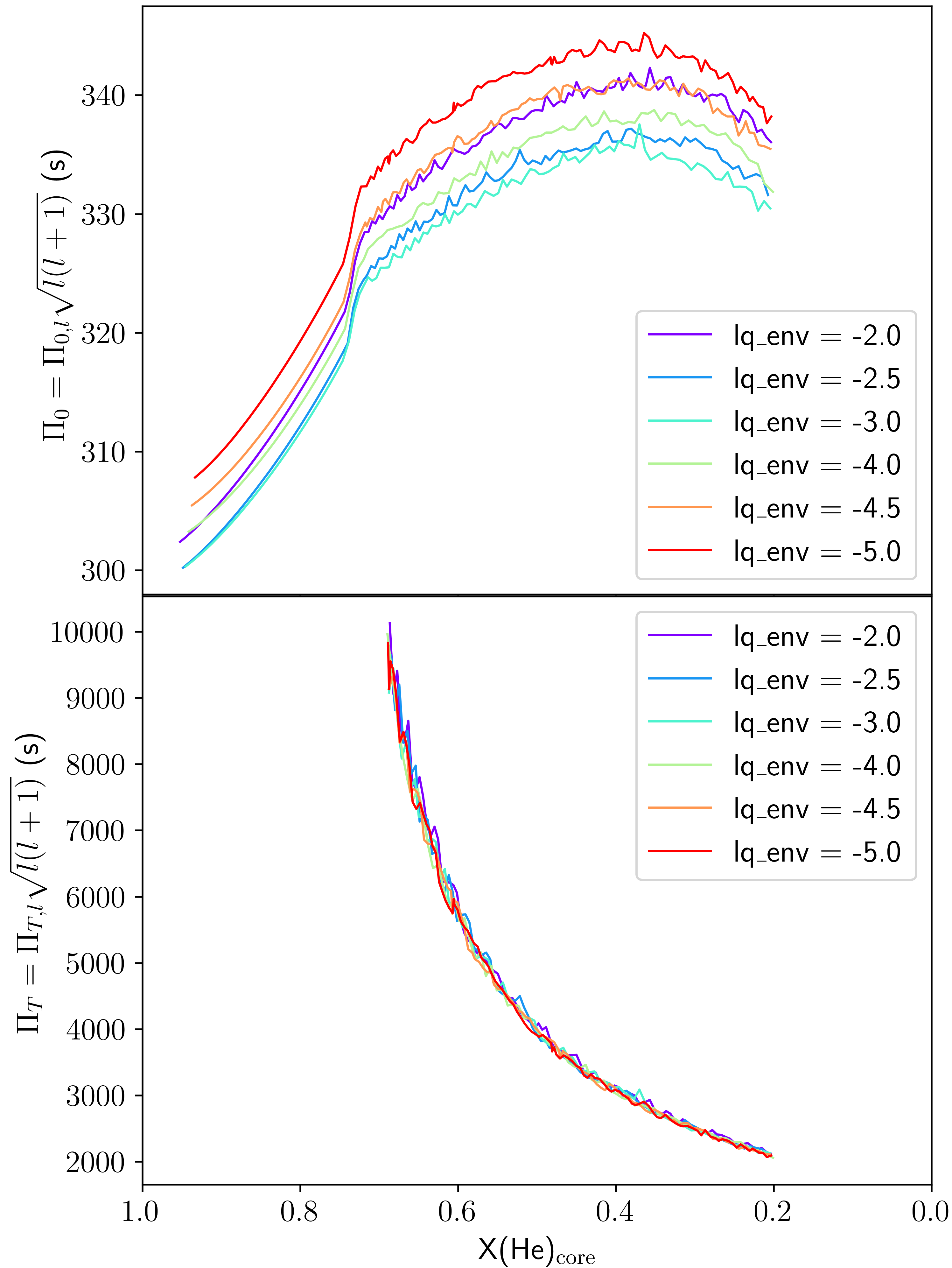}
\caption{\textit{Top panel}: Theoretical mean reduced period spacing between two consecutive modes, for evolutionary models of fixed $M_*=0.47 M_\odot$ and lq\_env~$=-2, -2.5, -3, -4, -4.5$ and $-$5.0. \textit{Bottom panel}: Theoretical mean reduced period spacing between two consecutive trapped modes, for the same evolutionary models.}
\label{fig:asymptotic_spacing}
\end{figure}

Figure~\ref{fig:asymptotic_spacing} (top panel) shows the theoretical asymptotic period spacing between two consecutive modes $\Pi_0$ as a function of X(He)$_{\rm core}$ for evolutionary models with $M_*=0.47M_\odot$ and lq\_env~$=-2.0, -2.5, -3.0, -4.0, -4.5$, and $-5.0$. The general increasing behavior of $\Pi_0$ as long as the star evolves is explained by the global dilatation of the star until X(He)$_{\rm core}\sim 0.35$, below which the core of the star begins to contract, quickly followed by the mantle at slightly lower X(He)$_{\rm core}$. Since the global dilatation includes the propagation cavity of g-modes, $\Pi_0$ first increases, and conversely decreases once the core and mantle contract. We note that throughout any X(He)$_{\rm core}$ down to the end of our evolutionary sequence at  X(He)$_{\rm core} \sim 0.2$, the envelope always dilates. We also observe that the period spacing $\Pi_0$ is of slightly higher values for thinner envelopes, as we noticed earlier in model pulsation spectra (Fig.~\ref{fig:Evol_lqenv}). Additionally, we observe that $\Pi_0$ shows a smooth curve before the onset of semiconvection at X(He)$_{\rm core}>0.7$, while afterwards, we instead find a coarser curve. This is directly due to the semiconvection treatment done in STELUM, which impacts the Brunt-Väisälä frequency, and thus the computed $\Pi_0$. Finally, let us note that the derived $\Pi_0$ values, from $\sim$ 305 s at the beginning of evolution and to about 340 s later in the evolution, correspond broadly with the ones observed in sdB stars \citep[e.g.][]{2021A&A...651A.121U}.

The bottom panel of Fig.~\ref{fig:asymptotic_spacing} shows the theoretical spacing between two consecutive trapped modes $\Pi_T$ as a function of X(He)$_{\rm core}$, for evolutionary models having $M_*=0.47M_\odot$ and lq\_env~$=-2.0, -2.5, -3.0, -4.0, -4.5$, and $-5.0$. We computed this quantity for modes trapped in the semiconvection zone, so the curve is null before its onset at X(He)$_{\rm core}\sim0.7$. Afterwards, we observe a steady decrease of the spacing between consecutive trapped modes $\Pi_T$, which is fully understood since the semiconvection zone grows in mass ($\log(q)$ size, thus expanding integration boundary $r_t$ of Eq.~(\ref{eq_spacing})) with decreasing X(He)$_{\rm core}$ (Fig.~\ref{fig:masses_evol}, green curve), and more pulsation modes can then be trapped in this semiconvection zone. 

\begin{figure}[h!]
\centering
\includegraphics[width=0.49\textwidth]{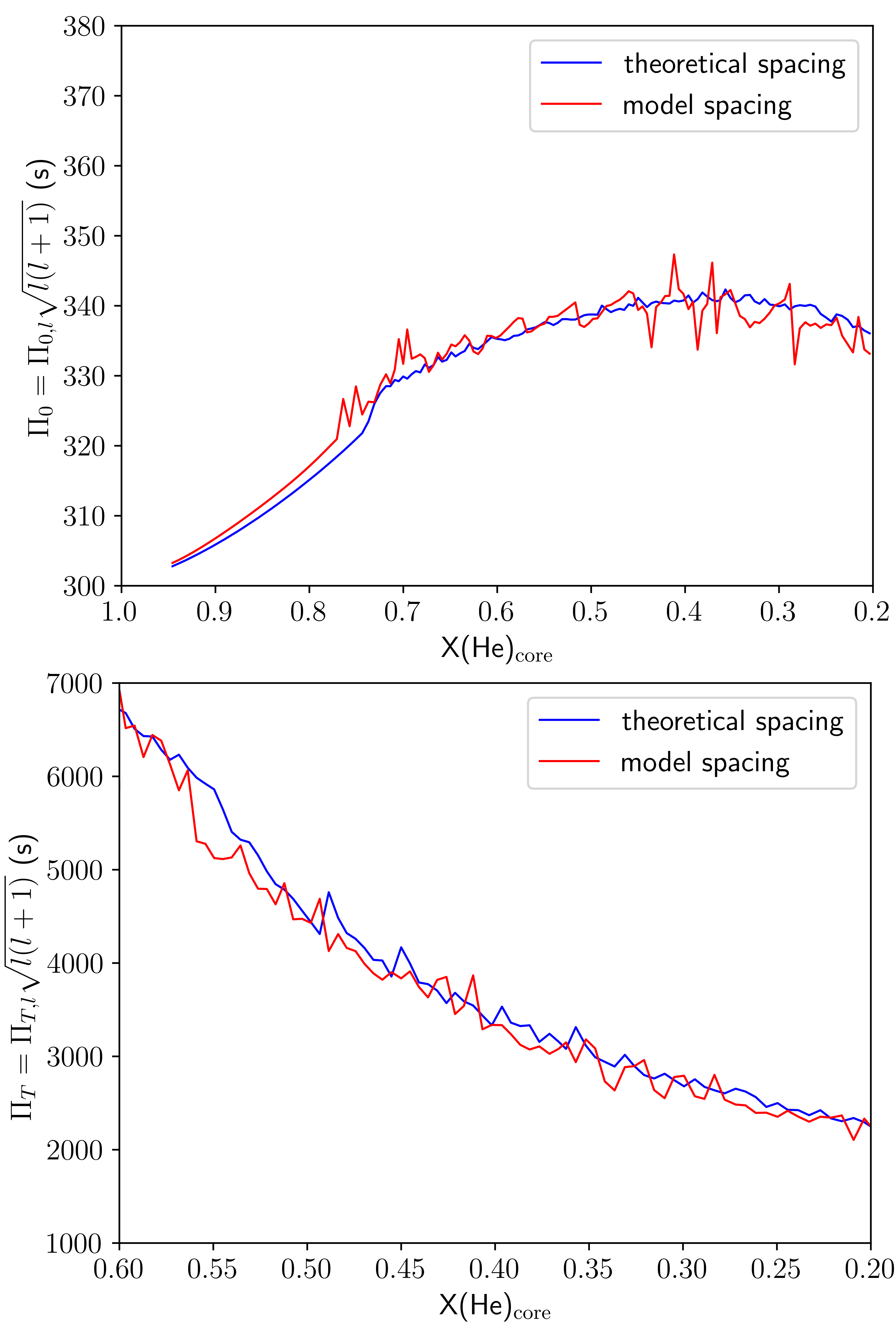}
\caption{\textit{Top panel}: Model (red) and theoretical (blue) mean reduced period spacing between two consecutive modes for an evolutionary model at $M_*=0.47 M_\odot$ and lq\_env~$=-$2. \textit{Bottom panel}: Model (red) and theoretical (blue) mean reduced period spacing between two consecutive trapped modes for an evolutionary model at $M_*=0.47 M_\odot$ and lq\_env~$=-$2.}
\label{fig:spacing_th_obs}
\end{figure}

In Fig.~\ref{fig:spacing_th_obs} (top panel), for an evolutionary model at $M_*=0.47 M_\odot$ and lq\_env~$=-$2  the model (red) and theoretical (blue) we compare the mean period spacing between two consecutive modes $\Pi_0$, and on the bottom the model (red) and theoretical (blue) mean period spacing between two consecutive trapped modes $\Pi_T$. We find a very good agreement between the model period spacing $\Pi_0$ computed from pulsation spectra and the theoretical one computed from a numerical integration of Eq.~(\ref{eq_spacing}) (Fig.~\ref{fig:spacing_th_obs}, top panel), with a relative difference of the order of 1\%, and always less than 2\%. This is also the case for the model period spacing between two consecutive trapped modes $\Pi_T$ and the theoretical one (Fig.~\ref{fig:spacing_th_obs}, bottom panel), which shows a relative difference averaging between 3\%-7\%, and always less than 15\%.

\section{Summary and conclusions}
\label{cc}
The main goal of this paper is to compute the theoretical pulsation spectra for g-modes in our current models of sdB stars. We have at our disposal, computed with the STELUM package, static (parameterized) and evolutionary models. Static models have been developed to perform quantitative seismic modeling of sdB stars \citep[][and references therein]{2019A&A...632A..90C} and are in their fourth-generation development. Their main peculiarity in the context of this work is the decoupling between their chemical and thermal structures in the core: while they are assumed to be fully mixed in C, O, and He up to the core-mantle transition (the lq\_core parameter), their thermal structure is set independently. In 4G models, the thermal structure is set according to the Schwarzschild criterion, which can lead in some models to the presence of a radiative region (where the temperature gradient is radiative) just below the transition from the core to the mantle or to a ``splitting of the core in two convection zones'' (where we have a radiative region encapsulated between two convection zones in the core). To remedy this last situation, which is considered to be unphysical, we recently developed the 4G+ models, whereby the full region below lq\_core is imposed to be convective, namely, with a temperature gradient forced to be adiabatic. In our evolutionary models, this situation is handled by the appearance of a semiconvective partially mixed region, obtained by the interaction between convection, overshooting, and gravitational settling. 

We computed the g-mode pulsation spectra for the 4G and 4G+ models in Sect.~\ref{static} and for the evolutionary models in Sect.~\ref{evol}. We first showed (Fig.~\ref{fig:chem_transi_influence}) that the core-mantle transition region is of the prime importance on the computed spectra for the mid- and high-radial-order modes, which are the ones we observe in sdB stars. Meanwhile, the core-mantle transition is precisely the zone having the most uncertain chemical and thermal structures, with well distinct chemical profiles and temperature gradients in the 4G, 4G+, and evolutionary models. As a consequence, the pulsation spectra of the 4G, 4G+, and evolutionary models are also well distinct (see Figs.~\ref{fig:4G_hecore}, \ref{fig:4GP_hecore}, and \ref{fig:Evol_hecore} for the main ones). Overall, we have found three types of pulsation spectra for mid- and high-radial-order modes: completely flat spectra, as presented in terms of the period spacing as a function of the period; spectra with deep minima of the period spacing interposing between modes of more regular period spacing; and spectra showing more ``wavy'' patterns. The flat spectra correspond to the expectations from asymptotic theory, which is valid for radiative and chemically homogeneous (at least, without significant chemical gradients) propagation cavities. This is observed in a few instances, for example in the 4G+ models at high X(He)$_{\rm core}$. In the opposite case, when we have stronger chemical gradients in the middle of a radiative propagation cavity, we have the appearance of minima in the period spacing, which correspond to trapped modes. The trapping cavity varies: for spectra showing ``wavy'' patterns, we identified the trapping region to be the width of the core-mantle transition. In the 4G models, we observed deep minima in various instances: at high X(He)$_{\rm core}$, these modes are trapped in the radiative region just below the core-mantle transition at lq\_core (until this radiative part disappears, erased by the growth of the convective core), while at low X(He)$_{\rm core}$, the modes are trapped in the radiative arch region encapsulated between the two convective regions of the core, originating from the core splitting phenomenon (this trapping only appear in the case of massive enough cores). For small cores (lq\_core~$=-0.10$), we observed rather deep minima in spectra at low X(He)$_{\rm core}$. We showed that these modes are trapped in the width of the core-mantle transition at lq\_core, corresponding to an extreme case of ``wavy'' spectra. Finally, trapped modes are also present in evolutionary models and we showed in this case that modes are trapped in the overshooting region below the core-mantle transition at lq\_core before the onset of semiconvection and in the semiconvective region afterwards.

When comparing to observations, we also have various situations. As listed in the introduction, we have a few stars in which deep minima in the regular period spacing sequence have been formally identified: in the original {\sl Kepler} field, 3 stars exhibit such a g-mode pulsation pattern (KIC 10553698A, \citealt{2014A&A...569A..15O}; KIC 10001893, \citealt{2017MNRAS.472..700U}; and KIC
11558725, \citealt{2018MNRAS.474.4709K}), for 16 g-mode sdB pulsators in total. Asymptotic sequences have been reported in the 13 others (\citealt{2011MNRAS.414.2885R}), although it is hard to judge based on the figures presented in this paper if the pulsation spectra are flat for high-radial-order modes or if a wavy pattern is present. Deep minima associated to trapped modes have also been identified for one star observed by K2 (EPIC 211779126; \citealt{2017A&A...597A..95B}). Let us note that for K2 and TESS data, it is more difficult to formally identify trapped modes in series of observed modes with contiguous radial orders, given their overall lower quality compared to {\sl Kepler} original field data. \citet{2021A&A...651A.121U} nevertheless succeeded in deriving the mean period spacing for five g-mode sdB pulsators observed by TESS. They  gave  a series of figures that indicated the observed period spacings as a function of the periods (tentatively identified as $\ell=1$ or $\ell=2$). It seems apparent that for at least four stars out of five, the g-mode spectra exhibit a wavy pattern for mid- and high-radial orders, rather than a flat pulsation spectrum. Finally, in a very recent development, \citet{2024arXiv240717887S} analyzed the pulsation spectrum of TIC 441725813, a new hybrid pulsator with a rich g-mode component, with 25 frequencies that can be associated to $\ell=1$ g-modes, and 15 frequencies to $\ell=2$ g-modes. The observed diagramme-échelle of these frequencies (see their Fig.~11) shows obvious ridges along the asymptotic values for $\ell=1$ and $\ell=2$, which is the translation of the wavy patterns in period-period spacing diagrams such as the ones presented in this paper.  

To conclude, we encounter various situations for g-mode spectra in sdB stars, both in models and in observations. We showed that asteroseismology clearly has the potential to discriminate between these various situations, by constraining the thermal and chemical structure of sdB stars. This particularly concerns the region above the He-burning core, which is the most impactful on the g-modes of mid- and high-radial orders that are observed in sdB stars. We emphasize that He-burning cores and their transition to the radiative mantle are expected to be the same for all core-He burning stars, including red clump stars. The approach we developed over the years to perform quantitative seismic modeling of sdB stars makes use of static models, which are much more flexible than evolutionary models to explore thoroughly the parameter space to identify one (or several) optimal seismic model(s) that best fit the observed pulsation periods. All seismic modeling we carried out so far for g-mode sdB pulsators has been based on the 3G or 4G models (which are the same with respect to their core implementation). We will now also use the new 4G+ models presented in this paper to perform such quantitative modeling and explore different thermal structures for the core. The recent 3D developments point to a thermal and chemical structure for He-burning cores having a steep transition from the convective core and its convective overshooting layer to the radiative region, similarly to the structure implemented in the 4G+ models. We may be in position to test this with asteroseismology. Finally, we point out that a fifth generation (5G) of static models already exists and is currently tested intensively. These new models aimed at seismic probing of sdB stars implement a new approach (based on Akima spline representations) to construct a variety of chemical transitions from the fully convective core to the radiative mantle and are able to mimic the partial mixing of elements caused by classical semiconvection. This implementation has been successfully used for white dwarf asteroseismology (and was first developed in this context; \citealt{2018Natur.554...73G,2022FrASS...9.9045G}). With 4G, 4G+, and upcoming 5G models to perform quantitative seismic modeling, we hope to be able in the coming years to perform more efficient seismic modeling of sdB pulsators and determine which He-burning core description best reproduces the observed spectra.

\begin{acknowledgements}
We warmly thank the referee for their insightful comments that improved the manuscript. N.G. is a grant holder from FRIA du Fonds de la Recherche Scientifique - FNRS. V.V.G. is F.R.S.-FNRS Research Associate. S.C. acknowledge support from the Centre National d'Etudes Spatiales (CNES, France), focused on the missions {\sl Kepler} and TESS. M.F. is a Postdoctoral Researcher of the Fonds de la Recherche Scientifique – FNRS. This work was granted access to the HPC resources of CALMIP supercomputing center under the allocation 2024-P0205. This publication is supported by the Communauté française de Belgique in the context of a FRIA grant.
\end{acknowledgements}

\clearpage


\begin{appendix}

\end{appendix}

\end{document}